\documentclass[twocolumn]{aastex631}
\usepackage{enumitem}
\usepackage{amsmath}
\usepackage{derivative}
\usepackage{hyperref}
\defcitealias{rmc2009}{MCM09}

\newcommand{\mdot}{\dot{M}}
\newcommand{\Rp}{\mathrm{R}_P}
\newcommand{\ergs}{\mathrm{ergs}\ \mathrm{s}^{-1}\ \mathrm{cm}^{-2}}

\newcommand{\rxuv}{R_{\rm{XUV}}}
\submitjournal{APJ}

\begin{document}

\title{Wind-AE: A Fast, Open-source 1D Photoevaporation Code with Metal and Multi-frequency X-ray Capabilities}
\shorttitle{Photoevaporation with Metals \& X-rays}

\author[0000-0002-7520-5663]{Madelyn I. Broome}
\affiliation{University of California Santa Cruz,
1156 High Street, 
Santa Cruz, CA 95060, USA}
\email{mabroome@ucsc.edu}

\author[0000-0001-5061-0462]{Ruth Murray-Clay}
\affiliation{University of California Santa Cruz,
1156 High Street,
Santa Cruz, CA 95060, USA}

\author{John R. McCann}
\affiliation{University of California Santa Cruz, 1156 High Street, Santa Cruz, CA 95060, USA}

\author[0000-0002-4856-7837]{James E. Owen}
\affiliation{Imperial Astrophysics, Imperial College London,  South Kensington Campus, London SW7 2AZ, UK}
\affiliation{Department of Earth, Planetary, and Space Sciences, University of California, Los Angeles, CA 90095, USA}
 
\begin{abstract}
Throughout their lives, short period exoplanets ($<$100 days) experience X-ray and extreme-UV (XUV) stellar irradiation that can heat and photoionize planets' upper atmospheres, driving transonic outflows. This photoevaporative mass loss plays a role in both evolution and observed demographics; however, mass loss rates are not currently directly observable and can only be inferred from models.
To that end, we present an open-source fast 1D, XUV multi-frequency, multispecies, steady-state, hydrodynamic Parker Wind photoevaporation relaxation model based on Murray-Clay et al. (2009). The model can move smoothly between high and low flux regimes and accepts custom multi-frequency stellar spectra. While the inclusion of high-energy X-rays increases mass loss rates ($\dot{M}$), metals decrease $\dot{M}$, and the net result for a typical hot Jupiter is a similar $\dot{M}$, but a hotter, faster, and more gradually ionized wind. We find that mulitfrequency photons (e.g., 13.6-2000eV) are absorbed over a broader range of heights in the atmosphere resulting in a wind-launch radius, $\rxuv$, that is of order 10 nanobars for all but the highest surface gravity planets. Grids of H/He solar metallicity atmospheres reveal that, for typical hot Jupiters like HD 209458b, $\rxuv\approx1.1-1.8R_P$ for low-fluxes, meaning that the energy-limited mass loss rate, $\dot{M}_{\mathrm{Elim}}(R)$, computed at $R=R_P$ is a good approximation. However, for planets with low escape velocities, like many sub-Neptunes and super-Earths, $\rxuv$ can be $\gg R_P$, making it necessary to use $\dot{M}_{\mathrm{Elim}}(R=\rxuv)$ to avoid significantly underestimating mass loss rates. For both high escape velocities and large incident fluxes, radiative cooling is significant and energy-limited mass loss overestimates $\dot{M}$.
\end{abstract}


\section{Introduction} \label{sec:intro}

Close-in planets are highly irradiated by X-ray ($>100$ eV) and extreme UV (EUV, in this paper 13.6-100 eV) photons from their host stars. These high energy photons can ionize atoms in a planet's upper atmosphere (above the planet's optical transit radius) and heat the layer in the atmosphere where they are absorbed. This heating creates a pressure gradient which drives an outflow known as a Parker wind. 

These outflows are observable in transit \citep[for a review, see][]{dos_santos_review_2023}, but observational limitations make it difficult to directly map observations to mass loss rates \citep{schreyer_2024}. It is, therefore, necessary to model atmospheric escape in order to predict mass loss rates. Constraining mass loss rates not only allows for the prediction of observables, but is important in tracing the mass loss histories of exoplanets and understanding their present-day demographics \citep[e.g.,][]{eve_radius_gap_2021,rogers_2021,kubyshkina_2022,poppenhaeger_2021,rogers_2024}. 

However, most currently available models are time-dependent hydrodynamic codes, which are run for many sound-crossing times to reach a steady-state solution, at which point the mass-loss rates can be measured. This approach requires pre-computing large grids of mass-loss rates, which can then be interpolated onto evolutionary models \citep[e.g.,][]{owen_planetary_2012,kubyshkina_extending_2021,rogers_2021}. As model sophistication grows, so does the number of parameters required for any grid, meaning that it is not computationally feasible to use these time-dependent approaches to fully map out the parameter space before they are coupled to any evolutionary calculation. Alternatively, relaxation methods which directly solve for the steady state are much more computationally efficient \citep{rmc2009}, so much so that they can be directly coupled to an evolutionary calculation dynamically allowing more parameters to be included and, in principle, more complete physics in the mass loss model. 
In this paper, we present a relaxation code with a more complete description of the physics of atmospheric escape than was included in \citet{rmc2009}.

We take particular care to model the X-ray physics and metal-X-ray interactions as many exoplanets likely have super-solar atmospheric metallicities \citep{thorngren_massmetallicity_2016,kempton_metalrich_2023,kirk_jwst_2025} and the presence of metals in the upper atmosphere can allow X-rays to contribute significantly to the heating and ionizing of the wind \citep{garcia_munoz_physical_2007}. This is the result of two properties of X-rays: (1) X-ray photons are energetic enough that they can induce collisional secondary ionizations via the high-energy photoelectron released when the X-ray photon ionizes a species and (2) X-ray photons are energetic enough to ionize the K-shell (innermost) electron of certain metals common to exoplanet atmospheres (e.g., C, N, O, Ne, Mg, Si, and S), giving these metals a larger photoionization cross-sections at high energies relative to those of the more abundant H and He and allowing X-rays to be absorbed higher in the atmosphere where they can contribute to driving the wind. 

Treating these X-ray properties allows us to model smoothly across the high and low stellar XUV flux regimes without changing any of the assumptions in our model. The outflow is energy-limited and is predominantly driven by EUV photons in the low XUV flux limit \citep[e.g.,][]{garcia_munoz_physical_2007,lammer_origin_2014,erkaev_euv-driven_2016,owen+alvarez2016}. In the high XUV flux limit, the contribution of the X-rays is more significant, and the outflow and ionization is balanced by recombination at the base, leading this regime to be referred to as recombination-limited \citep{owen_planetary_2012,cecchi-pestellini_xrays,ates}. Since the ratio of the XUV to bolometric luminosity decreases more than three orders of magnitude over a star's lifetime \citep[e.g.,][]{jackson_coronal_2012,xuv_over_time,affolter_planetary_2023} and the ratio of EUV to X-ray luminosity also varies \citep[e.g.,][]{chadney_xuv-driven_2015,king_2021}, being able to model smoothly between the high and low flux regimes will allow our model to be used to model the evolution of mass loss rates over Gigayear timescales.

While mass loss has been observed around both low \citep[e.g.,][]{vidal-madjar_HD209_discovery,ehrenreich_GJ436b,zhang_four_2021} and high \citep[e.g.,][]{wasp12b_ca_mg,lecavelier_hd198733b_2010,edwards_lt9779b_202} XUV flux stars, observations do not provide model-independent mass loss rates. For example, for typical systems, Lyman-$\alpha$ \citep[e.g.,][]{vidal-madjar_HD209_discovery,ehrenreich_GJ436b} Doppler broadening cannot be used to directly infer mass loss rates because ISM absorption and geocoronal emission obfuscate the Lyman-$\alpha$ line-center, making it impossible to extract the outflow velocity below the sonic point \citep{owen2019_review}. Helium 10830\AA\ transits \citep[e.g.,][]{spake2018_he,more_He_detection} are similarly limited by the need for full non-LTE models of metastable helium to back mass loss rates out of the He 10830\AA\ transits \citep{spake2018_he,allan2019,linssen_spectral_2023,biassoni2024}. Metal absorption lines may be more direct proxies of the mass loss rate \citep[e.g.,][]{fossati_wasp12_mg_ca,yan_halpha_2022,huang_hydrodynamic_2023,sunbather}, but inferring the observability of these lines still requires a model that predicts the velocity and ionization fraction of metals in an outflow \citep[e.g.,][]{sunbather}. 

For these reasons, a photoionization-driven atmospheric escape model that includes X-ray physics and metals is a necessary tool for predicting mass loss rates. A variety of valuable 1D models for photoionization-driven escape exist and are explored in more detail in Appendix \ref{appendix:comparisons} \citep[e.g.,][]{yelle_aeronomy_2004,tian_2005,garcia_munoz_physical_2007,owen+wu2017,pwinds,ates,malsky_coupled_2020,koskinen_mass_2022,Spinelli2023,huang_hydrodynamic_2023,matthaus,kubyshkina_precise_2024}. Nevertheless, the ability to quickly forward-model the mass loss rates and outflow structures of multispecies planetary atmospheres irradiated by both high and low flux multi-frequency XUV stellar spectra is valuable for parameter studies \citep[this paper, ][]{loyd_hydrogen_2025}, predicting observables \citep{pai_asnodkar_pepsis_2024}, and modeling evolution. 

To that end, we present \texttt{Wind-AE}\footnote{Pronounced, ``windy". Stands for Wind Atmospheric Escape.}. \texttt{Wind-AE} is a fast 1D, steady-state, forward model for a photoionization-driven transonic Parker Wind based on the relaxation model from \citet{rmc2009}, with XUV multi-frequency and multispecies capabilities and self-consistent modeling of the upper atmosphere below the wind. 

This open-source code is based in \texttt{C} with a Python wrapper that ramps smoothly between solutions that span the range of planetary parameter space (mass, radius, semi-major axis), stellar spectral parameter space (stellar mass and radius, XUV flux, bolometric luminosity, spectrum) and metallicity space (metals, ionization states, and metallicity). The relaxation method reliably finds a solution to two-point boundary value problems but is very sensitive to the proximity of the initial guess to the goal solution in parameter space. The parameter space ramping algorithm allows us to negotiate this sensitivity by stepping strategically through parameter space in order to reach the goal solution. We have the ability to specify metals present, as well as metallicities, but do not include diffusion or drag---an appropriate assumption for metals whose masses are below the crossover mass, which holds for all models in this paper \citep[see][for a model with full diffusion capabilities]{matthaus}. 
Nevertheless, \texttt{Wind-AE} fills a niche not only with its speed, but with the inclusion of metals and full X-ray ionization physics, as well as the ability to customize stellar spectra and model both the high- and low-flux limit. 

In Section \ref{sec:methods} we give an overview of the methods we use to model multispecies and multi-frequency outflows. We then explore the impact of these additions on the outflow structure and mass loss rate of HD 209458b, a Neptune-like planet, and a mini-Neptune in Section \ref{sec:additions_results}.
Finally, in Section \ref{sec:results} we produce high and low flux mass loss grids and discuss the parameter-space limitations of our model in Section \ref{sec:discussion}.  

Additionally, in Appendix \ref{appendix:comparisons} we benchmark \texttt{Wind-AE} against existing 1D models for HD 209458b and GJ 1214b in the low EUV/XUV flux limit \citep{garcia_munoz_physical_2007,salz_simulating_2016,pwinds,ates,kubyshkina_precise_2024} and for a 1$M_J$, $1.7R_J$ planet and WASP 121b in the high XUV flux limit \citep{owen_planetary_2012,huang_hydrodynamic_2023}.

\section{Methods} \label{sec:methods}
We have built upon the 1D relaxation code presented in \citet{rmc2009} by adding multi-frequency and multispecies capabilities, as well as updating the way the lower boundary is treated. In \S\ref{methods:assumps} we introduce our model assumptions when solving the mass, momentum, and energy conservation and ionization balance equations (which are given in their generic species- and frequency-independent forms in \S\ref{methods:equations}). Heating and cooling terms are described in \S\ref{methods:equations:heatcool}. The physical and numerical impact of adding multi-frequency and multispecies capabilities are explored in \S\ref{methods:multi}, along with a more detailed discussion of our spectrum smoothing algorithm (\S\ref{methods:multi:spectrum}) and the X-ray physics (\S\ref{methods:multi:secondary_ion}) which motivate the multi-frequency and multispecies versions of the ionization equation and the photoionization heating rate in the energy equation (\S\ref{methods:multi:equations}).

\subsection{Species- and Frequency-independent Hydrodynamic Steady-state Parker Wind Equations} \label{methods:equations}
\begin{figure}[h]
    \centering
    \includegraphics[width=0.9\linewidth]{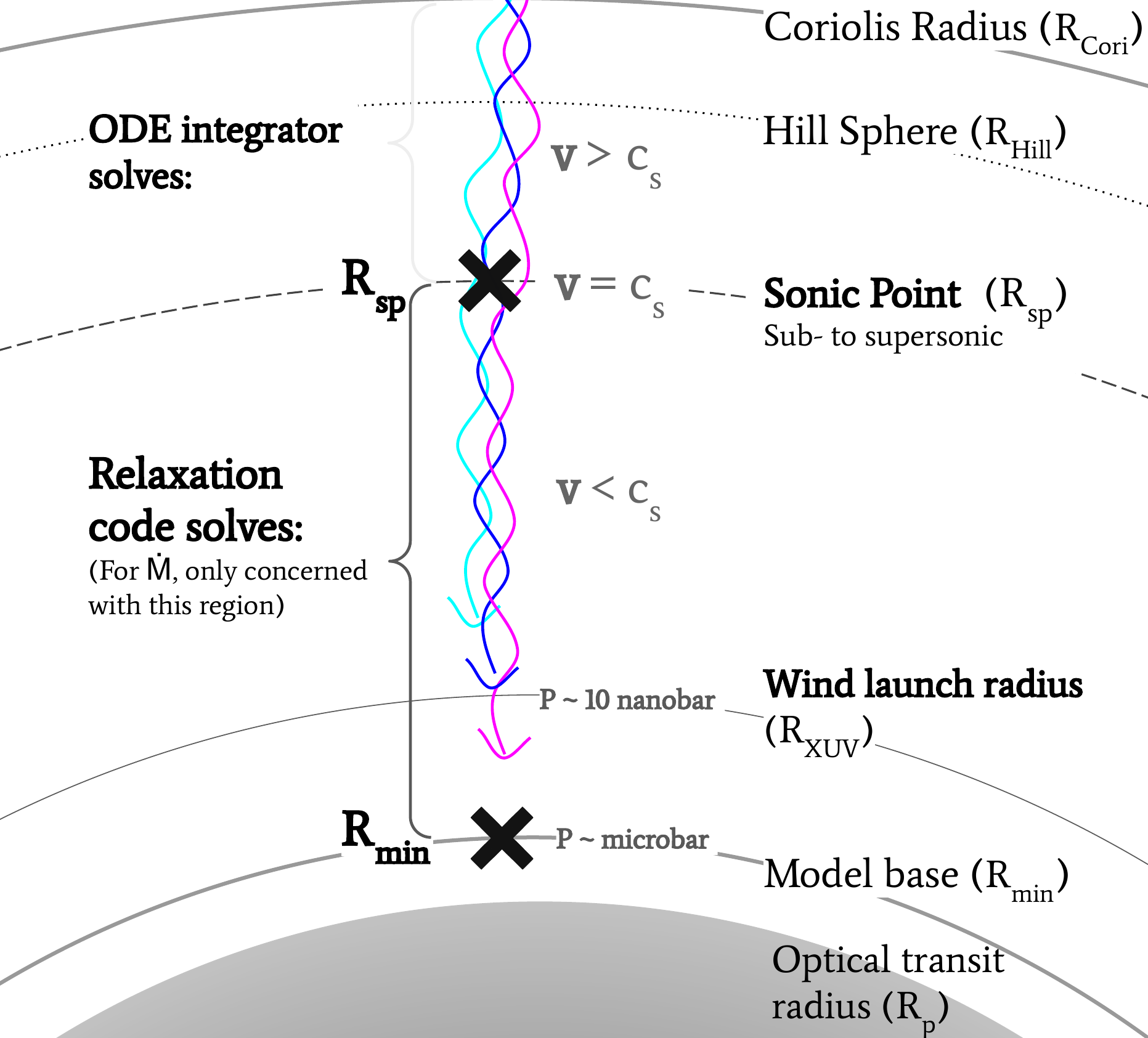}
    \caption{\textbf{Diagram of wind structure} - A thermally-driven, Parker-wind-like outflow is driven by photoionization heating, primarily deposited near the wind launch radius ($R(\tau_{\rm{XUV}}) = 1$).  Our relaxation code solves for the structure of this outflow by integrating between two boundary conditions, the minimum radius of the simulation ($R_{\rm{min}}$) and the sonic point ($R_{\rm{sp}}$),  identified by large black `x's. Shorter wavelengths of incident stellar irradiation, like x-rays, are represented by the magenta wave and penetrate deeper into the atmosphere than the longer wavelength dark purple (higher-energy EUV) and cyan (lower-energy EUV). Magnitude of the local velocity, $v$, relative to the local sound speed, $c_s$, is given in the middle column. Important planetary radii in the wind's structure are identified in text in the righthand column (gray semicircle, $R_P$; heavy solid, $R_{\rm{min}}$; light solid, $\rxuv$; dashed, $R_{\rm{sp}}$; dotted, $R_{\mathrm{Hill}}$; heavy solid, $R_{\mathrm{cori}}$).}
        \label{fig:structure}
\end{figure}

We use finite difference equations and \textit{Numerical Recipes}'s relaxation method solver \texttt{solvde} \citep{numericalrecipes} to solve the substellar 1D spherically-symmetric steady-state mass, momentum, energy, and ionization balance equations between upper and lower boundary points (Fig.\ \ref{fig:structure}). Given the critical point, the ``sonic point" ($R_{\rm{sp}}$), in the Parker wind transonic solution, the outflow structure becomes a two-point boundary problem, with the upper boundary being $R_{\rm{sp}}$ and the lower boundary ($R_{\rm{min}}$, typically $\sim$microbar; see Figure \ref{fig:structure}) being a point lower in the atmosphere than the wind launch radius. We define ``wind launch radius" ($\rxuv$, a.k.a., photoionization base) to be the lowest radial extent of substantial photoionization energy deposition (see Appendix \ref{appendix:bcs} for more details). It is, therefore, also the radius where the monoatomic gas starts to accelerate, driven by the pressure gradient generated by ionization heating\footnote{Operationally, we compute $\rxuv$ as the radial point at which photoionization heating begins to dominate over PdV cooling}.

Relaxation is an efficient method for solving two-point boundary value problems \citep{numericalrecipes} and it is sufficient to model just the relaxation region, $\rxuv< r < R_{\rm{sp}}$, in order to calculate mass loss rates. This is because photons absorbed above the sonic point do not contribute significantly to the heating of the atmosphere. The reasons for this are twofold: (1) past the sonic point the wind becomes very optically thin, meaning that it absorbs far fewer photons relative to the optically thicker region below the sonic point and (2) past the sonic point the flow is supersonic--losing causal contact with the gas below it--thus, the photons absorbed past the sonic point do not contribute to the heating and mass loss rate of the planet. We do find that computing self-consistent column densities at the sonic point for each atomic species does result in improved mass loss rate estimates, though, so it is recommended to integrate out to the Coriolis radius in order to compute those values. Unless otherwise specified, \texttt{Wind-AE} automatically does so as part of the ``polishing" step. Since our model is 1D and flow lines are assumed to be radial, the flow structure is only valid out to the Coriolis turning radius, $R_{\rm{cori}}$. We estimate $R_{\mathrm{cori}}$ as the radius where the outflow velocity, integrated starting at the sonic point and ignoring pressure acceleration, is deflected by one radian due to the Coriolis force. Past $R_{\mathrm{cori}}$, the velocity of the wind is predominantly set by the gravity of the star and secondarily by interactions with the stellar wind \citep{schreyer_2024}.  Our method cannot capture any flow and shock structures that result from 3D hydrodynamic effects \citep{McCann_morphology}.  In particular, our model will underestimate the density in the region where outflowing gas is shocked by its interaction with the stellar wind.  At yet larger radii, the flow is confined into a tube-like geometry often referred to as a ``tail" \citep{Owen2023,schreyer_2024}.  Due to this geometric confinement, the density of gas in the tail is larger than computed in our spherically-symmetric model.
See Appendix \ref{appendix:bcs} for more detail on boundary conditions.

Our 1D slice is placed along the line connecting the substellar point and the star - an approach which has been shown to give good agreement with mass loss rates from 3D simulations \citep[e.g.,][]{Owen2020}. Along this line, however, tidal gravity and stellar flux have the maximal impact on increasing the mass-loss rate, so, extrapolating the mass loss rate at that location to the rest of the planet's surface would overestimate the total mass loss. Thus, for the results in the main body of this paper, we multiply $\mdot$ by a generic reduction factor of 0.3 that encodes adjustments for spherical geometry and horizontal heat redistribution \citep{rmc2009}. The equation for surface-averaged mass loss rate then becomes $\dot{M}=0.3\cdot4\pi R_{\rm{sp}}^2 \rho(R_{\rm{sp}}) v(R_{\rm{sp}})$, in agreement with 3D model results and similar to approaches where the incident flux is divided by 4. 

Mass continuity is given by
\begin{equation} \label{eq:mass_continuity}
    \frac{\partial}{\partial r}(r^2 \rho v)=0,
\end{equation}
where the gas density is $\rho$, gas velocity is $v$, and distance from the planet's center is $r$. This equation takes on no species or frequency dependence because we do not model drag, but rather assume a constant mass fraction for each species throughout the wind. Here, species refers to each unique element, including all of the ionization states of that element. We, therefore, assume that the species are co-moving in the outflow, so share the same velocity.
Momentum conservation in a frame rotating with the planet's orbital frequency is 
\begin{equation} \label{eq:mom_conserv}
    \rho v \pdv{v}{r} = - \pdv{P}{r} - \rho \nabla \phi.
\end{equation}
For the gravitational potential including stellar tides, $\phi$, we use the full form of 
\begin{equation} \label{eq:tidal_grav}
    \phi = -\frac{G M_p}{r} - \frac{GM_*}{r_*} - \frac{1}{2} \Omega^2r_{\perp}^2 ,
\end{equation}
where, for the substellar point, $r_* = a-r$ and $r_{\perp}=a\big(\frac{M_*}{M_*+M_p}\big) -r$, where $a$ is the semimajor axis and $r$ radius from the planet's center. Here, $G$ is the gravitational constant, $P$ is the gas pressure ($P=\mu k_B T/\rho$), $\Omega$ is the rotation rate of a frame centered on the center of mass of the star-planet system (i.e., the planet's orbital frequency), and $M_P$ and $M_*$ are the planet and stellar mass, respectively. Our tidal gravity term incorporates the transition to Roche lobe overflow, but our code is not designed to model Roche lobe overflow.

In its generic form, energy conservation is,
\begin{equation} \label{eq:energy_conserv}
    \rho v \pdv{}{r} \Bigg[\frac{kT}{\mu(\gamma-1)}\Bigg] = \frac{kTv}{\mu} \pdv{\rho}{r}+\Gamma +\Lambda.
\end{equation}
The left-hand side represents the change in the internal thermal energy of the fluid, where $k$ is the Boltzmann constant, $\gamma=5/3$ for a monatomic ideal gas, and $\mu$ is the mean molecular/atomic weight (Eq.\ \ref{eq:mutrans}) which changes self-consistently as a function of radius as the ionization fraction changes. On the right-hand side, the first term tracks $PdV$ cooling (work due to adiabatic expansion of the gas). The heating rate per volume, $\Gamma$, is due to radiative heating by bolometric stellar photons absorbed below the wind and to photoionization heating by XUV photons absorbed within the wind. Photoionization calculations now include primary and secondary ionizations and thus gain a dependence on species and frequency (\S\ref{methods:multi:equations}). The cooling rate per volume, $\Lambda$, now contains not only Lyman-$\alpha$ cooling as in \citet{rmc2009}, but also atomic metal line cooling (Appendix \ref{appendix:metal_line}) inside of the wind and radiative bolometric cooling below the wind. The multispecies and multi-frequency versions of these equations and the relevant assumptions are discussed in \S\ref{methods:equations:heatcool}.

Since much of the energy budget of the wind is set by photoionization, it is also necessary to solve for the ionization balance in the wind which is, generically,
\begin{equation} \label{eq:ion_eq_generic}
    \mathcal{I} = \mathcal{R} - \mathcal{A}.
\end{equation}
The photoionization rate, $\mathcal{I}$ is balanced by the two right-hand terms, which are the radiative recombination rate $\mathcal{R}$ (for which we adopt the uncoupled on-the-spot approximation, meaning that we do not consider the possibility that the resulting photons ionize other species \citep{friedrich_c2ray_2012}) and the rate at which ions are advected away, $\mathcal{A}$.  \citet{rmc2009} showed that collisional ionization is negligible for a pure-H hot Jupiter atmosphere and we find the same for all planets modeled in this paper.
This generic form is expanded into the species and frequency dependent form in \S\ref{methods:multi:equations}.

Using the species-dependent finite difference forms of these four equations (Eq. \ref{ap-eq:v}---\ref{ap-eq:Ys}), we are able to use the relaxation method to solve for the structure of a hydrodynamic steady-state Parker wind up to the sonic point:
\begin{enumerate}
    \item total mass density, $\rho(r)$                 
       (see Eq.\ (\ref{ap-eq:rho}))
    \item temperature, $T(r)$                                          (see Eq.\ \ref{ap-eq:T})
    \item velocity, $v(r)$                                          (see Eq.\ \ref{ap-eq:v})
    \item per-species neutral fraction, $\Psi_s(r)$            (see Eq.\ \ref{ap-eq:Ys})
    \item per-species column density, $N_{\mathrm{col},s}(r)$    (see Eq.\ \ref{ap-eq:Ncol})
\end{enumerate}
We track $N_{\mathrm{col}}$ since the neutral fraction is calculated separately for each species. Thus, we need to track the column density of individual species in order to compute the optical depth to photoionizing radiation, $\tau$ \citep{friedrich_c2ray_2012}.

To navigate the sensitivity of the relaxation method, we have created ramping algorithms that take a series of smaller, adaptive steps in parameter space and re-converge boundary conditions as needed. This allows us to smoothly ramp between solutions that may be too far apart in parameter space to converge to in a single jump. 

Our boundary conditions are the temperature, mass density, and per-species neutral fraction at $R_{\rm{min}}$ and the per-species column density at $R_{\rm{sp}}$. While we have the ability to set these BCs explicitly, unless otherwise indicated, in this paper, we post-facto compute per-species column density self consistently from the density between the sonic point and Coriolis radius and pre-compute the lower boundary conditions by assuming that the energy budget between the optical transit radius, $R_P$, and the wind-launch radius, $\rxuv$, (where $R_P<R_{\rm{min}}<\rxuv$) is dominated by a balance of bolometric heating and cooling and that the temperature structure is isothermal between $R_P$ and $R_{\rm{min}}$. These assumptions are only for the purposes of computing more physically-informed lower boundary conditions. We do not enforce an isotherm in the region below the wind ($R_{\rm{min}} < r < \rxuv $) and allow the balance between the bolometric heating and cooling, PdV cooling, and other cooling and heating terms to set the temperature structure therein. For detailed derivations of the mass density, temperature, and radius of the base of the simulation, see Appendix \ref{appendix:bcs}. Note that if the sonic point is outside of the exobase of the planet, the transonic Parker wind solution is not valid and a flag is raised in our model. These planets would not be undergoing mass-loss hydrodynamically and would switch to thermal mass-loss via Jeans escape. For the purpose of ensuring that all of the outflows in this paper are in the fluid regime, we estimate the exobase using the Knudsen number for hardbody collisions of hydrogen ($\sim10^{-15}$) which gives a very conservative estimate of the exobase because in an ionized flow, the Coulomb cross section Knudsen number would yield a higher altitude exobase. 

The ionization fraction, velocity and temperature information at $r>R_{\rm{sp}}$ can be of interest when inferring the observability or coupling to other atmospheric escape models, so we also included the ability to integrate the solution outward beyond the sonic point to the Coriolis radius, using Numerical Recipes' \texttt{odeint} and the Bulirsch-Stoer (\texttt{bsstep}) adaptive stepsize ODE integrator \citep{numericalrecipes} with tolerance 10$^{-13}$. 

Integrating outward to the self-consistently computed Coriolis radius, self-consistently converging the column density at the sonic point, and computing the lower boundary density, radius, and temperature, as well as adjusting the molecular-to-atomic-wind transition radius (\S\ref{methods:equations:heatcool}) constitute the process we call ``polishing".

\subsection{Model Assumptions} \label{methods:assumps}
When computing the ionization balance we include photoionization (including secondary ionization from collisions with photoelectrons), advection, and recombination. Since we are primarily concerned with the wind's laumch, we do not model charge exchange as it has a secondary effect on the net ionization where the wind is launching and requires a more expensive photochemical model \citep[e.g.,][]{garcia_munoz_physical_2007,huang_hydrodynamic_2023}. We also do not model drag or diffusion as these require expensive multifluid models \citep{matthaus}. Diffusion of atomic species within the wind should be negligible for species below the cross-over mass \citep{crossovermass} as postfacto calculations have confirmed all atomic species in this paper are. If the mass of a species is greater than the crossover mass, its upwards diffusion rate is slower than those of the lighter species and it will diffuse throughout the wind and experience drag. Species with masses less than the crossover mass, on the other hand, can be considered entrained in the outflowing gas. This is assumed for all species (neutral and ionized) in the models presented here. Thus, those elements present at the lower boundary of our model (typically 1 $\mu$bar) will maintain the same relative abundance throughout the whole upper atmosphere, though their ionization states will change. 


When computing heating and cooling we include photoionization heating, PdV, Lyman-$\alpha$, recombination cooling,  and, when oxygen and/or carbon species are included in the simulation, OI, OII, OIII, CII, and CIII line cooling as relevant. These terms represent the first-order heating and cooling sources for the planets in the parameter space we present here. Our model is fully customizable and other metals and ionization states can be added by users to explore the role of different coolants in the future. We do not model conductive, free-free/bremsstrahlung, or collisional cooling, but compute them post-facto to confirm their irrelevance to the cases we present here. For the puffy planets with low escape velocities and for planets in the high stellar XUV flux limit, conduction and free-free cooling can be a significant energy term; so, we reserve the implementation of free-free and conductive heating and cooling for future updates of \texttt{Wind-AE} and present in this paper only planets for which conduction is not significant ($<1\%$ of the energy budget).

\subsection{Multi-frequency \& Multispecies } \label{methods:multi}

The photoionization heating rate per unit volume, $\Gamma_{\rm{ion}}(r)$, and ionization rate per unit volume, $\mathcal{I}(r)$, both contain a dependence on frequency $\nu$ and species $s$ in the form of the photoionization cross-section $\sigma_s(\nu)$ and number of secondary ionization, $\eta_s(\nu)$. 
Broadly speaking, the photoionization cross-sections, $\sigma_{s}(\nu)$, for most species, $s$, are maximal at the species's ionization edge and decrease  with increasing energy. Thus, the optical depth $\tau_{s}(\nu)=N_{\mathrm{col},s}\sigma_{s}(\nu)=1$ surfaces for a species like HI will occur deeper in the atmosphere for higher-frequency photons. 

However, the picture is not always so simple. X-rays not only have the ability to ionize more than one species per photon (see \S\ref{methods:multi:secondary_ion} for a discussion of secondary ionizations), and certain species, such as C, N, O, Ne, Mg, Si, and S experience a spike in $\sigma_s$ at high energies (e.g., 120 eV for C). For these species, X-rays above that energy threshold can ionize the innermost, K-shell electron in the atom. The first five of these seven species are predicted to be among the most abundant species in exoplanet atmospheres \citep[e.g.,][]{kempton_jwst_review}. Add to this the findings of \citet{thorngren_massmetallicity_2016}, which suggest that most exoplanets should have super-solar metallicities, and the interactions between X-rays and metals become essential to accurately modeling where photoionization heating begins to dominate and the wind launches.  We reserve a full discussion of metals and high metallicity physics for a forthcoming paper.

The density and pressure at which the wind launches has a significant impact on mass loss rates and on the temperature and velocity of the wind, so accurate modeling of the photoionization base is one of the benefits of multi-frequency, multispecies modeling. 

\begin{figure}
    \centering
    \includegraphics[width=\linewidth]{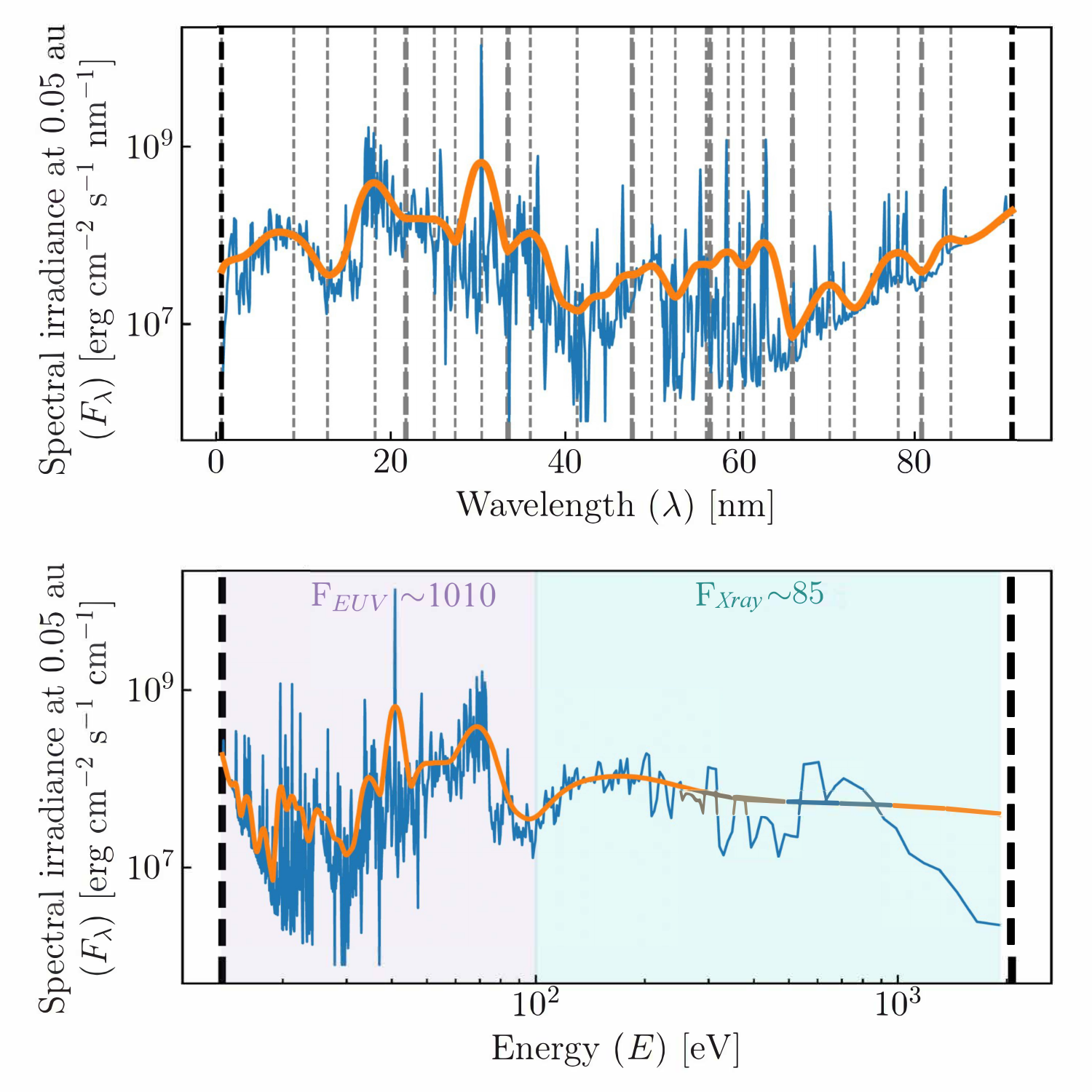}
    \caption{\textbf{Savitzky-Golay smoothed and binned \texttt{FISM2} solar spectrum scaled to 0.05 au} - Flux at 0.05 au vs. wavelength in nm (top) and vs. energy in eV (bottom). The solid purple and cyan highlights correspond to the EUV (13.6-100eV) and X-ray ($>$100eV) portions of the spectrum, respectively, and the approximate fluxes of each portion are labeled at the top of the bottom plot in $\ergs$. Bin edges (heavy vertical dashed lines) are automatically set at ionization edges for species present in a given simulation for maximum accuracy in calculating ionization rates (here pure-H). Thin vertical dashed lines are the critical points in the smoothing (Appendix \ref{appendix:spectrum}). We crop our spectra in this investigation at 2000 eV because contributions from higher energies are negligible and most photons that high energy have $\tau_\nu=1$ surfaces below the base of the wind and do not contribute to driving the wind. The XUV smoothed spectrum for a pure-H planetary atmosphere (above) results in 59 wavelength bins and the EUV in 66.}
    \label{fig:spectrum}
\end{figure}

\subsubsection{Multi-frequency Spectrum} \label{methods:multi:spectrum}
In order to capture the most important features of a multi-frequency spectrum, while simultaneously lowering computational cost, we implement multi-frequency EUV photons by modeling a smoothed solar spectrum with smoothing bin edges located at the ionization energies of species present in the wind. We employ a custom Savitzky-Golay binning and smoothing algorithm that requires that the smoothed spectrum's normalized flux to be accurate to the flux of the high resolution spectrum at ionization energies and the relevant K-shell ionization energies of each species included in a simulation (Fig.\ \ref{fig:spectrum}). A Savitzky-Golay method is of particular use for fitting polynomials to a spectrum because it locally conserves ionizing energy since the peaks of the spectrum are smoothed and distributed locally. The resultant smoothed spectrum produces identical mass loss rates and outflow structures (Fig.\ \ref{ap-fig:hires}). Although the spectrum is dominated by lines, our primary concern in calculating the ionization balance and energetics of the outflow is the integral of the product of the flux and the cross-section. Since the cross-section varies smoothly with energy, it is acceptable to smooth the spectrum, provided that energy conservation is maintained.. See Appendix \ref{appendix:spectrum} for a complete discussion of the smoothing algorithm algorithm.

Since XUV stellar spectra are not available for many stars and instrumental observational limitations mean that many ``full" stellar spectra are reconstructions, our default spectrum is a flux-scaled version of the \texttt{FISM2} \citep{fism2} solar spectrum, though it is possible to implement unique stellar spectra in our model. 

For all of the results presented in this paper we use a scaled solar spectrum. Many atmospheric escape models employ scaled solar spectra to simulate the spectra of stars of a similar type \citep[e.g.,][]{garcia_munoz_physical_2007,salz_tpci_2015,koskinen_mass_2022,huang_hydrodynamic_2023,kubyshkina_precise_2024} and it has been shown that the SED shape affects the upper atmosphere ionization structure and therefore the outflow structure \citep{guo_influence_2016,biassoni2024,kubyshkina_precise_2024}.  We find the same when we benchmark against existing 1D models (Appendix \ref{appendix:comparisons}) and the difference can be especially significant for an M-dwarf \citep{loyd_hydrogen_2025} vs. solar spectrum, as M-dwarfs have higher relative X-ray flux than FGK stars. We reserve an exploration of a the impacts of a highly-XUV-active star for a forthcoming investigation of HD 189733b, but do explore the high XUV-flux recombination limit modeled by \citet{owen_planetary_2012} in Appendix \ref{appendix:comparisons}.

For the remainder of this paper, when we refer to flux, we will use the following designations:
$F_*$ is the total bolometric flux, $F_{\rm{tot}}$ is a generic total flux over any high-energy spectral range,  $F_{\rm{XUV}}$ is the total flux over 13.6-2000 eV, and $F_{\rm{EUV}}$ is always the flux in the range 13.6-100 eV, all at the semi-major axis of the planet. These flux values may occasionally be normalized to different energy ranges and we will identify when we do so. For example, HD 209458's EUV flux is typically quoted in the literature as 450 $\ergs$, which is the total flux over 13.6-40 eV \citep[e.g.,][]{garcia_munoz_physical_2007,rmc2009,koskinen_escape_2013,salz_tpci_2015}. We use this value to normalize our scaled solar spectrum, which makes $F_{\rm{EUV}}$=1010 $\ergs$ and $F_{\rm{XUV}}=1095\ \ergs$.

\subsubsection{X-rays and Secondary Ionizations}  \label{methods:multi:secondary_ion}  
In the case of a star with low XUV flux (typical of an older star), X-rays penetrate deeply into the atmosphere and are absorbed at $\tau$(X-ray) = 1. The $\tau$(X-ray) = 1 surface is typically at pressures $>10^{-9}$ bar, a region which \citet{yelle_aeronomy_2004}, \citet{garcia_munoz_physical_2007}, and \citet{huang_hydrodynamic_2023} indicate is dominated by molecules. When low-flux X-rays fall in this region, the majority of the energy the X-rays deposit is radiated away by molecular line cooling \citep[e.g.,]{yelle_aeronomy_2004,garcia_munoz_physical_2007}. In that case, the wind is instead launched at $\tau$(EUV) = 1, which is higher in the potential well and at a lower density. These ``low flux" winds, therefore, tend to be predominantly EUV driven. 

In the case of a star with high XUV flux, the heat deposited by X-rays at $\tau$(X-ray) = 1 is significant enough to contribute to the dissociation of those molecules into atoms. These atoms cannot cool as efficiently as molecules and, as a result, the heat deposited by X-rays is no longer radiated away. Thus, the $\tau$(X-ray) = 1 layer---which is deeper in the potential well than the $\tau$(EUV) = 1 surface and also denser---is able to reach the temperatures necessary to launch a wind. The result is a denser wind and a higher mass loss rate. Thus, because young stars ($<$100 Myr old) are expected to have such high relative flux of ionizing XUV photons \citep[e.g.,][]{chadney_xuv-driven_2015,king_euv_2020}, X-rays are expected to be major contributors to the period of most photoevaporative significant mass loss for planets \citep[e.g.,][]{owen_planetary_2012,cecchi-pestellini_xrays,kubyshkina_2022}.

Properly modeling X-ray physics requires addressing the unique ionization properties of X-rays. First, in the low flux limit for typical planets, the ionization cross sections, $\sigma(\nu)$, for hydrogen and helium peak at their ionization energies (13.6 and 24.59 eV, respectively) and drop off  with frequency. Thus, for an H-He atmosphere, the $\tau(\nu)=1$ surface where the X-rays are absorbed is too deep in the atmosphere to contribute to heating and driving the wind. This follows from the definition of $\tau$, where $\tau(\nu)=\int_{\infty}^{r} (\sum_s n_{0,s}\sigma_s(\nu)) \mathrm{d}l$ and $n_{s,0}$ is the number density of the lowest ionization state of species, $s$, where $s$ need not be a neutral atom in the case of our simulation. However, the decrease in $\sigma(\nu)$ with frequency is nonmonotonic for some metals (C, N, O, Mg, Si, and S), giving them an outsized impact on the optical depth of at high frequencies despite their much lower relatively abundance (Appedix Fig. \ref{ap-fig:sigma}).

These species' innermost K-shell electron can be ionized by X-ray photons with energies as low as 124 eV \citep{kshell} (Appendix Fig.\ \ref{ap-fig:sigma}), resulting in these metals having comparable EUV cross sections to H and He, but a much larger X-ray ionization cross sections. Several of those six species are among the most abundant species predicted in exoplanet atmospheres and even in the relatively small abundances of a 1$\times$solar metallicity atmosphere \citep{asplund_chemicalsun_2009,penzlin_formation_2024}, the K-shell ionization cross section of metals weighted by abundance can be an order of magnitude higher than the abundance-weighted hydrogen ionization cross section. It is possible, then, that we may be missing the heating/ionizing contribution of X-rays to planetary outflows--even in the low flux limit--if we do not take into account metal opacities \citep[e.g.,][]{cecchi-pestellini_xrays,ates}.

Further complicating the X-rays picture, as seen in \citet{gillet_self-consistent_2023}, highly energetic X-rays have the ability to ionize more than one atom/ion, which changes the energy and ionization budget throughout the wind depending on the local fraction of the gas that is already ionized \citep[e.g.,][]{habing1971,xray_secondary,dalgarno_electron_1999}. On the whole, when X-rays are absorbed in the atomic upper atmosphere they are a significant source of flux which may be able to contribute to higher mass loss rates.

At the same time, metals contribute to a higher mean atomic weights, which, when coupled with metal line cooling, may result in lower mass loss rates than a pure-H model. This effect has been seen in disk photoevaporation models \citep{ercolano} and some atmospheric escape models \citep[][]{huang_hydrodynamic_2023,sunbather}. Since X-rays and metals are predicted to have opposite effects, it becomes necessary to model the two together to understand the net effect of including both metals and X-rays on mass loss rate.

Carefully tracing the distribution of the energy of incident photons is important for capturing the contributions of X-rays and metals.
First, we must determine what fraction, $\epsilon_{s,\nu}$, of the incident photons at each frequency, $\nu$, is absorbed directly by each species, $s$. We refer to the photon energy minus the energy of the initial ionization as $E_{0,s}$. Then, for X-ray and high energy EUV photons, we must determine what fraction of the energy, $E_{0,s}$, carried by the photoelectron released during the initial ionization of species $s$ will contribute to heating the gas ($f_{\mathrm{heat}}$), what fraction ($f_{\mathrm{ion,tot}}$) will contribute to collisionally ionizing other species in the wind, and what fraction will collisionally excite H and be released as Lyman-$\alpha$ radiation ($f_{\mathrm{excite}}$) \citep{xray_secondary}.

Starting with the fraction of the incident stellar XUV flux at frequency, $\nu$, that is initially absorbed by a species $s$, then, $\epsilon_{s,\nu} = n_{0,s} \sigma_{s,\nu}/ \sum\limits_{s}n_{0,s} \sigma_{s,\nu}$, where the $\nu$ subscript indicates a dependence on frequency, $n_{\mathrm{col},s}$ is the volumetric number density of a given species, and $\sigma_{s,\nu}$ is the species' photoionization cross section \citep{Osterbrock}. This relation is valid in both optically thin and optically thick regimes \citep{Osterbrock}.
Note that the X-ray photoionization cross sections for certain species (C, N, O, Mg, Si, S) are an order of magnitude larger than the species' EUV ionization cross sections because, for those species, even soft X-rays are able ionize the K-shell electron. In those cases, the electron is removed from the innermost shell rather than the outermost and the ionizing energy required to do so is on order of hundreds of eV. 
K-shell X-ray photoionization cross sections are accounted for using the analytic approximations from \citet{kshell}. Other photoionization cross sections are computed from the \citet{verner-recombo} database's coefficients and analytic fits.

X-rays further complicate ionization physics by allowing for secondary ionizations by energetic photoelectrons. While it is sufficient to assume that EUV photons are able to ionize only one species, more energetic X-ray photons can carry keVs of energy. For example, in the case of a 200 eV X-ray photon ionizing HI, the primary photoelectron released during that initial ionization of HI will have energy $E_{0,\mathrm{HI}} = 200-13.6$ eV = 186.4 eV. Hence, the primary photoelectron has enough energy to ionize several more species in the atmosphere. Thus, each X-ray photon can yield one primary and multiple secondary ionizations.

The number of secondary ionizations, $\eta$, that a species $s$ experiences is parameterized using the prescription for a H-dominated atmosphere from \citet{xray_secondary}, with updated coefficients from \citet{dalgarno_electron_1999}. We opt for this parameterization rather than a constant $\eta$ and heating efficiency, because where the photoelectron energy is distributed--whether into secondary ionizations or into heat--is a function the local ionization fraction, $\chi(r)$ \citep{xray_secondary,gillet_self-consistent_2023} and, therefore, will have significant impact on the wind structure. We take $\chi(r)$ to be the total ionization fraction ($\sum_s n_{\rm{ion},s}(r) / \sum_s n_s (r)$) instead of the hydrogen ionization fraction quoted in \citet{xray_secondary} ($n_{\rm{HII}}(r)/n_{\rm{H}}(r)$) and find the difference to be negligible even in high metallicity cases (though cases where H is no longer the dominant species will require further investigation). This approach is also taken by \citet{guo_influence_2016}.
While more sophisticated prescriptions for the distribution of primary photoelectron energy exist \citep[e.g.,][]{salz_tpci_2015,cecchi-pestellini_stellar_2006}, these require radiative transfer calculations or call on CLOUDY \citep{CLOUDY} adding computational expense. We compare our wind structure using the \citet{xray_secondary} and \citet{dalgarno_electron_1999} prescription with the results of radiative-transfer based escape models in Appendix \ref{appendix:comparisons}.

If the local background ionization fraction, $\chi(r)$, is high---meaning the most abundant species is largely ionized---a larger fraction, $f_{\mathrm{heat}}$(r), of $E_{0}$ will go into heating the gas and a smaller fraction, $f_{\mathrm{ion}}$, will go into ionizing the gas.
This distribution of energy is the result of elastic collisions between the photoelectrons and the high number of thermal electrons. These collisions transfer most of $E_0$ to the thermal electrons, resulting in the heating of the gas. The opposite relation holds true if $\chi$ is low: the energetic photoelectron is more likely to encounter a neutral atom, $f_{\mathrm{ion}}$ is higher, and $f_{\mathrm{heat}}$ is lower. 

\citet{xray_secondary}'s empirical equations for $f_{\mathrm{ion}}$ and $f_{\mathrm{heat}}$ as a function of $\chi$ are good approximations for X-ray stellar photons with energies $>$ 100eV. The effect on the wind structure of lowering that energy floor to, e.g., 40 eV, is minimal. Above 40 eV, we model secondary ionizations. Below 40 eV, we do not model secondary ionizations and instead assume $E_0$ goes entirely into heating.

\subsubsection{Species-dependent Versions of Ionization and Heating Equations} \label{methods:multi:equations}
Given the effect of secondary ionizations and heating when multiple species are present, we use a species-dependent version of Equation \ref{eq:ion_eq_generic} for each species $s$:
\begin{align} 
\mathcal{I}_s(r) &= \mathcal{R}_s (r) - \mathcal{A}_s(r)\\
     &= n_{\mathrm{ion},s}(r) n_e (r)\alpha_{\text{rec},s}(r) + \frac{1}{r^2}\frac{\partial}{\partial r}(r^2 n_{\mathrm{ion},s} v) \label{eq:ion_eq_species}
\end{align}
where  $n_{\mathrm{ion},s}$ and $n_e$ are the number densities of the higher ionization state of species $s$ and \textit{total} electron number densities, respectively. We take $n_e=\sum_s (n_{\mathrm{ion},s} \zeta_s+n_{0,s}(\zeta_s-1))$, where $\zeta_s$ is the ionization number or the number of electrons removed from species $s$. The total number density is $n_{tot,s}=n_{0,s}+n_{\mathrm{ion},s}$, where $n_{0,s}$ is the number density of the lowest state of species $s$. 

The advection term, $r^{-2}\partial(r^2 n_{\rm{ion},s}v)/\partial r$, in Equation \ref{eq:ion_eq_species} for convenience can be written $-n_{tot,s} v\frac{\partial \Psi_{s}}{\partial r}$ by continuity, where $\Psi_{s}=n_{0,s} / n_{tot,s}$ is the number fraction of species $s$ in the lowest state (a.k.a, the ``neutral" fraction). This is the form we adopt in \texttt{Wind-AE}.
The recombination coefficient, $\alpha_{\text{rec},s}$ is temperature dependent and calculated using the recombination coefficient algorithm from CLOUDY \citep{CLOUDY}.

We define $s$ as the species that is ionized either directly by a stellar photon or secondarily when a stellar photon of energy $E_{\nu}$ first ionizes species $j$ and that ionization releases a primary photoelectron carrying energy $E_{0,j}=E_\nu - \mathrm{I}_j$, a portion of which goes into collisionally ionizing species $s$.  Our updated local ionization rate per unit volume equation for species $s$ irradiated by an SED with the frequency range [$\nu_{\rm{min}},\nu_{max}$] becomes, 
\begin{equation} \label{eq:ion_rate}
    \mathcal{I}_{s}(r) \equiv \sum_{\nu=\nu_{\rm{min}}}^{\nu_{max}}\left(\mathcal{I}_{1,s,\nu}(r) + \mathcal{I}_{2,s,\nu}(r)\right)\\
\end{equation}
where $\mathcal{I}_{1,s,\nu}$ is the primary ionization rate per unit volume as a function of radius and $\mathcal{I}_{2,s,\nu}$ is the secondary, with
\begin{align} 
     \mathcal{I}_{1,s,\nu}(r) &=
    \epsilon_{s,\nu}(r)  \Phi_\nu e^{-\tau_{\nu}(r)} \sigma_{s,\nu} n_{0,s}(r) \\
    \label{eq:secondary}
    \mathcal{I}_{2,s,\nu}(r) &= 
    \sum\limits_{j=1}^{N_{\mathrm{species}}}\mathcal{I}_{1,j,\nu}(r)\eta_{s,j,\nu}(r)
\end{align}
The photon flux per spectrum frequency bin is $\Phi_\nu$ (in units of $cm^{-2}$ $s^{-1}$), $\tau_\nu$ is the total optical depth for all species, and $\sigma_{s,\nu}$ is the ionization cross section as a function of frequency and species.

The number of secondary ionizations that species $s$ experiences when impacted by a photoelectron emitted from species $j$ is 
\begin{equation} \label{eq:eta}
    \eta_{s,j,\nu} = \frac{ E_{0,j}f_{\mathrm{ion},s}(r,E_{0,j})}{\mathrm{I}_s},
\end{equation}
where $\mathrm{I}_s$ is the ionization energy of species $s$. The fractional distribution of photoelectron energy, $E_{0,j}=E_\nu -\mathrm{I}_j$, when $E_{0,j}>40$eV into heat, hydrogen excitations, and secondary ionizations is given by
\begin{align*}
    &f_{\rm{heat}}(r) = 0.9971\left[1-\left(1-\mathcal{X}(r)^{0.2663}\right)^{1.3163}\right]
    \\&f_{\rm{excite,H}}(r) = 0.4766\left(1-\chi(r)^{0.2735}\right)^{1.5221}
    \\&f_{\rm{ion,s}}(r,E_{0,j})=(1-f_{\rm{heat}}-f_{\rm{excite,H}})\frac{ n_{0,s}\sigma_{\rm{col},s}(E_{0,j})}{\sum_m[n_{0,m}\sigma_{\rm{col},m}(E_{0,j})]}
\end{align*}
where $\sigma_{\rm{col}}$ is the collisional ionization cross section \citep{dere_ionization_2007}. This is equivalent to $f_{\rm{ion,s}}=n_{0,s}R_s(E_{0,j})/\sum_m[n_{0,m}R_m(E_{0,j})]$, so, for numerical speed, we use the spline tables for $R_s$, the secondary ionization rate coefficients, from \citet{dere_ionization_2007}. The above equation follows from the primary photonelectron energy fraction given by \citet{xray_secondary} and ionization rate coefficients given by \citet{dere_ionization_2007}. This form allows us to easily add secondary ionizations to our ionization rates by tracing the number of secondary ionizations each photoelectron released produces.
Recall that when $E_{0,j}<40$eV, $f_{\rm{heat}}=1$ and $f_{\rm{ion,s}}$ and $f_{\rm{excite,H}}$ are 0.  

The photoionization heating term for the energy conservation equation (Eq. \ref{eq:energy_conserv}) then becomes,
\begin{multline}
\label{eq:heating}
    \Gamma_{\mathrm{ion}}(r) = \sum_s \sum_{\nu=\nu_{\rm{min}}}^{\nu_{\rm{max}}} \mathcal{F}_{s,\nu}(r)e^{-\tau_\nu(r)} \sigma_{s,\nu} n_{0,s}(r) ,
\end{multline}
where, $\mathcal{F}_{s,\nu}(r)$ is the energy flux of photoelectrons that goes into heating the gas and
\begin{equation}
    \mathcal{F}_{s,\nu}(r) =   (E_{\nu}-I_s) \left[\epsilon_{s,\nu}(r) \Phi_\nu \right] f_{\rm{heat}}(r).
\end{equation}

\begin{figure}
    \centering
    \includegraphics[width=\linewidth]{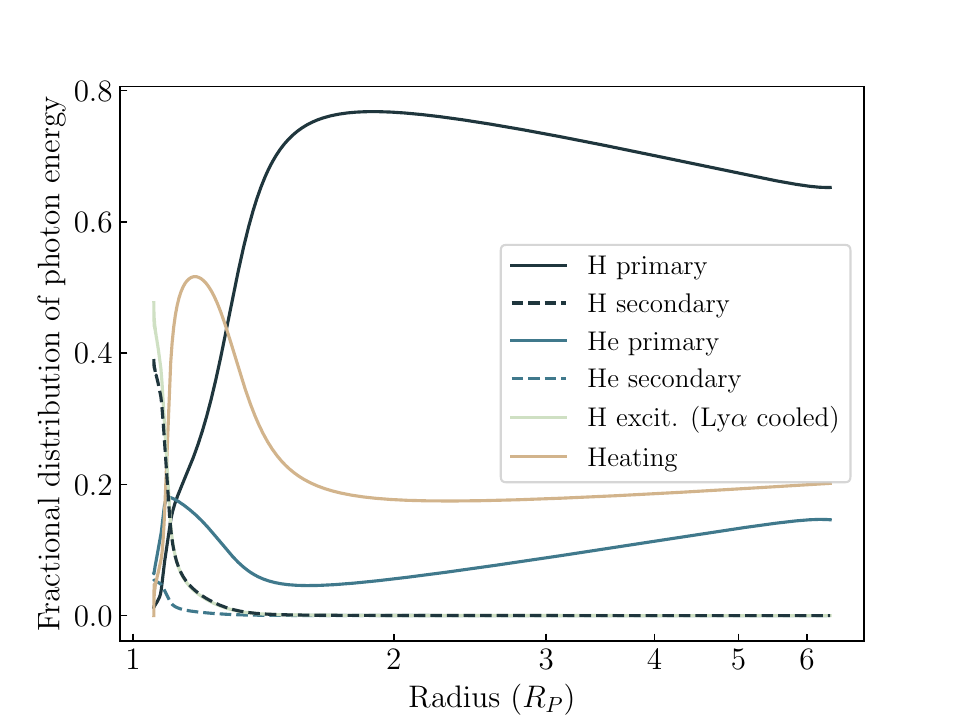}
    \caption{\textbf{Energy Deposition Fraction for HD 209458b} - Fraction of total incident stellar energy into ionizing hydrogen (black), helium (navy), heating (tan), and hydrogen excitation (light green) of which, in our model, 100$\%$ is assumed to escape as Lyman-$\alpha$ radiation. Total XUV flux over 13.6-2000 eV is 1095 $\ergs$.}
    \label{fig:energy_frac}
\end{figure}

Excited H releases a Lyman-$\alpha$ photon which we assume escapes, resulting in radiative cooling. We find that, for HD 209458b for an H, He atmosphere that these excitations account for a total of 14$\%$ of the total incident energy (Fig.\ \ref{fig:energy_frac}). Monte-Carlo radiative transfer calculations confirm that a little more than half of Lyman-$\alpha$ photons generated by H excitations due to collisions with non-thermal electrons escape upward to space and more than 99\% leave the simulation domain through either the upper or lower boundary without being thermalized (see Appendix \ref{appendix:lyaesc}). 
We expect Lyman-$\alpha$ photons that exit through the lower boundary to be effectively lost and not to contribute to heating the outflow.  Nevertheless, we conduct a sensitivity test by modeling HD 209458b with 0$\%$ of non-thermally-excited Lyman-$\alpha$ photons escaping and 100$\%$ converted into heat. We find that effect on the mass loss rate and structure of the planet is negligible.

The \citet{xray_secondary} approximation for the secondary ionization rate of metals is not expected to be appropriate for high metallicities ($>50 Z$) and future iterations of this model will include appropriate high metallicity physics. However, for low metallicities, we are able to model any species for which photoionization cross sections and collisional ionization rate coefficients as a function of frequency are available.

\subsection{Heating \& Cooling Terms} \label{methods:equations:heatcool}
Below the photoionization base, the atmosphere is molecular and the temperature structure can often be approximated by an isotherm at the skin temperature \citep{guillot2010}. Often, this region is treated as having a constant isothermal base temperature inferred from equilibrium temperature models (which is an oversimplification; \citealt{parmentier_irradiated_2014,parmentier_irradiated2_2015}) or estimates for similar planets and this temperature is used as a lower boundary condition at some radius higher than $R_P$. More sophisticated and expensive models \citep[e.g.,][]{huang_hydrodynamic_2023} perform full lower atmosphere photochemistry calculations to obtain more physical and accurate temperature and density structure and we compare to such models in Appendix \ref{appendix:comparisons}.

We elect for an approximation somewhere between the two approaches. We directly compute the skin temperature (Eq.\ \ref{ap-eq:tskin}) as the balance between bolometric heating and cooling and use the assumption of an isotherm and a constant mean molecular weight to compute the microbar radius as our simulation base given a measured optical transit radius (see Appendix \ref{appendix:bcs}). Because we do not treat molecules or photochemistry in our model, we set a constant mean molecular weight,\footnote{For the results shown in this work $\mu_{\rm{mol}}=2.3m_H$ which is the mean molecular weight of H$_2$ and He in solar abundances.} $\mu_{mol}(r)$, for all points below the wind and force $\mu$ to transition to the numerically-solved mean atomic weight, $\mu(r)$. We do so smoothly, by multiplying $\mu_{mol}$ by a complementary error function normalized to 1 (Eq.\ \ref{ap-eq:erfc}) that drops off as the wind becomes optically thin (Appendix Fig.\ \ref{ap-fig:mu}), the molecules photodissociate, and ionization heating begins to dominate, launching the wind. As a result, we can write $\mu(r)$ in units of $g$ for all $r$ in the simulation as
\begin{align}\label{eq:mutrans}
\mu(r) &= \mu_{\rm{mol}}(r) \mathrm{erfc^\prime}(x) + \mu_{\rm{atom}}(r)[1-\mathrm{erfc^\prime}(x)] \\
    &=2.3 m_H\mathrm{erfc^\prime}(x) + \frac{m_H\cdot \left[1-\mathrm{erfc^\prime}(x)\right]}{\sum_s Z_s \frac{m_H}{m_s}(2-\Psi_s(r))},
\end{align}
where the mean molecular weight of the species explicitly modeled in our outflow, $\mu_{\rm atom}(r)$, includes electrons, ions, and neutral atoms and hence changes as a function of radius as the gas ionization state evolves.
As we currently do not model multiple ionization states per atom, we can simplify the traditional definition of $\mu_{\rm atom}(r)$ to be in terms of $\Psi_s$, the ``neutral" fraction (fraction of the species in the lowest ionization state). Here, $m_H$ is the atomic mass of hydrogen, $m_s$ is the atomic mass of species $s$ in the wind, and $Z_s$ is the mass fraction of that species. The complementary error function is discussed in more detail in Appendix \ref{appendix:bcs}.

We use the same error function to force the optical ($\kappa_{\rm{opt}}=0.004$) and IR opacities ($\kappa_{\rm{\rm{IR}}}=0.01$) \citep{guillot2010} in the bolometric heating (Eq. \ref{eq:boloheat}) and cooling equations (Eq. \ref{eq:bolocool}) to drop off at the same transition point as the molecular-to-atomic transition (Appendix \ref{appendix:bcs}). The result is that the bolometric heating and cooling that dominated below the wind give way to photoionization heating and atomic line cooling within the wind. This simplification does not affect hot Jupiters, but will likely be important for puffy sub-Neptunes (\S\ref{discuss:highgrav}). Future work explicitly modeling the physics in this region is merited (for more on the impacts of lower atmosphere modeling, see, Appendix Figures \ref{fig:huang} \& \ref{ap-fig:gillet})

Above the photoionization base ($\sim10$ nanobars), the wind is generally atomic and the energy budget (Eq. \ref{eq:energy_conserv}) is set by a balance of photoionization heating, advective heating, cooling due to $PdV$ work (gas expansion), recombination cooling, Lyman-$\alpha$ cooling, and metal line cooling. For the planets in this paper, as modeled in other approaches \citep[e.g.,][]{kubyshkina_young_2018,ates,koskinen_mass_2022,sunbather,huang_hydrodynamic_2023}, it is reasonable to assume that this is optically-thin line cooling, but other regions of parameter space where this may not be an appropriate assumption merit further investigation. Also, most likely, H$_{3+}$ cooling (\citealt{yelle_aeronomy_2004}; \citealt{garcia_munoz_physical_2007}) and other molecules \citep{yoshida_molecules_2022,yoshida_molecule_2024} also play a significant role in cooling in the molecular region below the wind, absorbing and radiating away the highest energy photons of the XUV irradiation. 

The cooling term, however, is now $\Lambda(r)=\Lambda_{\rm{Ly}\alpha}(r)+\Lambda_{\rm metal}(r)+\Lambda_{\rm{bolo}}$(r), where 
\begin{equation}\label{eq:lya_cool}
\Lambda_{\mathrm{Ly}\alpha}(r) = -7.5\times 10^{-19} n_e n_{\mathrm{HI}} e^{-11834K/T}
\end{equation}
and $n_e$ is the number density of total electrons, $n_{\mathrm{HI}}$ is the number density of neutral H, $T$ is temperature in Kelvin, and all vary as a function of radius. $\Lambda_{Ly\alpha}$ is radiative cooling from the Lyman-$\alpha$ line of atomic hydrogen, $\Lambda_{\rm metal}$ is radiative cooling from metal lines, and $\Lambda_{\rm{bolo}}$ is bolometric cooling from thermal emission in the molecular layer of the atmosphere, where we assume that radiative cooling can be treated with an average infrared opacity \citep{guillot2010}
rather than needing to be modeled line by line. All cooling rates are per unit volume and have units of ergs s$^{-1}$ cm$^{-3}$. In Appendix \ref{appendix:lyaesc} we verify that most of the Lyman-$\alpha$ photons escape the wind into space and the vast majority leave the domain of our simulation through either the upper or lower boundary before being thermalized, verifying that thermal Lyman-$\alpha$ cooling is well treated by Equation (\ref{eq:lya_cool}).

In the atomic outflow, when OII, OIII, CII, and/or CIII are present in the simulation, we have found the corresponding lines to be non-negligible coolants at high fluxes and \citet{sunbather} find Fe II and Ca II line cooling significant in high metallicity cases (which are not treated in this paper and thus these lines are neglected for the time being). There is some debate as to whether Mg is a net heater or coolant \citep{huang_189733b_2017,fossati_wasp12_mg_ca}, so here we choose to initially work with O and C, and our model can handle further line coolants to explore their role in the future.
Line cooling is generally most impactful at high fluxes. However, even at low fluxes, some Lyman-$\alpha$ photons are able to escape the outflow and cool the wind, making it the second most significant contribution to the cooling of the wind after $PdV$ cooling. 

A fully complete description of metal line cooling would require additionally tracking the energy levels of each metal ion. Such an approach would likely require at least tens of more fields per atom/ion, which would quickly become computationally infeasible. However, the spontaneous decay timescales of most energy levels that produce strong cooling in the optical/UV are extremely short, with only some metastable levels (e.g. OI/OIII 1D$_2$) approaching a few hundred seconds \citep[e.g.][]{Wiese1996}. These decay timescales are typically much shorter than the flow timescale; therefore, we can approximate that the level populations in an individual atom/ion are in \emph{local} collisional statistical equilibrium.  Furthermore, given the temperatures we find in our simulations are typically $\lesssim 10^4$~K we can make the nebular approximation \citep[e.g.][]{Osterbrock} and ignore excitations from excited permitted levels, along with recombinations to excited states for determining the level populations. This allows us to simplify the level populations of each ion/atom and model metal line cooling using a two-level atom model, where following \citet{Schulik2025}:
\begin{equation} \label{eq:metal_line_cool}
        \Lambda_{\textrm{metal}}(r) = -\sum_{s} n_e n_{\rm{ion},s} \left[ A_s\frac{ \exp\left(-\frac{T_{\rm{line},s}}{k_b T}\right)}{n_e \left(1+\frac{n_{c,s}}{n_e}\right)}\right],
\end{equation}
where $n_{\rm ion,s}$ is the number density of OII, OIII, CII, and/or CIII, and the constants $A_s$, $T_{\rm{line},s}$, and $n_{c,s}$ are derived from \citet{CLOUDY} and \texttt{CHIANTI} \citep{chianti1,chianti8}.
These constants are derived by fitting this two-level atom model to the cooling rates numerically calculated using the \texttt{CHIANTI} database in the temperature range 500-20,000~K. In the \texttt{CHIANTI} calculations we include all the energy levels and transitions; however, we still assume a statistical collisional equilibrium \emph{within} each atom/ion\footnote{We note that while \texttt{CHIANTI} is a collisional plasma code, since our outflow is photoionzation and advection dominated we only use \texttt{CHIANTI} to compute the level populations of individual atoms/ions in a collisional statistical equilibrium and switch off collisional ionziation terms, as appropriate within the framework of the nebular approximation.}. We find this approach accurately models the true cooling function to within an accuracy of a few percent. The values of the species-dependent constants $A_s$, $T_{line,s}$, and $n_{c,s}$ are listed in Appendix \ref{appendix:metal_line}, along with comparisons of our two-level model to the fully numerically computed cooling rate. We note that this is exactly the same approximations and approach as those made by previous work to determine Equation~\ref{eq:lya_cool} to model Lyman-$\alpha$ cooling. 

The final cooling term is the bolometric cooling, which is discussed in more detail in Appendix \ref{appendix:bcs}:
\begin{equation} \label{eq:bolocool}
    \Lambda_{\rm{bolo}}(r) = 2\sigma_{SB} T_{\rm{skin}}^4 \rho(r) \kappa_{\rm{IR}} \mathrm{erfc^\prime}(x).
\end{equation}
Here $\sigma_{SB}$ is the Stefan-Boltzmann constant, $T_{\rm{skin}}$ is the skin temperature which we take to be $T(R_{\rm{min}})$, and $\kappa_{\rm{IR}}$ is the IR opacity. In the atomic/ionized wind, line cooling dominates; however, in the molecular region between $R_{\rm{min}}$ and $\rxuv$, where it is optically thick to most XUV photons and, in the low flux limit, the highest energy X-ray photons are radiated away by molecular line cooling, our assumption that bolometric cooling dominates provides a reasonable approximation. If the photodissociation front extends beyond the wind launch radius and molecules survive into the wind, transitioning at $\rxuv$ as we do in this paper would no longer be appropriate. We explore the limits of these assumptions in the low escape velocity limit in \S\ref{discuss:lowgrav}, but for the majority of planets in this investigation we can safely assume that the molecules have photodissociated and/or thermally dissociated below $\rxuv$. To approximate this behavior, we multiply $\kappa_{\rm{IR}}$ and $\kappa_{\rm{opt}}$ (the optical opacity) by the same complementary error function (Eq.\ref{ap-eq:erfc}) as is used in Equation \ref{eq:mutrans}, which lowers these opacities to zero so that the bolometric cooling does not unphysically dominate in the atomic wind. This transition is visible at $\sim$1.1 $R_P$ in Figure \ref{fig:energy}. 

\label{sec:additions_results}
\begin{figure*}
    \centering
    \includegraphics[width=0.9\textwidth]{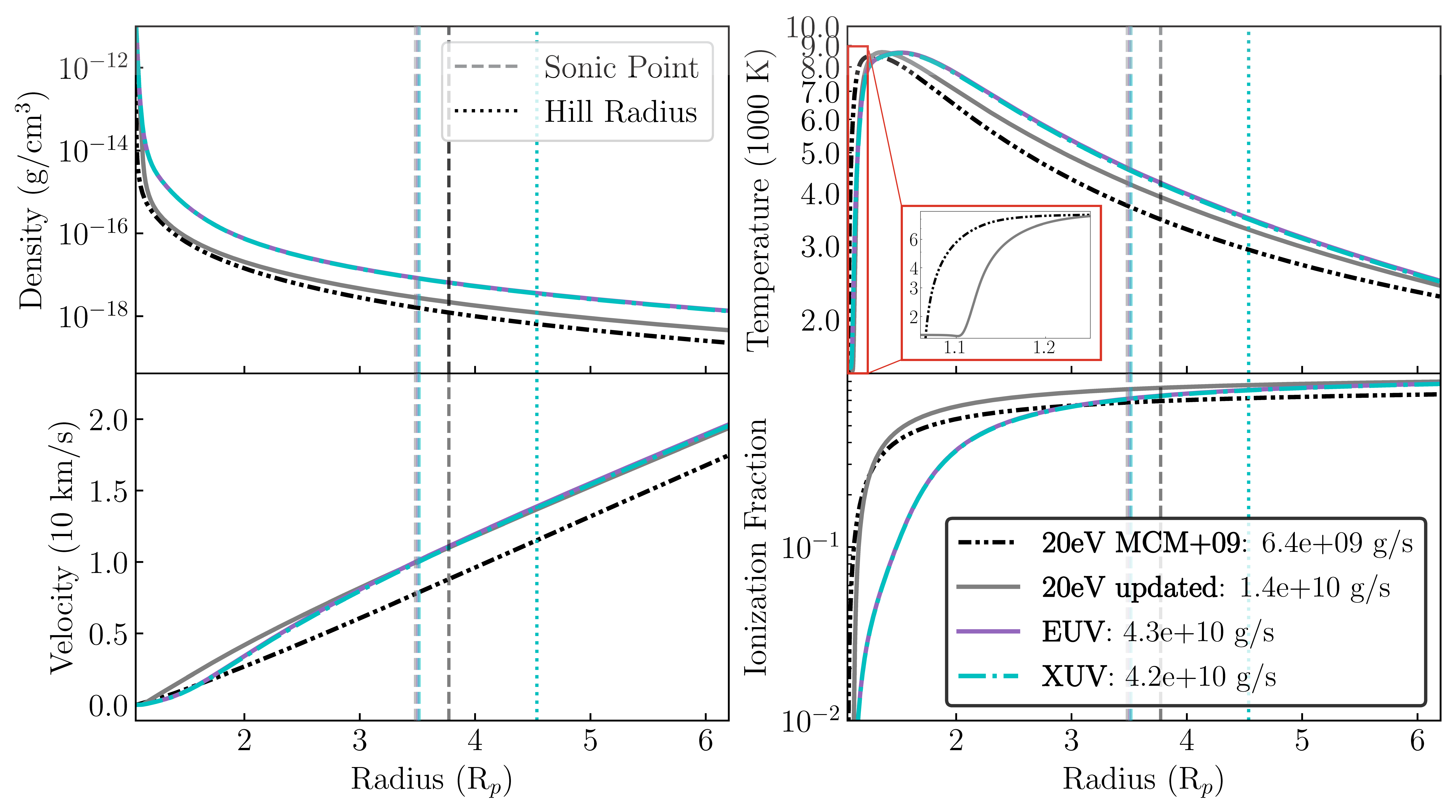}
    \caption{\textbf{Multi-frequency Profiles for a pure-H HD 209458 b} - Black dash-dotted is the original 20eV monofrequency and pure-H \citet{rmc2009} model ($R_{\rm{min}}=1.037\ R_P$, $\rho(R_{\rm{min}})=2.7\times10^{-11}$ g cm, T($R_{\rm{min}}$)=1000 K). Gray solid is also monofrequency 20eV, but with our updated physical lower BC and bolometric heating and cooling at the base ($R_{\rm{min}}$=1.057 $R_P$, $\rho(R_{\rm{min}})=1.8\times10^{-11}$ g cm$^{-3}$, T(R$_{\rm{min}}$)=1534 K). The remaining plots all use the updated BCs. Purple is the EUV multi-frequency (13.6-100 eV) version. Because no metals are present, X-rays (cyan, dash-dotted, XUV 13.6-2000 eV) contribute relatively little to the profiles or mass loss rates of a pure-H atmosphere so the solutions overlie the EUV. Stellar spectra in all simulations are normalized to 450 $\ergs$ between 13.6 and 40 eV (in keeping with \citet{rmc2009}). Our model is not valid past the Coriolis radius (upper limit of x-axis). The sonic point and Hill sphere are given by dashed and dotted lines respectively.}.
    \label{fig:multifreq}
\end{figure*}

The same is done for for the bolometric heating term, 
\begin{equation}\label{eq:boloheat}
    \Gamma_{\rm{bolo}}(r) = F_{*} \rho(r) \left(\kappa_{\rm{opt}}\mathrm{erfc^\prime}(x) + \frac{1}{4} \kappa_{\rm{IR}}\mathrm{erfc^\prime}(x)\right), \\
\end{equation}
where $F_*$ is the single-band bolometric flux which is an independent variable that is not dependent on the chosen SED shape or integrated XUV flux ($F_{\rm{tot}}$). In this double-gray approximation, the total heating then becomes $\Gamma(r)=\Gamma_{\mathrm{ion}}(r)+\Gamma_{\rm{bolo}}(r)$\footnote{Full derivation in Appendix \ref{appendix:bcs}.}. 
The photoionization heating term that dominates in the wind, $\Gamma_{\mathrm{ion}}$, is also now species-dependent and also incorporates the contribution to heating by highly energetic primary photoelectrons released during the ionization of each initial species by a high energy X-ray photon (Eq.\ \ref{eq:heating}). The species and frequency dependent $\Gamma_{\mathrm{ion}}$ equation is discussed in detail in \S\ref{methods:multi}.

\section{Results of Introducing Multi-frequency and Multispecies Assumptions} 
\subsection{Multispecies \& Multi-frequency} \label{results:multi}
Though metals are required to understand the true impact of X-rays on an outflow, for clarity we first consider the difference between a wind launched by a multi-frequency spectrum and a mono-frequency spectrum in a low-to-moderate flux pure-H atmosphere, illustrated in  Fig.~\ref{fig:multifreq}. In the following section, we build up incrementally from the results presented in \citet{rmc2009} to elucidate the individual contributions of multi-frequency X-rays, metals, and boundary conditions. There has, naturally, been more detailed work since \citet{rmc2009} and we compare to those models in Appendix \ref{appendix:comparisons} and Section \ref{sec:sumcompare}. 

If we first implement multi-frequency X-rays without metals, as expected, the effect of adding X-ray photons (cyan), compared to a spectrum that extends only to the EUV (purple), is negligible since metals are not present in the atmosphere (Fig.\ \ref{fig:multifreq}). In addition to exploring the impact of mono- vs. multi-frequency stellar spectra, Figure \ref{fig:multifreq} also highlights the impact of our lower boundary condition assumptions. 

\begin{figure}
    \centering
    \includegraphics[width=\linewidth]{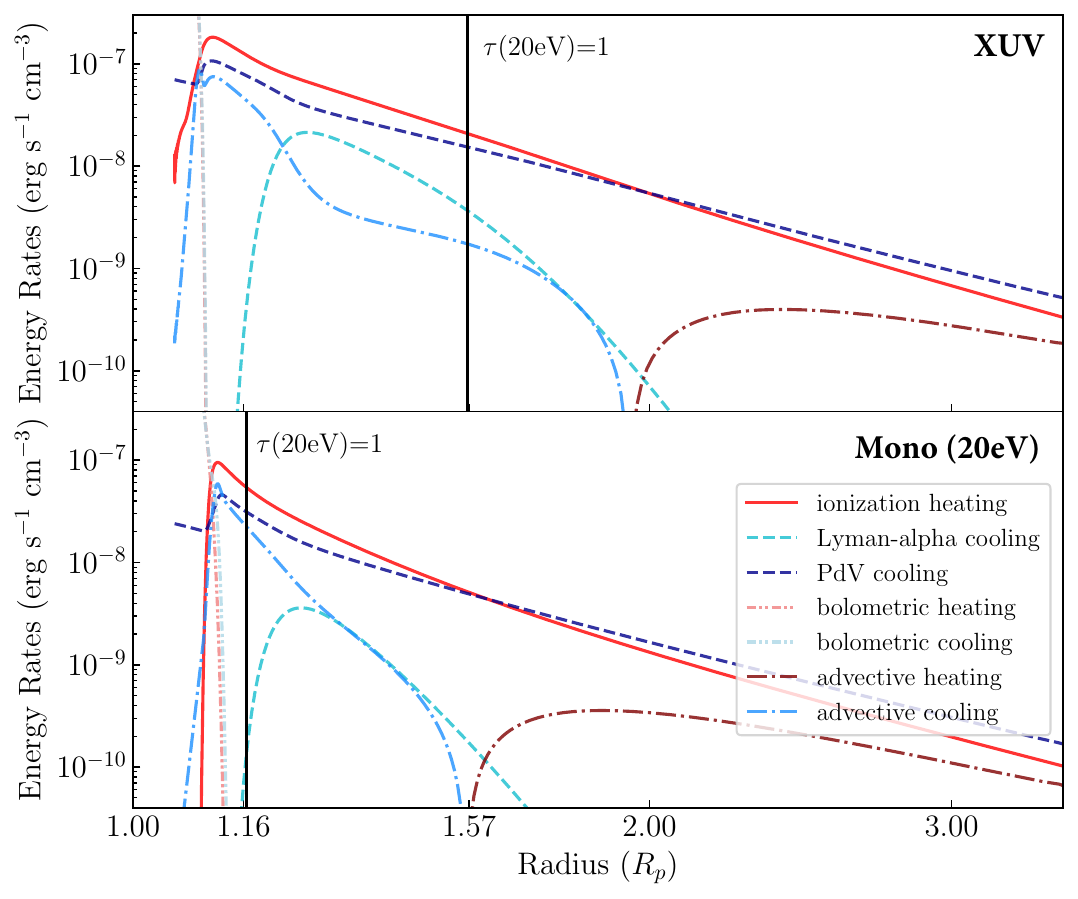}
    \caption{\textbf{Energy Plot for pure-H XUV vs. Monofrequency HD 209458 b} - Energy structure of multi-frequency XUV solution (top panel, cyan solution in Fig.\ \ref{fig:multifreq}) and 20eV monofrequency solution (bottom panel, gray solution in Fig.\ \ref{fig:multifreq}). The $\tau(20\mathrm{eV})=1$ surface for the XUV multispecies is at a higher radius because the deeper penetration of high energy XUV photons puffs up the atmosphere, resulting in higher densities and optical depths at higher radii than in the monofrequency solution.}
    \label{fig:energy}
\end{figure}

The differences in density, temperature, and ionization fraction structure between the two monofrequency models stem from differences in the structure of the bolometrically-heated region below the wind, which sets the radius at which the ionizing photons that drive the wind are deposited. \citet{rmc2009} does not treat this region and does not include either the molecular or bolometric correction and simply solves for that region assuming the same physics as the rest of the wind. Our updated boundary conditions yield a somewhat puffier atmosphere below the wind. We refer to this as increased ``bolometric puffing", a process that is also modeled analytically and discussed in \citet{james_and_hilke}. Absorption of the 20 eV photons occurs at approximately the same pressure and density in both models; however, increased bolometric puffing increases the radius at which that density and pressure occur. This launches a wind from higher in the planet's potential well, where less work needs to be done to reach the planet's escape velocity. As a result, the velocity of the outflow is higher (Figure \ref{fig:multifreq}, gray versus black dash-dotted lines).   

\citet{rmc2009} found that the wind structure and mass loss rate are relatively insensitive to changes in the temperature at the base of the simulation ($T(R_{\rm{min}})$); this remains true.
In general, anything that results in a wind being launched from shallower within the potential well, such as the bolometric puffing will results in a faster, hotter outflow with a higher $\dot M$ (Fig.\ \ref{fig:multifreq}). 

The wind launch radius, $\rxuv$, is often cited as 1 nanobar pressure \citep[e.g.,][]{rmc2009}. With the inclusion of multi-frequency XUV irradiation and a solar-like spectrum, however, we find that the winds for most planets launch closer to 10 nanobars, irrespective of planet mass. This is a result of the higher energy photons in a multi-frequency spectrum being deposited at a lower radius where the density is $\sim10\times$ higher than the radius at which monofrequency 20eV photons are absorbed.

In other words, solutions for hot Jupiters with moderate escape velocities and ionizing fluxes like the one in Fig.~\ref{fig:multifreq} are not very sensitive to including a pseudo-molecular region below the wind with a higher mean molecular weight and bolometric heating and cooling (\S \ref{sec:methods}). However, inclusion of this layer can change the lower boundary dramatically for planets with lower escape velocities at the planet's radius, $R_P$, including many sub-Neptunes \citep{james_and_hilke}. This agrees with the findings of e.g., \citet{allan2019}, who found that mass loss rates were sensitive to surface gravitational potential.

While hot Jupiters typically have wind launch radii $\rxuv\sim 1.1-1.3 R_P$, planets with lower escape velocities may have $\rxuv \gtrsim 2-8 R_P$. Many works model mass loss using the energy-limited mass loss rate $\dot{M}_{\mathrm{Elim}} = \varepsilon F_{\rm{XUV}} \pi R^3 / G M_P$ where $\varepsilon$ is the efficiency, $M_P$ is the planet's mass, and $R$ is chosen to be $R_P$, the IR photosphere radius \citep[e.g.,][]{erkaev2007,kubyshkina_overcoming_2018}. Therefore, for low escape velocity planets, the energy-limited mass loss rate evaluated with $R_P$ equal to the planet's optical transit radius can dramatically underestimate the mass loss rate (Fig.\ \ref{fig:grids2}). This finding is consistent with the literature \citep[e.g.,][]{erkaev_euv-driven_2016,kubyshkina_2022,huang_hydrodynamic_2023}; we discuss the outflow behavior in the high and low escape velocity limits in more detail in \S\ref{sec:discussion}.

The remainder of the solutions in Fig.\ \ref{fig:multifreq} all use the physical lower BCs detailed in Appendix \ref{appendix:bcs}.
The frequency-dependent forms of equations \ref{eq:ion_rate} and \ref{eq:heating} allow us to trace the distribution of the photon energy throughout the wind. Here we choose to scale a \texttt{FISM2} \citep[daily average from 01-01-2009,][]{fism2} solar XUV SED to $F_{\rm{XUV}}=1095$ $\ergs$ (normalized such that the integrated flux between 13.6-40 eV is 450 $\ergs$) which is the flux of HD 209458, a late F or early G type star, experienced by planet b. 

\begin{figure*}
    \centering
    \includegraphics[width=0.9\textwidth]{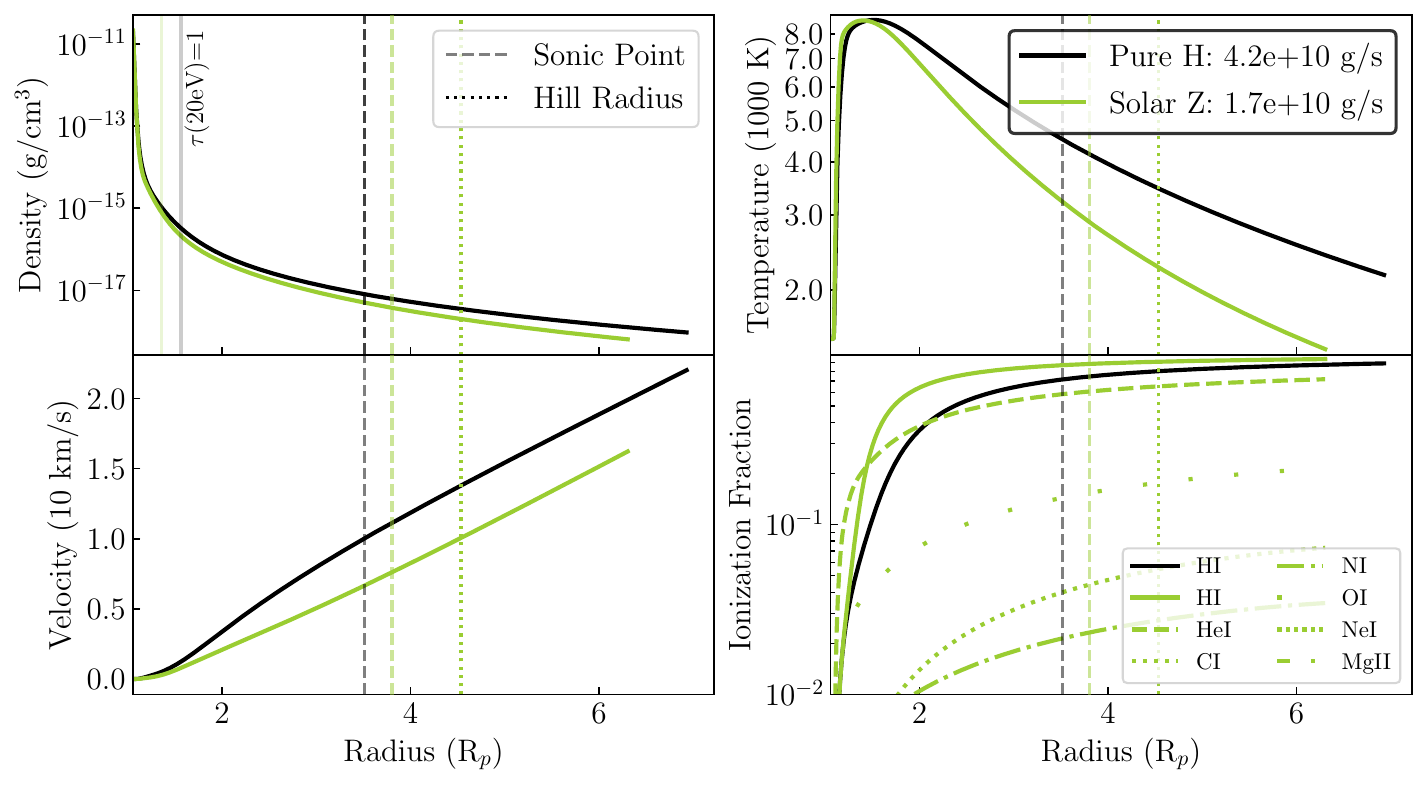}
    \caption{\textbf{HD 209458b Multispecies} - Pure hydrogen (black) and solar abundances of HI, HeI, CI, NI, OI, NeI, MgII (green) XUV (11.26-2000eV) solutions. Vertical lines: Sonic points (dashed), Hill radius (dotted), $\tau$(20eV)=1 surfaces (light solid), and solutions end at the Coriolis radius for each simulation.}
    \label{fig:multispecies}
\end{figure*}

Changing the energy distribution of incident photons from a single frequency to multiple has the most significant impacts on the density and ionization structure of the wind (Fig.\ \ref{fig:multifreq}). To see why this is the case, consider the energy between 13.6 and 20 eV in the scaled solar spectrum. The 13.6-20 eV photons represent 40$\%$ of the total flux in the spectrum. Placing these photons all at 20eV overestimates the resulting heating and underestimates the ionization fraction.  Notice that, relative to the monofrequency solution, the EUV solution in Figure \ref{fig:multifreq} has a higher mass loss rate by a factor of 3 and an equal wind velocity. 

The total flux, $F_{\rm{tot}}$, of stellar photons is normalized in both the EUV and XUV cases such that the flux of photons with energies in the EUV range 13.6-40 eV is equivalent to the flux at 20 eV in the monofrequency case. This means that simulations quoted as XUV (spanning 13.6-2000eV) have a higher $F_{\rm{tot}}$ than those quoted as EUV (spanning 13.6-100eV, unless otherwise noted\footnote{Existing mass loss rates and outflow profiles in the literature frequently quote the integrated EUV flux as the flux between 13.6 and 40 eV (see Appendix \ref{appendix:comparisons}), so when comparing to those solutions we normalize our $F_{\rm{tot}}$ such that the integrated flux between 13.6 and 40 eV matches the quoted values.}). The changes in temperature profile between the monofrequency results in Figure \ref{fig:multifreq} in gray and the multi-frequency results in cyan and purple can be attributed in equal part to the presence of multi-frequency photons and higher $F_{\rm{tot}}$.

 Figure \ref{fig:energy} shows the source of these multi-frequency-induced structural changes in the wind. When compared to a monofrequency 20eV solution with the same lower boundary physics and conditions, the energy of a 20 eV photon is absorbed much higher in the atmosphere. This phenomenon is the result of the higher energy photons penetrating deeper into the atmosphere, puffing it up, raising the scale height of the atmosphere and causing lower energy photons to be absorbed higher in the atmosphere. Thus, the flux of a multi-frequency spectrum is deposited over a much broader physical range of $\tau(\nu)=1$ surfaces. This can explain the broadened peak in both the energy deposition (Fig.\ \ref{fig:energy}) and the temperature (Fig.\ \ref{fig:multifreq}) with radius.

The addition of metals somewhat tightens this radius range of XUV energy deposition (Fig.\ \ref{fig:multispecies}) and shifts it deeper into the potential well. When we add both X-rays and metals in solar abundance (Z$_{\odot}$) to the planet HD 209458b (Fig.\ \ref{fig:multispecies}), we see that the higher ionization cross sections of metals mean that some of the X-rays are absorbed in the atomic layer and contribute to the wind, rather than being deposited deep below it. Our spectral range is lowered to 11.26-2000eV to capture the first ionization energy of CI and we do not lower to the ionization energy of MgI (7.65 eV) as MgI is immediately ionized to MgII, whose ionization energy is 15.04 eV. Thus, the integrated flux is 1185 $\ergs$. Although we reserve a full exploration of the effect of metallicity for a future investigation, even in solar abundances, the metals have a significant role to play in the structure of the wind (if less so in the mass loss rate for moderately-irradiated hot Jupiters). The mass loss rate is similar for both the single species and multispecies planets because, taken independently, the metals and X-rays have opposite effects on $\mdot$.

The metals increase the mean atomic weight, lowering the scale height, increasing the density and opacity of the outflow (Fig.\ \ref{fig:multispecies}), but, more importantly, lowering the sound speed, $c_s$ which raises the sonic point of the outflow (an isothermal Parker wind has its sonic point at $R_{\rm{sp}} = GM_p/(2c_s^2)$). Therefore, multiple species lower the mass loss rate due to their effect on $\mu$.
Conversely, X-ray photons penetrate deeper into the atmosphere than EUV photons do. Because the density is higher at these greater depths, X-rays can drive an outflow that is more massive, slower, and more ionized thanks to secondary ionizations.

\subsection{Tidal Gravity and Comparison to Energy Limit} \label{results:tides}
Returning to the canonical example of HD 209458b---this time with an H-He atmosphere---we also plot $\mdot$ as a function of semi-major axes spanning from 0.03 to 0.7 au. A changing semi-major axis, $a$, not only influences the amount of flux at the planet's location, but also affects the stellar tidal gravity (Eq. \ref{eq:tidal_grav}). We plot a high ($L_{\rm{XUV}}/L_{\rm{tot}} = 10^{-4}$) and low XUV luminosity ($L_{\rm{XUV}}/L_{\rm{tot}} = 10^{-6}$) series to approximate the behavior of young and old stars, respectively (\S \ref{results:size}). It is intentional that we do not include lower fluxes for the ``old star" lower flux case, because while those stellar fluxes corresponding to $a>0.4$ au can launch winds, conductive cooling becomes significant at those extremely low fluxes. We do not model conductive cooling, so we exclude all simulations for which post facto calculations of the conductive cooling rate per unit volume as a function of radius show it to be $\geq 5\%$ of the total heating rate per unit volume.

\begin{figure*}
    \centering
    \includegraphics[width=0.9\textwidth]{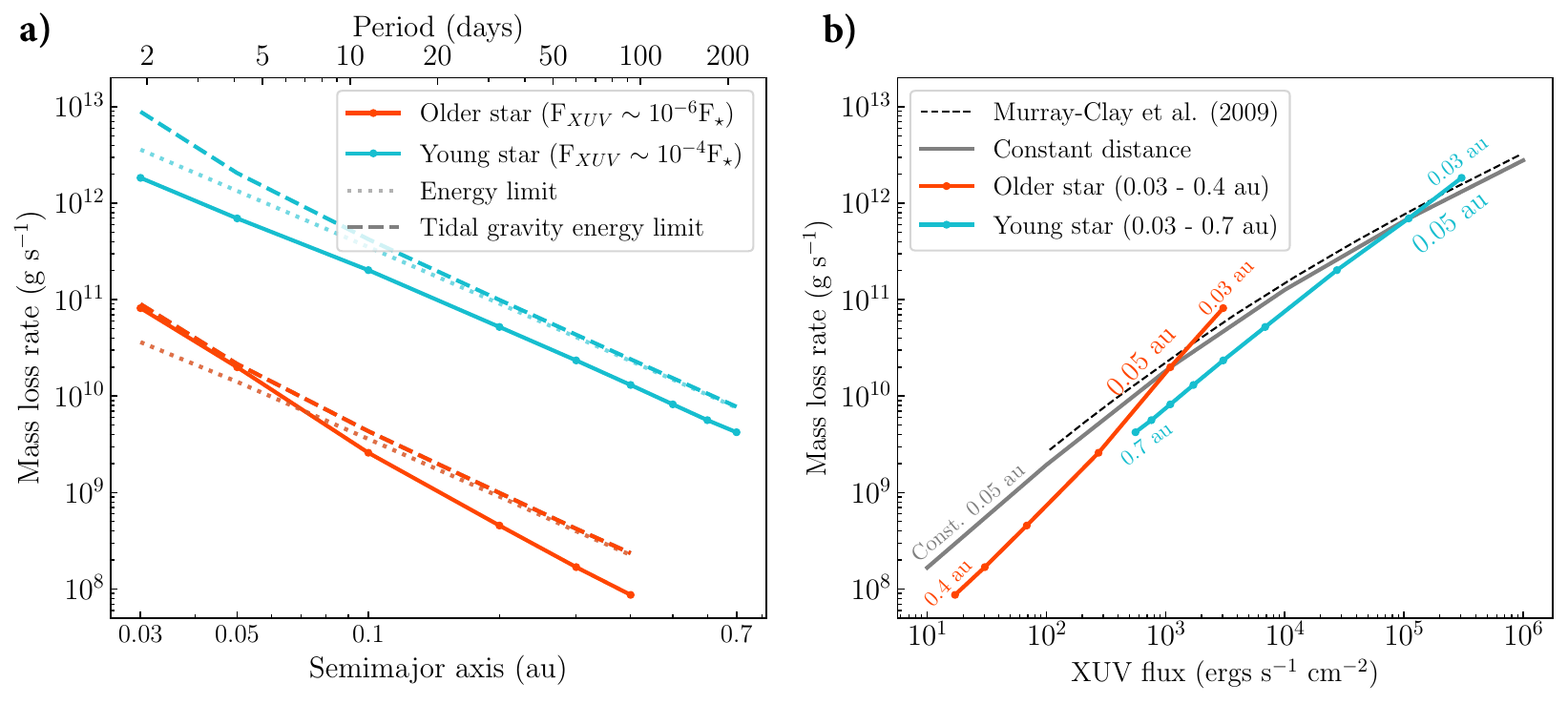}
    \caption{\textbf{$\mdot$ and tidal gravity} - (a) $\mdot$ for a planet with parameters of HD 209458b and a multi-frequency XUV, H-He atmosphere is plotted as a function of semi-major axis, $a$, in au (solid lines) for a high XUV activity ``young" HD 209458 ($L_{\rm{XUV}}\sim10^{-4}L_{*}$, cyan) and lower XUV flux older HD 209458 ($L_{\rm{XUV}}\sim10^{-6}L_{*}$, orange). Lower/moderate flux is $F_{\rm{XUV}}$=1095 ergs s$^{-1}$ cm$^{-2}$ and high $F_{\rm{XUV}}=1.095\times10^{5}$ $\ergs$ at 0.05 au and the stellar spectral shape does not change. Flux scales with $a^{-2}$. Traditional energy-limited $\mdot$ (dotted) and \citet{erkaev2007}'s tidal-gravity-corrected energy limit $\mdot$ (dashed) with our modeled efficiencies (average 0.37, computed point by point as described in Section \ref{results:tides}) are plotted for both flux regimes. 
    (b) Mass loss is plotted as a function of XUV flux. We reproduce \citet{rmc2009}'s Fig. 7 (black dashed) for a pure H, 20 eV monofrequency simulation, in which semi-major axis is held constant while flux at 0.05 au is varied. The $\mdot$ values are multiplied by the geometric reduction factor of 0.33 for consistency with our current results. The gray line is an update of the previous constant distance line, now with an XUV multi-frequency spectrum and an H-He atmosphere. The computed $\mdot$s for young and older stars from the left are overplotted. The difference in slope is the result of tidal gravity. We assume all planets are tidally-locked, meaning that we use a constant geometric reduction factor.}
    \label{fig:tidal}
\end{figure*}

Since $F_{\rm{tot}}\propto a^{-2}$ and the tidal gravity term is $\propto a^{-3}$, the tidal gravity is significant for close-in planets, but negligible at larger semi-major axes. The transition is illustrated by the lower flux old star in Figure \ref{fig:tidal}. Along with our model's $\dot{M}$, we plot the energy limited $\dot{M}_{\mathrm{Elim}}(\varepsilon)=\varepsilon F_{\rm{tot}}\pi R_P^3 / G M_P$, where $\varepsilon$ is the efficiency with which stellar photon energy is converted to heat, and the tidally-corrected energy limited 
\begin{equation}\label{eq:Mdotelim}
    \dot{M}_{\mathrm{tidal}}(R,\varepsilon)=\frac{\varepsilon F_{\rm{tot}}\pi R^3 }{ G M_P}\left(1 - \frac{3}{2\xi} - \frac{1}{2\xi^3}\right)^{-1}
\end{equation} where $\xi=R_{\mathrm{Hill}}/R_P$, the ratio of the planet's Hill sphere and $R$, where $R$ is traditionally taken to be $R_P$ \citep{erkaev2007}.

\begin{figure*}
    \centering
    \includegraphics[width=0.9\textwidth]{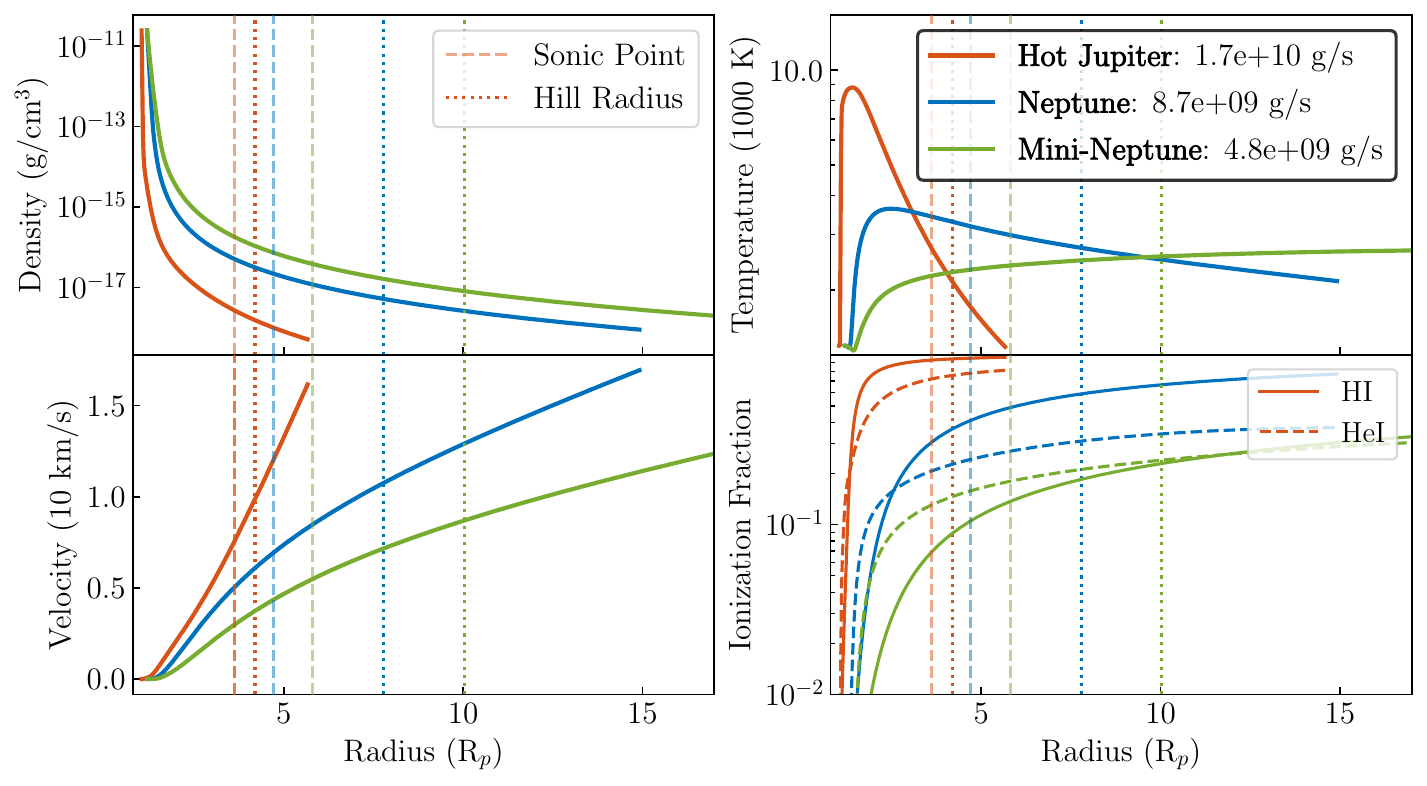}
    \caption{\textbf{Hot Jupiter, Neptune, and Mini-Neptune H-He Outflows} - Wind structure of H-He (0.8:0.2 mass fraction) atmospheres for a 0.7 $M_J$, 1.4 $R_J$ (222.6 $M_{\Earth}$, 15.7 $R_{\Earth}$) planet (orange) with escape velocity of at $R_P$ of $v_{\rm{esc}}=4.6\times10^{6}$ cm s$^{-1}$, a Neptune-sized 17 $M_{\Earth}$, 3.9 $R_{\Earth}$ planet (blue) with $v_{\rm{esc}} = 2.4\times10^{6}$ cm s$^{-1}$, and a mini-Neptune with 5 M$_\Earth$, 2 R$_{\Earth}$ (green) with $v_{\rm{esc}}=1.77\times10^{6}$ cm s$^{-2}$. All three are located at 0.05 au around a 1 $M_\odot$ and 1 $L _\odot$ star, with total flux at 0.05 au of F$_{tot}=1095$ $\ergs$ (13.6-2000eV). All are plotted in terms of their $R_P$, but all have different $R_P$ values. The super-Earth's small thermal inversion below the wind (where larger planets have an isotherm) is the result of bolometric heating/cooling naturally falling off before photoionization heating becomes significant and the wind launches.}
    \label{fig:three_planet}
\end{figure*}

 The energy limited mass loss approximation is an upper limit on a planet's mass loss rate that assumes that the incident stellar flux is converted with efficiency, $\varepsilon$, into heating the wind and driving an outflow.  Planets with lower incident stellar flux are generally said to be in the ``energy-limited" regime, meaning that the mass loss rate is directly proportional to flux and adding additional flux will result in a higher $\dot{M}$ - to a point. Once a planet has sufficiently high flux, the limiting factor on $\dot{M}$ is the rate at which species can recombine and the planet is said to have entered the ``recombination-limited" mass loss regime.
 We directly compute $\varepsilon$ for each point in this plot by computing the frequency-averaged efficiency of energy deposition into heat (as opposed to ionizations), taking into account secondary ionizations.  Radiative cooling is not included in our efficiency. We find that our values range between $\varepsilon\sim 0.3-0.4$. 

 For the ``old" star with the current day $L_{\rm{XUV}}$ of HD 209458, $\dot M$ approaches the tidally-corrected energy limit for very close-in planets, but diverges for planets at larger separations (Fig.\ \ref{fig:tidal}a) \citep{schulik_heliumtriplet_2024}. 
Despite being two orders of magnitude higher flux, the ``young" star does not approach the energy limit, because the Lyman-$\alpha$ cooling is more significant for higher flux stars because the planets' upper atmospheres are able to reach temperatures closer to the $\sim10,000$ K at which Ly-$\alpha$ cooling efficiency peaks.

The dependence of $\dot{M}$ on tidal gravity also results in an update to Figure 7 of \citet{rmc2009}, which plotted $\dot{M}$ as only a function of changing $F_{20eV}$ (which was the total flux concentrated at 20 eV). With the inclusion of XUV multi-frequency radiation and an H-He atmosphere with $F_{\rm{tot}}$ normalized such that the flux in the EUV range of 13.6 - 40 eV the same as the $F_{20eV}$ in \citet{rmc2009}, if the planet's distance is held constant at 0.05 au, we reproduce a similar result (Fig.\ \ref{fig:tidal}b). The characteristic relationship found by \citet{rmc2009}, where $\dot{M}\propto F_{20eV}^{0.9}$ in the energy-limited regime (low fluxes) and $\dot{M}\propto F_{20eV}^{0.6}$ in the recombination-limited regime (high fluxes), holds when distance is held constant. Tidal gravity is present in both models, but its impact is not noticeable without changing the semi-major axis. When the semi-major axis is also changed, the magnitude and slope of the flux-$\dot{M}$ relationship changes significantly as a result of tidal gravity. Our derived lower boundary temperature, $T(R_{\rm{min}})$, also has a semi-major axis dependence (Appendix \ref{appendix:bcs}), but tidal gravity is the dominant contributor to the mass loss rate. 

The aiding effect of tidal gravity is evident where the orange and cyan lines intersect with the gray in Figure \ref{fig:tidal}b. At 0.05 au, the tidal gravity is equivalent to the case in which distance is held constant. At 0.03 au, the tidal gravity contribution elevates the mass loss rate above that of the equivalent flux at 0.05 au. We can turn off tidal gravity in our simulations, but elect to keep it on, as it has a significant effect on mass loss rates for planets whose potential well would otherwise be too deep to launch a wind.

\subsection{As a Function of Planet Type} \label{results:size}
The additions of multi-frequency photons and multiple species highlighted the importance of where in the potential well the wind launches. Planet mass and radius, then, are unsurprisingly important since the planet's escape velocity sets the amount of energy required to lift the wind out of the potential well.

\begin{figure*}[h]
    \centering
    \includegraphics[width=\textwidth]{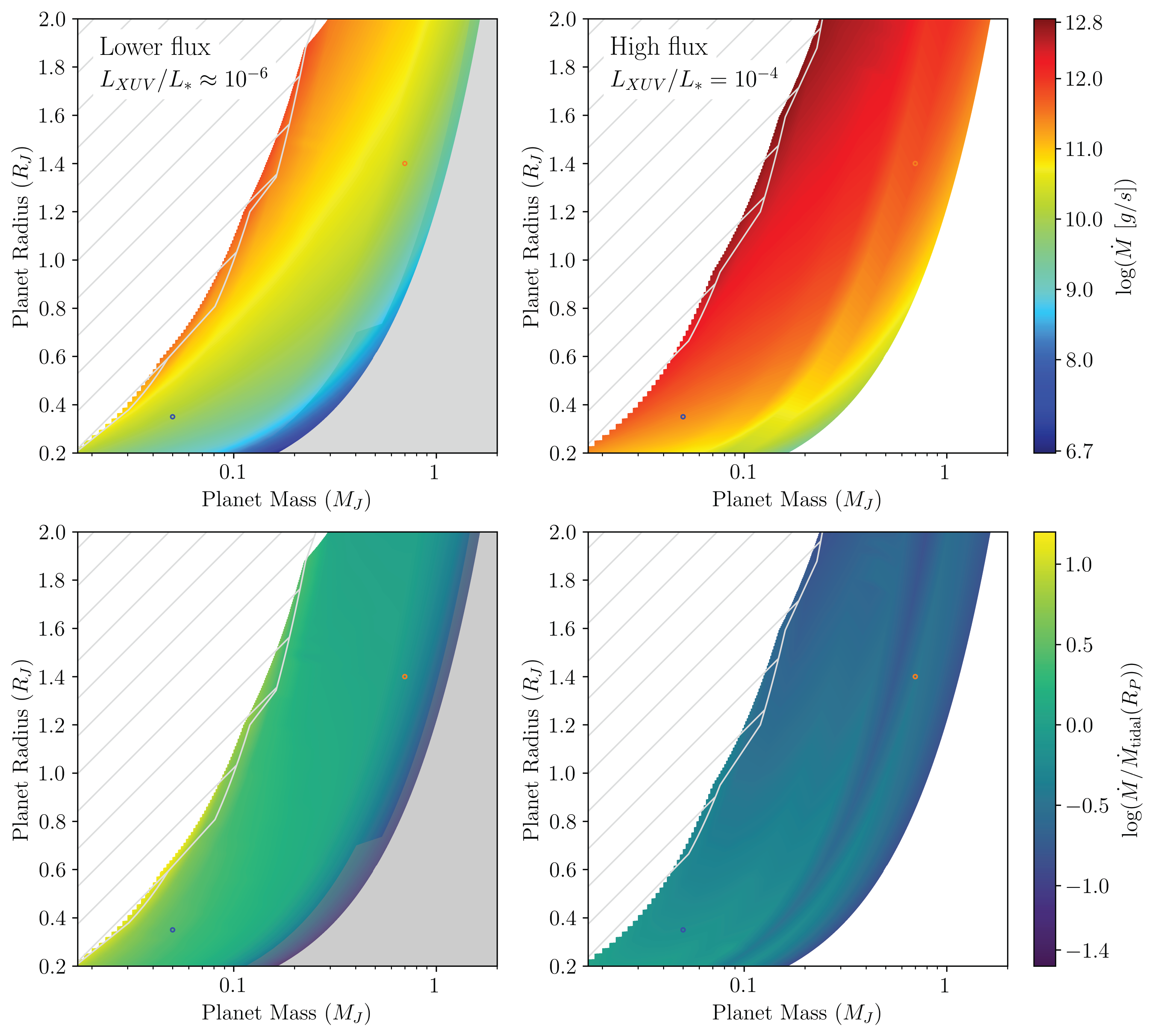}
    \caption{H-He $\dot{M}$ grids for Low and High Fluxes - 2D-linearly-interpolated grid of 663 H-He planets located at 0.05 au around a solar mass, solar luminosity star. Fluxes correspond to an old star ($F_{\rm{XUV}}\approx10^{-6}F_*$) of 1095 $\ergs$ (left column) and a younger star ($F_{\rm{XUV}}\approx10^{-4}F_*$) of 109500 $\ergs$ (right column). Locations of a Neptune- and HD 209458b-analog are plotted as blue and orange points, respectively. The grayed out region represents where conductive cooling---which we do not model---would be significant. The hatched region are planets for which $R_{\rm{sp}}\leq R_{\rm{Hill}}$ (see \S\ref{discuss:lowgrav}). The top row is colormapped to the log of the mass loss rate in g/s and the bottom to the log of the ratio of $\dot{M}$ with the tidally-corrected energy limited $\dot{M}(R_P)$ (Equation \ref{eq:Mdotelim}) with efficiencies calculated on a point-by-point basis in the grid (low-flux $\left<\varepsilon\right>approx 0.37$, high-flux $\left<\varepsilon\right>\approx 0.33$). See \S\ref{discuss:highgrav} for a discussion of the bifurcation near 0.1$M_J$ in the high flux grids.}
    \label{fig:mdots}
\end{figure*}

For an approximately 5-Gyr-old solar mass and luminosity star, we model three generic planets of varying sizes located at 0.05 au. The first is a hot Jupiter of 0.7 $M_J$ and 1.4 $R_J$, the second is a Neptune-mass and -radius planet, and the third is a super-Earth with 5 M$_\Earth$ and 2 R$_{\Earth}$ (Figure \ref{fig:three_planet}).

We compute the total XUV flux (13.6 - 2000 eV) from the stellar age-XUV flux relation in \citep{xuv_over_time}. For a 4.5-Gigayear-old star like our Sun, the ratio of XUV to total bolometric luminosity has been estimated to be $L_{\rm{XUV}}/L_{bol} \sim 10^{-6}$.  We take it to be $F_{\rm{tot}}= 1095$ ergs s$^{-1}$ cm$^{-2}$, appropriate for the low-activity Sun.  These models, and all others henceforth, model a hydrogen-helium atmosphere where the mass fraction in H is 0.8 and in He is 0.2. The escape velocity at $R_P$ of the hot Jupiter is the highest of the three planets ($v_{\rm{esc}}$ = 4.6$\times10^{6}$ cm s$^{-1}$). The Neptune-like planet has $v_{\rm{esc}}$ = 2.4$\times10^{6}$ cm s$^{-1}$ and the super-Earth has $v_{\rm{esc}}$ = 1.8$\times10^{6}$ cm s$^{-1}$.

For the hot Jupiter, the scale height is smaller and photons are absorbed closer to the planet's surface (optical transit radius) than on smaller planets, resulting in a wind that launches from a relatively lower $\rxuv$ (Fig.\ \ref{fig:three_planet}). This radius is deeper in the potential well, so the wind requires more heating in order to escape the planet's gravitational pull, so the outflow has a much higher maximum temperature of 8800 K than the Neptune-like planet (3600 K) and the super-Earth (2900 K). One consequence of this much higher temperature is that the Lyman-$\alpha$ cooling is more significant for the hot Jupiter, which, along with the rapid drop in density with radius results in the steep drop off in temperature. The small scale height means that the hot Jupiter also absorbs the incident multi-frequency photons over a smaller range of heights in the atmosphere, which gives the much narrower temperature peak and the faster ionization with radius. The rapid ionization with radius is also a result of the distribution of stellar photon and photoelectron energy between heating and ionization. 

A hotter wind is almost always a faster wind and the three planets are consistent with this relationship. Because the smaller planets absorb photons over a broader range of radii, require less heating to launch a wind, and thus have a slower decrease in density with radius, they also ionize more slowly as a function of planetary radius. For all planets, at the lowest radii in the wind, He ionizes in greater fraction than H does. This is particularly pronounced for the super-Earth. 


\section{Mass Loss Grids for H-He Atmospheres} \label{sec:results}

\subsection{$\dot{M}$ as a Function of Planetary Parameters}
\begin{figure} 
    \centering
    \includegraphics[width=0.8\linewidth]{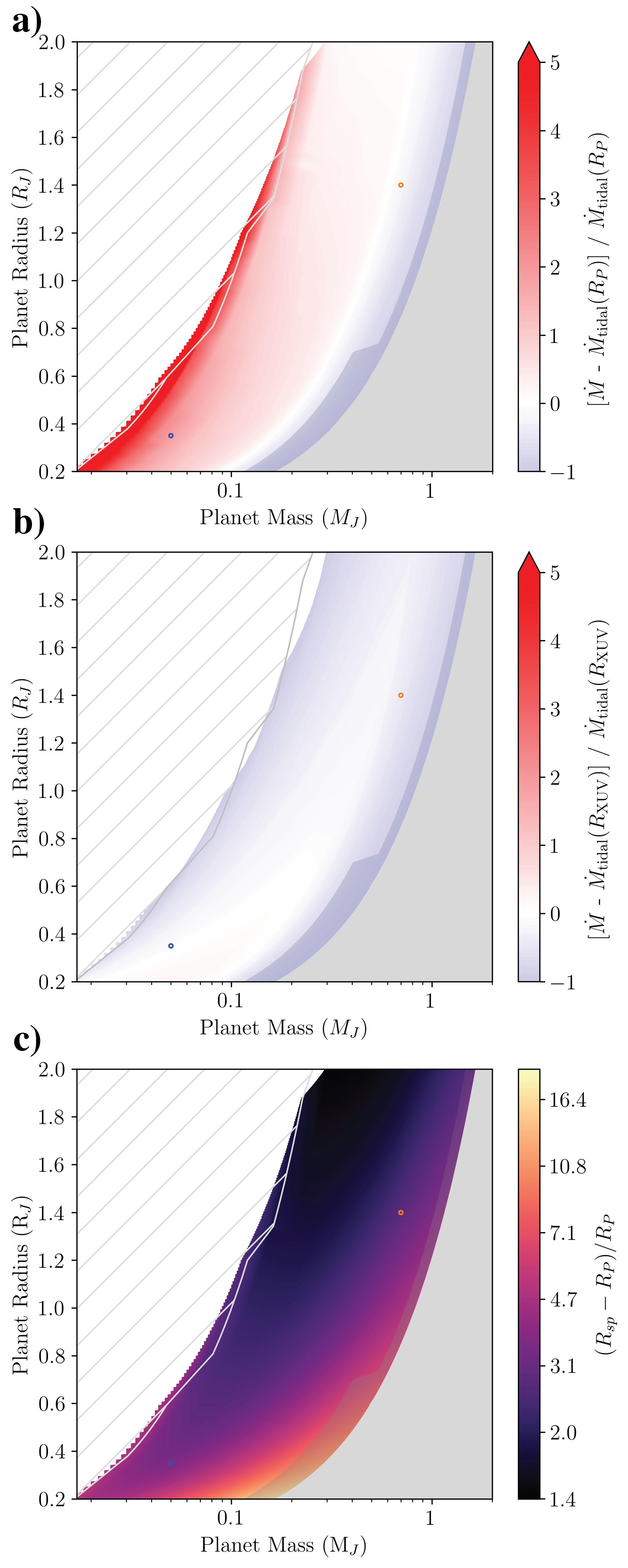}
    \caption{H-He $\mdot$ relative to energy limited mass loss rates - (a) Lower flux (1095 ergs s$^{-1}$ cm$^{-2}$) fractional difference ($\dot{M}-\dot{M}_{\mathrm{tidal}}(R_P))/\dot{M}_{\mathrm{tidal}}(R_P)$, from tidally-corrected energy limited mass loss rate computed at $R_P$. We use the same grid as Fig.\ \ref{fig:mdots}. Orange and blue dots correspond to an HD 209458b and Neptune-analog, respectively. Red indicates that $\dot{M}_{\mathrm{tidal}}$ underestimates the mass loss rate computed by our model and blue indicates that $\dot{M}$ is less than the energy limit, as is expected. (b) Same colorbar scaling as (a), here we plot $(\dot{M}-\dot{M}_{\mathrm{tidal}}(\rxuv))/\dot{M}_{\mathrm{tidal}}(\rxuv)$. (c) Height of the sonic point above $R_P$ in units of $R_P$. All $R_{\rm{sp}}$ are below the Hill radii, except at the lowest escape velocity edge (off the left edge of the plot), where the outflow transitions to boil-off or core-powered mass loss---evaluation of our code's performance in this regime is beyond the scope of this work.}\label{fig:grids2}
\end{figure}

Since the relaxation method is relatively fast, we are able to run a grid of 663 points inexpensively\footnote{About 1 hour run in ``embarrassingly-parallel" (a.k.a., 10 simultaneous sequential runs of $\sim$70 planets each) on an 8-core laptop with an M1 chip. Relative computation time comparisons are made in Appendix \ref{appendix:comparisons}.} The majority of the time per grid point is the ``polishing" process in which we enforce self consistency in the upper and lower boundary conditions. This speed was aided by our choice to create a non-linear grid that spans the total mass range of 0.009 - 1.66 $M_J$ (3 - 528 $M_{\Earth}$) and a radius range of 0.165 - 2.02 $R_J$ (1.85 - 22.65 $R_\Earth$).  The 53 steps in radius space are linear, but the masses are chosen such that the grid lines constitute lines of constant escape velocity, $M_{\rm{P,init}}/ 1.85 R_\Earth$, where $M_{\rm{P,init}}$ is the linear array of 12 initial masses that span from 3 $M_\Earth$ to 43 $M_\Earth$.  These mass and radius ranges are chosen to cover the range of observed masses and radii of super-Earths, sub-Neptunes, and hot Jupiters with semi-major axes of 0.05 au. The results are presented in Figure \ref{fig:mdots} for an ``old" star flux (left) and a ``young" star flux (right). 

The grid is limited by a number of numerical and physical limitations. The lower end of the mass range of super-Earths ($M_P \leq 3.5 M_\Earth$) is limited by the numerical failure to converge to a wind solution. Notably, this failure is not purely numerical in nature. These planets are so low escape velocity and optically thick that $\rxuv$ computed using the lower BC scheme in Appendix \ref{appendix:bcs} can be as large as 8$R_P$. 

In these extreme cases, when we compute the sonic point for a wind driven purely by bolometric radiation, $R_{\rm{sp,bolo}}\approx G M_P/(2c_s^2)$ for the sound speed, $c_s$, at the $R_{\rm{min}}$, we find that $R_{\rm{sp,bolo}}$ is of order $\rxuv$. This means that the bolometric heating is sufficient to launch a transonic wind and the XUV photons are being absorbed near or within the supersonic region of the wind. The XUV photons cannot contribute to heating and driving a wind, hence, these flows are likely not photoionization-driven winds, but rather are still undergoing boil-off or core-powered mass loss \citep[as in][]{rogers_2021,tang_fortney_2024}. Therefore, we do not include those solutions in these grids. We also indicate the region for which $R_{\rm{sp}}\leq R_{\rm{Hill}}$, the Hill sphere, with hatch marks in Figures \ref{fig:mdots} and \ref{fig:grids2}. Note that these hatches do not represent a strict limit for \texttt{Wind-AE}'s applicability (see \S\ref{discuss:lowgrav} for a complete discussion).

The lower-escape velocity limits seen in the grids are the result of the bolometric luminosity and XUV flux we have chosen. Because we created a grid based off of the parameters of the quintessential photoevaporating planet, HD 209458b, a hot Jupiter with $L_* = 1.78 L_{\odot}$ at 0.05~au and $F_{\rm{tot}}=1095\ \ergs$, super earths and low-mass sub-Neptunes are so highly irradiated proportional to their escape velocities, that they are on the verge of core-powered mass loss and potentially Roche-lobe overflow. While \texttt{Wind-AE} is able to solve for some of these planets (at the left edge of the grids), it struggles to converge beyond that because even our ``lower flux" grid has a moderately high bolometric and XUV flux.

We make a similar cut at the upper end of masses. Our maximum initial mass for $R_P=1.85 R_{\Earth}$ is 43 $M_\Earth$ to avoid unphysically dense planets. \texttt{Wind-AE} is capable of modeling these extremely dense planets, though at high enough escape velocities, it eventually suffers from numerical and physical difficulties discussed in \S\ref{discuss:highgrav}. However, we cut off our grid at the high escape velocity end because, for lower $F_{\rm{XUV}}$, conductive heating and cooling are significant and we do not model conduction in this paper (grayed out region in Fig.\ \ref{fig:mdots},\ref{fig:grids2}).  

We plot the fractional difference of our computed values of $\dot M$ from the energy-limited estimate, $(\dot M - \dot M_{\mathrm{tidal}})/\dot M_{\mathrm{tidal}}$ in Figure \ref{fig:grids2} with the heating efficiencies computed point by point as before (for the lower flux case $\left<\varepsilon\right>=0.37$, high flux $\left<\varepsilon\right>=0.33$). Given that $\dot{M}_{\mathrm{tidal}}(R,a, \varepsilon)$ is the theoretical upper limit on mass loss rates, it should not be possible for our mass loss rates to be higher for the same efficiency value, $\varepsilon$, as the amount of red in Figure \ref{fig:grids2}a seems to suggest. However, $\dot{M}_{\mathrm{tidal}}$ is derived assuming that all of the photon energy is deposited near $\rxuv\approx R_P$ and that the optical transit radius, $R_P$, is frequently used in the energy-limited mass loss equation. This assumption is a good one for hot Jupiters with moderate XUV instellation. However, as we have seen, for planets with lower escape velocity, multi-frequency photons and bolometric heating/cooling below the wind drive $\rxuv\gg R_P$. Likewise, when we plot the ratio between the tidally-corrected energy limited mass loss rate, $\dot{M}_{\mathrm{tidal}}(R_P,a, \varepsilon)$ and the $\dot{M}$ found by our model (Figure \ref{fig:mdots}, bottom panels), we see that for these high escape velocity planets, $\dot{M}$ is more than an order of magnitude below the energy limit. Thus, $\dot{M}_{\mathrm{tidal}}(R_P)$ will always underestimate the energy-limited mass loss rate for lower escape velocity planets and overestimate it for high escape velocities at moderate XUV fluxes. Instead, it is necessary to use $\dot{M}_{\mathrm{tidal}}(\rxuv)$. 

Indeed, when we compute $\dot{M}_{\mathrm{tidal}}(R=\rxuv,0.05 \rm{au}, \varepsilon)$, for the lower flux grid, we see that our mass loss rates are below this photoionization-radius-corrected mass loss rate (Fig.\ \ref{fig:grids2}b) and thus do not violate the maximum energy available. Note that the valley feature in Figure \ref{fig:grids2}b can also be seen in Figure 13 of \citet{owen_planetary_2012}, though we disagree with some of the aspects of the model therein and provide more detailed comparisons in Appendix \ref{appendix:comparisons}. 

The findings in Figure \ref{fig:grids2} differ slightly from the results of \citet{kubyshkina_overcoming_2018}, who found that the tidally-corrected energy limited mass loss rate computed at $\rxuv$ for a heating efficiency of 0.15 underestimates mass loss rates for the lowest density planets. The \citet{kubyshkina_overcoming_2018} grids span lower planetary radii and masses (1-10 $R_{\oplus}$, 1-39 $M_\oplus$ vs. our 1.85-22.65 $R_{\oplus}$, 3.8-528 $M_\oplus$) than our model can currently do without conductive cooling, but the lower left quadrant of, e.g., Figure \ref{fig:grids2} corresponds to a similar range of masses and radii.

The primary source of the discrepancy is the efficiency in Equation \ref{eq:Mdotelim}. When we compare our mass loss rates to $\dot{M}_{\rm{tidal}}(\rxuv,\varepsilon=0.15)$ as in \citep{kubyshkina_overcoming_2018}  instead of the escape-velocity-dependent $\varepsilon=0.26$-0.38 in Figure \ref{fig:grids2}b, we likewise find that $\dot{M}_{\rm{tidal}}$ overestimates the mass loss rates for low density planets. It is also likely that we find slightly higher $\rxuv$ on average than the $R_{\rm{eff}}$ computed by \citet{kubyshkina_overcoming_2018}. Both methods take the absorption radius to be the point of maximum ionization, but the input stellar spectrum in \citet{kubyshkina_overcoming_2018} is parameterized as two monofrequency EUV (20 eV) and X-ray (247 eV) bins. 
This is an appropriate parameterization for computational efficiency, but, as Figure \ref{fig:energy} shows for monofrequency EUV vs. multi-frequency XUV, will result in a slightly lower estimate for $\rxuv$.

These discrepancies further validate the ultimate conclusion of \citet{kubyshkina_overcoming_2018}: regardless of the relative accuracy of the energy limited approximation for lower flux planets, it still requires accurately estimating both $\rxuv$ and the heating efficiency a priori.

For planets in the high flux limit, however,  it is not necessary to find $\rxuv$ to compute a reasonable estimate for the tidally-corrected energy limited mass loss rate.
In Figure \ref{fig:grids2} we do not show the fractional difference for the high flux case, because the planets are in the radiation/recombination-limited mass loss regime, meaning that radiative cooling is energetically important. As such, $\dot{M}<\dot{M}_{\mathrm{tidal}}(R_P)$ for the extent of the high flux grid. 

Slices of an even higher flux grid are included in Appendix \ref{appendix:comparisons}, Figure \ref{fig:james_grid} where we compare to the high-
XUV-flux-limit mass loss rate grids of \citet{owen_planetary_2012}. In short, we find that the lack of PdV and Lyman-$\alpha$ cooling in the \citet{owen_planetary_2012} model results in very different mass loss rates across the grid, highlighting the importance of those two cooling sources even in recombination-limited outflows.

\section{Discussion} \label{sec:discussion}
\subsection{High Escape Velocity Limit} \label{discuss:highgrav} 

Atmospheric escape in the high escape velocity limit ($\gtrsim 5\times10^{6}$  cm s$^{-1}$) will be explored in more detail in future works, but we make note of a couple of interesting features here. The first few numerical and physical complications result from the small scale height that results from high escape velocity at large planet masses. 

A small scale height not only yields an atmosphere that is more compressed, it also means that most of the ionizing photons are absorbed very deep in the potential well. This means that it can be difficult for these planets to launch winds, since they require a large flux to achieve sufficient energy to launch. For the same reason, these outflows can reach $>4000$ K before launching (e.g., the temperature difference between planets of varying surface gravity in Figure \ref{fig:three_planet}). Inspection of these planets' energy distributions reveals that they experience significant Lyman-$\alpha$ cooling, even when they experience relatively low fluxes. Ly-$\alpha$ cooling is significant for these planets because their deep potential wells necessitate high temperatures to escape, so these atmospheres heat to high temperatures where Ly-$\alpha$ is efficient before the atmospheres outflows. This agrees with the analytic predictions of \citet{owen_atmospheres_2016}.

Even when \texttt{Wind-AE} can find a solution, for high escape velocity and lower flux planets, the default base boundary condition radius of the 1-$\mu$bar radius may not capture all of the $\tau(\nu)=1$ surfaces for the highest energy photons. 
Instead, the wind launches immediately at the base of the simulation and $\rxuv=R_{\rm{min}}$. In these cases, when computing our grids, we raise $T(R_{\rm{min}})$ several thousand Kelvin from the expected temperature in the bolometrically-heated region, $T_{\rm{skin}}$ to account for the fact that we are essentially beginning our model mid-wind launch (see Appendix \ref{appendix:bcs} for more details). Because we do not capture all of the photoionizing photons in our simulation, we lose up to 10\% of the flux out of the bottom of the simulation. We have validated that this procedure provides a good match for solutions that set the simulation base at higher densities and take substantially longer to converge, but do not lose flux out the bottom (Appendix \ref{appendix:bcs}, Fig. \ref{ap-fig:t_oscill}). Making the adjustment to $T(R_{\rm{min}})$ results in a similar wind structure (Fig.\ \ref{ap-fig:t_oscill})) and brings $\dot{M}$ to within a factor of, on average, 1.5 of the $\dot{M}$ for the simulation whose base is set deeper in the wind.
 
 Despite this, the transition from $\rxuv>R_{\rm{min}}$ to $\rxuv<R_{\rm{min}}$ is likely the source of the sharp transition near 0.1$M_J$ visible in the high flux grids (Fig.\ \ref{fig:mdots}) and the energy limited mass loss rate comparison plots (Fig.\ \ref{fig:grids2}a,b). This indicates that while this approximation brings $\dot{M}$ close to the expected value, it still systemically underestimates it slightly along the line of constant escape velocity that starts at 0.1$M_J$, though the general trend across the grid still holds. The other consequence of not being able to trace the deposition of the highest few energies of photons is that we compute a lower average heating efficiency than expected if all photons were captured.  
 Thus, we include a warning in \texttt{Wind-AE} when $\rxuv<R_{\rm{min}}$.

The reasons we opt for this simplification rather than raising the pressure of the base boundary for every high escape velocity simulation are twofold. In addition to the fact that winds are more compressed and are launched deeper in the planet's potential well, another consequence of a small scale height is that a small change in $R_{\rm{min}}$ results in a large change in pressure. This means that, following our prescription for the calculation of lower boundary conditions (Appendix \ref{appendix:bcs}), raising the pressure of $R_{\rm{min}}$ by, say, a factor of 10, results in a very small change in radius (e.g., lowering from $R_{\rm{min}}=1.04R_P$ to just 1.03$R_P$) and an order of magnitude change in $\rho(R_{\rm{min}})$. Large changes in $\rho(R_{\rm{min}})$ are numerically costly and likely to fail. Too, the resultant change in radius may still not be sufficient to capture the actual photoionization base, $\rxuv$. Thus, modeling the lower portion of the upper atmosphere in high escape velocity limit remains an area of ongoing exploration.

\subsection{Low Escape Velocity Limit} \label{discuss:lowgrav}
The low escape velocity limit presents its own set of challenges related to the lower boundary of the simulation (those at the border of or within the hatched region of Figures \ref{fig:mdots} and \ref{fig:grids2}). For low escape velocity planets, the large scale height means that \texttt{Wind-AE} finds $\rxuv$ to be extremely high in the atmosphere (up to 8$R_P$ for super-Earths modeled). This elevated $\rxuv$ occurs regardless of whether the bolometric heating/cooling and molecular layer prescription are turned on. When they are turned on, however, the  magnitudes of bolometric heating / cooling terms (which are a function of density) often naturally decline with radius (i.e., not as a result of the enforced decline of the complementary error function that enforces the molecular to atomic transition) before reaching $\rxuv$. This means that there is no longer an isotherm in the molecular region below the wind and the resulting thermal inversion can dip several hundred kelvin below $T_{\rm{skin}}$ (e.g., from 1200 K to 1000 K) (e.g., Fig.\ \ref{ap-fig:gillet}). 

While this thermal inversion is neither unphysical nor unexpected, it may require the inclusion of additional physics, making the ability of our model to be easily coupled to more sophisticated lower atmosphere models especially valuable in these cases. These inclusions may consist of more carefully treating the infrared and optical opacities used in the bolometric heating and cooling calculations. They may also consist of adding molecular line cooling terms, since the low, non-isothermal temperature in region below the wind \citep{matthaus,misener2024}, may mean that molecules do not thermally- or photodissociate and are present throughout the region and into the wind.


In some of the most extreme cases, the sonic point ($R_{\rm{bolo,sp}}$) for bolometrically-driven mass loss (i.e., core-powered mass loss or boil off) is of order $\rxuv$. This means that XUV photons may not be the predominant driver of the wind and, indeed, they may be absorbed in the supersonic portion of an outflow and be unable to contribute to the heating of the wind because a supersonic flow is not causally connected. This regime is usually referred to as core-powered mass loss or boil off, depending on the point in a planet's evolution that this phenomenon is occurring. \citet{james_and_hilke} mapped out this parameter space analytically, showing that both occur for expected planet properties; however, detailed numerical simulations like ours are required before the combined mass-loss rates are used in evolutionary models.
These underdense planets may also be in the regime where Roche-lobe overflow is a significant contributor to mass loss rates \citep[e.g.,][]{erkaev2007,jackson_roche_2017}.

We do not include planets where $R_{\rm{bolo,sp}}<R_{\rm{XUV}}$ in our grid, but we do include some planets for which $R_{\rm{sp}} > R_{\rm{Hill}}$, the Hill sphere (hatched region in Fig.\ \ref{fig:mdots}. Also see, e.g., Appendix Fig.\ \ref{ap-fig:gillet}, for a profile of such a planet.). While an isothermal Parker wind does not allow for the sonic point to be outside of the planet's Hill sphere, a non-isothermal solution with a positive temperature gradient past the Hill sphere (through the sonic point) does \citep[e.g.,][]{Lamers_1999}. We test our analytical boundary condition prescription against a more sophisticated multi-layer interior model which includes, e.g., more detailed bolometric radiative transfer (Tang et al., 2025) and find that we match the microbar radius computed by the interior model. This agreement indicates that these planets are indeed incredibly puffy and potentially on the verge of core-powered mass loss. Given that planets with low escape velocities in our grids have incredibly high mass loss rates, it is also likely that these mass-radius combinations (overlap of the hatched region and colored region in Fig.\ \ref{fig:mdots}, \ref{fig:grids2}) would not exist in nature for such highly irradiated planets, as they would have likely undergone a short and violent period of boil-off or core-powered mass loss early in their lives. Even if these mass-radius combinations did exist, for the level of bolometric and XUV irradiation our grids, they may not be stable for long, with the incredibly high mass loss rates stripping significant fractions of their remaining atmospheres. Detailed simulations of the transition between the core-powered and photoevaporation mass loss regimes \citep{james_and_hilke} merit further investigation.

\subsection{Additional Physics} \label{discuss:H3plus}
Within the molecular layer below the wind, there may be a thin region toward the top of the layer which is still optically thin to heating by the highest energy X-ray photons, but is likely cooled by H$_{3+}$ molecular line emission \citep{yelle_aeronomy_2004,garcia_munoz_physical_2007}. This means that heat deposited by high energy photons in the molecular region below the wind will be largely removed by radiative and/or bolometric cooling, meaning that the contributions of the highest energies X-rays to the outflows may be negligible. 

Explicitly modeling the H$_{3}^+$ cooling requires more nuanced and expensive photochemical calculations than are sensible in a relaxation model, but we can approximate the effects of H$_{3+}$ cooling by cutting off the high energy end of stellar spectrum at 165 eV and reducing the flux above 70 eV by 70$\%$ as found by Frelikh et al. (submitted). The differences in outflow structure and $\dot{M}$ are negligible for a 2.02 $M_{\Earth}$ 5.3 $R_\Earth$ H-He planet when compared to the same planet irradiated by a full XUV spectrum (13.6-2000 eV) when both planets have fluxes normalized to 13.6-40 eV. Since the highest energy X-rays are already being absorbed below the wind in an H-He atmosphere, they already contribute minimally to the direct heating of the outflow and contribute negligibly to puffing the atmosphere below the wind when compared to the contributions of bolometric heating. So, the absence of the highest energy photons ($>$165 eV) when ``removed" by our pseudo H$_{3+}$ cooling has minimal impact.  For winds containing metals, this choice will be more important.  

At the top of the hypothetical X-ray-heated and H$_{3+}$-cooled layer, molecules are ultimately thermally dissociated and the outflow is atomic, so we elect to treat the molecular layer holistically as a region with constant mean molecular weight, whose energy budget is predominantly set by bolometric heating and cooling. As we noted in the previous section, though, these assumptions break down for small planets which may have a different thermal or molecular-atomic fraction structure.

\subsection{Summary of Comparison to Other Work}\label{sec:sumcompare}

To first order, the results of our benchmarking can be broken down as follows. In the high flux limit, differences in the outflow energetics prove important and in the low flux limit, differences in SED shape and lower boundary modeling have the largest impact. The primary source of differences between our model and the high X-ray flux model \citet{owen_planetary_2012} is our inclusion of Lyman-$\alpha$ and $PdV$ cooling where the aforementioned model does not. As a result, we find higher mass loss rates for planets with higher escape velocities and lower for lower escape velocities than the high X-ray flux grids in \citet{owen_planetary_2012}.

An advantage of our model is the ability to model both the high (as above) and low incident stellar flux limit with the same physics. In the low flux limit, we explore several case studies. In keeping with previous findings that emphasize the impact of stellar spectra shape, profile differences between our model and \citet{garcia_munoz_physical_2007} (HD 209458b, H-He), \citet{kubyshkina_precise_2024} (GJ 1214b, H-He), and \citet{ates} (HD 209458b, H-He) can be attributed primarily to the differences in the input stellar spectra. Likewise, the good agreement with \citet{salz_simulating_2016} (HD 209458b, H) we attribute to the similarities in the spectra. 

The other major source of differences is the lower boundary conditions. Our approximation of the region below the wind as molecular and dominated by bolometric heating / cooling represents an improvement over models that scale directly from the planetary transit radius and this approximation results in a much higher altitude $\rxuv$ and significantly different outflow profile \citep[][, 0.05 $M_J$, 0.55 $R_J$, H-He]{gillet_self-consistent_2023}. However, in comparison to a model with more sophisticated lower atmosphere modeling \citep[][, WASP 121b, metals]{huang_hydrodynamic_2023}, we find a much lower $\rxuv$. Notably, when we match the spectrum and lower boundary conditions to the incredibly comprehensive and more expensive \citet{huang_hydrodynamic_2023}, we are able to closely reproduce their profile and mass loss rates in less than 1/100th of the time.

\section{Summary}
We built upon the 1D photoionization-driven steady-state hydrodynamic transonic Parker wind relaxation code developed by \citet{rmc2009} by adding metal and X-ray physics, as well as a python wrapper that aids in ramping through the relaxation code solution space, in performing post-facto calculations, in smoothing input stellar spectra, and in visualizing solutions. A fast forward model allows us to perform broad parameter space studies smoothly across such parameters as stellar flux (Figs.\ \ref{fig:mdots}, \ref{fig:energy_frac}), planet size (Fig.\ \ref{fig:three_planet}), semi-major axis (Fig.\ \ref{fig:tidal}), spectral ranges (Figs.\ \ref{fig:energy}, \ref{fig:multifreq}), presence of metals (Fig.\ \ref{fig:multispecies}), and, in the future, metallicity. 

The speed and versatility of \texttt{Wind-AE} allows it to be applied to explore a number of open problems. 
\texttt{Wind-AE}'s python interface and open-source nature allows it to be easily coupled with chemical network codes that model atmosphere behavior below the optically-thin upper atmosphere and with interior models and XUV spectrum evolution codes in order to model planetary evolution due to atmospheric escape.\\[1\baselineskip]

\begin{acknowledgments}
The authors thank the referee for thoughtful comments.
This paper makes use of the following Python libraries: \texttt{numpy} \citep{numpy}, \texttt{matplotlib} \citep{matplotlib}, and \texttt{scipy} \citep{2020SciPy-NMeth}.

 MIB, RMC, and JHM acknowledge funding support from NSF CAREER grant 1663706. MIB and RMC acknowledge support from NASA'S Interdisciplinary Consortia for Astrobiology Research (NNH19ZDA001N-ICAR) under grant number 80NSSC21K0597 and from NASA XRP grant number 80NSSC23K0282. JEO is supported by a Royal Society University Research Fellowship. This project has received funding from the European Research Council (ERC) under the European Union’s Horizon 2020 research and innovation programme (Grant agreement No. 853022). We acknowledge use of the Lux supercomputer at UC Santa Cruz, funded by NSF MRI grant AST 1828315. We would like to thank Chenliang Huang for sharing data, Linda Lewis for time evolution discussions, Jorge Fernandez-Fernandez, Emerson Tao, and Artem Aguichine for beta testing \texttt{Wind-AE}, and Ethan Schreyer and Sarah Blunt for Hill sphere discussions.
 
\end{acknowledgments}

\software{\texttt{Wind-AE} is available on Github (\hyperlink{https://github.com/mibroome/wind-ae}{github.com/mibroome/wind-ae}). Data for all plots can be found at \hyperlink{https://github.com/mibroome/wind-ae/Notebooks}{github.com/mibroome/wind-ae/Notebooks}.}

\bibliography{ref}{}
\bibliographystyle{aasjournal}

\appendix 

\section{Comparisons to Existing Results} \label{appendix:comparisons}
Here we benchmark our model against a selection of the other numerical models in the literature. For all of the comparisons herein, we continue to use a scaled solar spectrum from a low activity period of the Sun and use the physical boundary conditions for the outflow discussed in \S\ref{sec:methods} unless otherwise specified. It is important to note that a number of models use different geometric surface averaging schemes for the mass loss rate, so instead of reporting $\dot{M}=0.3\cdot4\pi R_{\rm{sp}}^2 \rho(R_{\rm{sp}}) v(R_{\rm{sp}})$ as in the main body of the text, here we report $\dot{M}_{/f}=\dot{M}/f=4\pi R_{\rm{sp}}^2 \rho(R_{\rm{sp}}) v(R_{\rm{sp}}) / f$ where $f$ is a factor between 1 and 4. This allows us to make direct comparisons to mass loss rates found models in the literature that use different geometric averaging factors. We consider the mass loss rates to be in agreement if they fall within 30$\%$ of each other.

\begin{figure*}
    \centering
    \includegraphics[width=\textwidth]{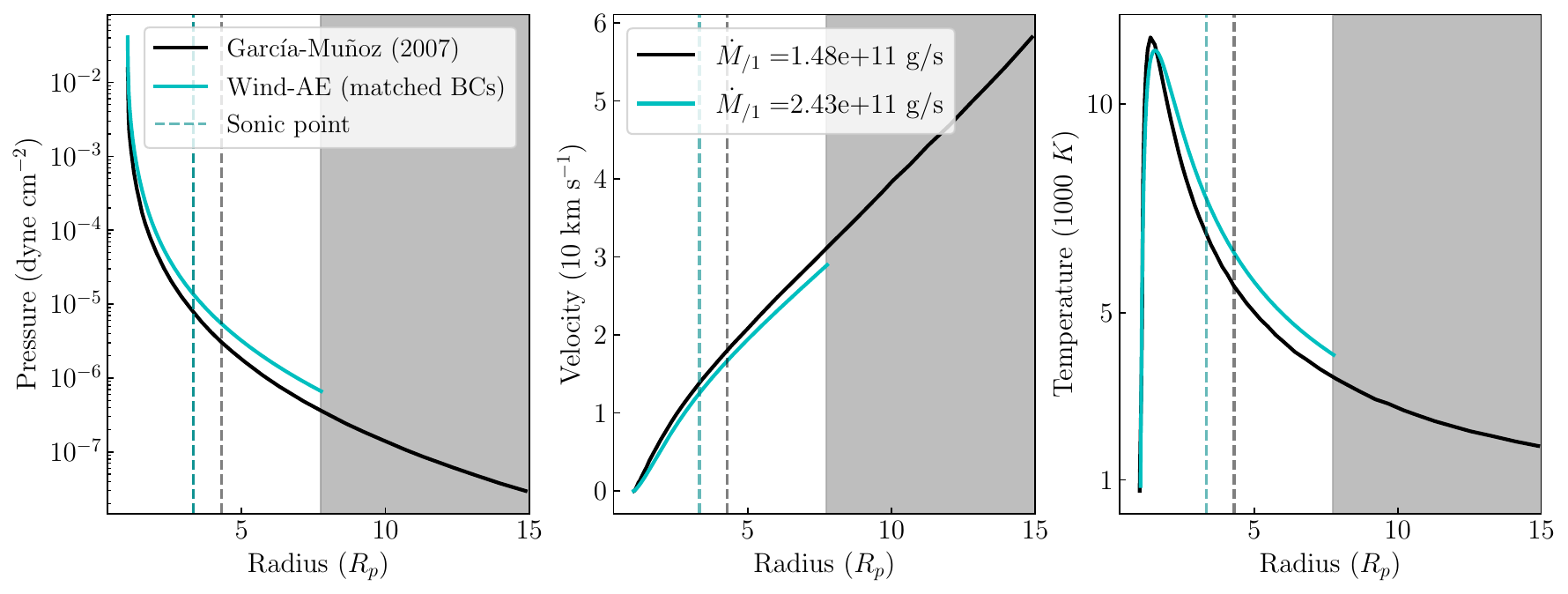}
    \caption{ \textbf{HD 209458b Pure-H \citet{garcia_munoz_physical_2007}}- We reproduce the SP solution of \citet{garcia_munoz_physical_2007} Figs. 3-4 for an EUV-irradiated pure-H atmosphere (black). Wind-AE solution plotted (cyan) is for a pure-H atmosphere with no Lyman-$\alpha$ cooling, a similar if not identical high activity solar spectrum ($F_{\rm{tot}}=2474$ $\ergs$ integrated over 13.6-165 eV, which loosely simulates H3+ cooling), and BCs matched to those of the black solution ($R_{\rm{min}}$=1.03 $R_P$, T=730 K, $\rho$=5.82$\times 10^{-13}$ g cm$^{-3}$). Sonic points are dashed lines and the solid gray is the Coriolis radius past which our model is not valid.}
    \label{fig:garcia_munoz}
\end{figure*}

One of the advantages of \texttt{Wind-AE}'s relative speed is that we are able to reproduce the results of more expensive models including those coupled with full photochemical lower atmosphere models. \citet{garcia_munoz_physical_2007} is one such well-known model.  \citet{garcia_munoz_physical_2007} is a 1D multi-frequency EUV hydrodynamic model that uses a photochemical network to model photoionization of hydrogen down to $r=R_P$ and includes conductive and H3+ cooling, but not Lyman-$\alpha$ cooling.
For a pure-H HD 209458b, we find good agreement in the outflow structure and mass loss rate (Fig.\ \ref{fig:garcia_munoz}) when we employ the same high activity EUV solar spectrum used by \citet{garcia_munoz_physical_2007} \citep{euvac}, turn off Ly-$\alpha$ line cooling, and match lower boundary conditions (Fig.\ \ref{fig:garcia_munoz}). We normalize the spectrum to $F_{\rm{tot}}=2474$ $\ergs$ over the range of 12-165 eV, where the upper bound is chosen to simulate the effect of H$_3^+$ cooling which radiates away the highest energy photons. 

Notably, we get relatively poor agreement with the same flux, same wavelength range, but different solar spectrum (the lower activity \texttt{FISM2} spectrum used in the main body of the text). This is consistent with the findings of, e.g., \citet{guo_influence_2016,kubyshkina_precise_2024,schulik_heliumtriplet_2024}, who report that SED shape can have a significant influence on the outflow physics. In particular, we find that a higher flux or different SED shape in low energy ($<$60 eV) EUV photons, results in a shallower temperature gradient and warmer outflow at high radii. This appears to be the result of a larger flux of low energy photons being absorbed at $\tau(EUV)=1$ surfaces at high fluxes.

\begin{figure*}
    \centering
    \includegraphics[width=\textwidth]{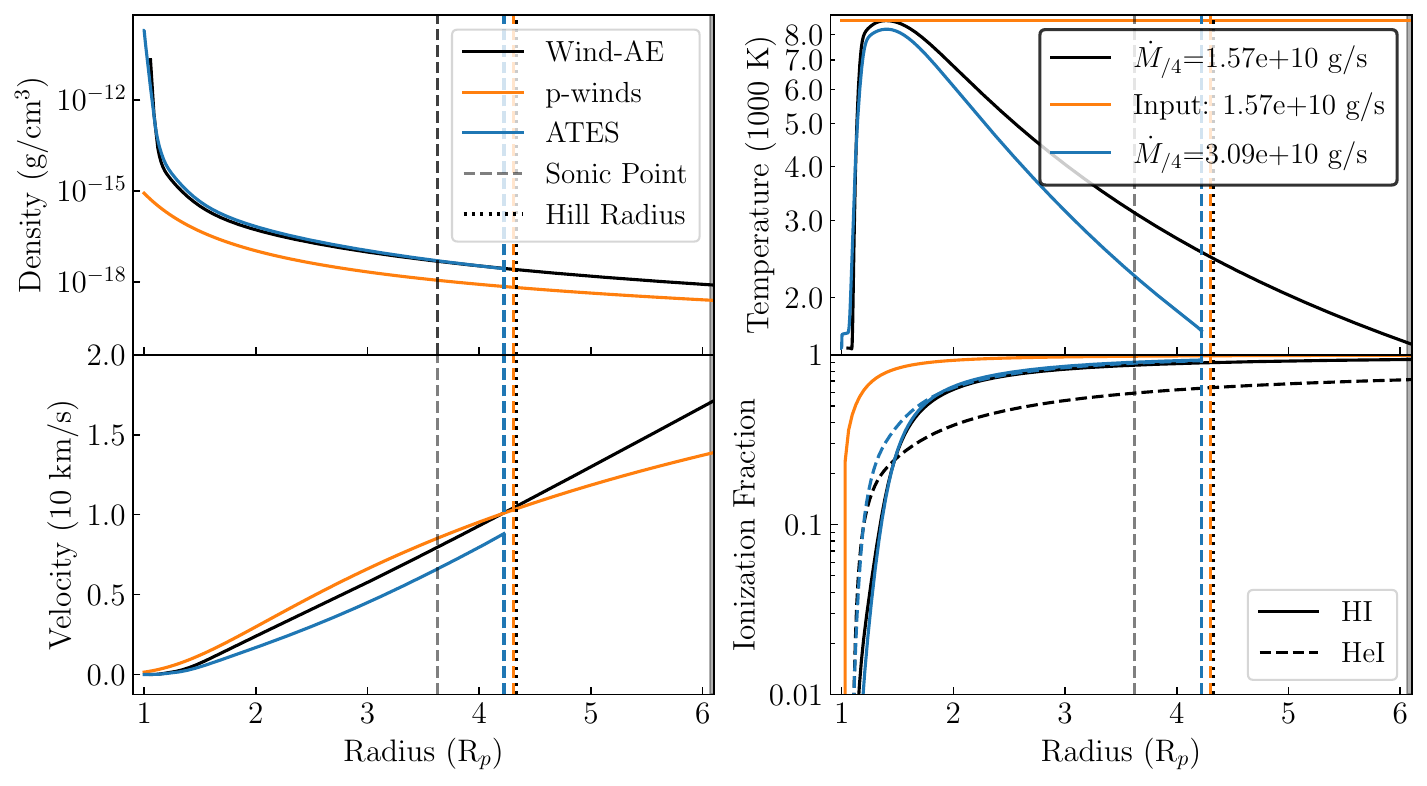}
    \caption{\textbf{HD 209458b H-He comparisons with ATES and \texttt{p-winds}} - \texttt{Wind-AE} H-He (black), ATES H-He \citep[blue,][]{ates}, and \texttt{p-winds} pure-H \citep[orange,][]{pwinds} profiles for HD209458b. Inputs to \texttt{p-winds} are $T=8616$ and $\dot{M}=1.57\times10^{10}$ g s$^{-1}$ (the $\dot{M_{/4}}$ computed by \texttt{Wind-AE}). Sonic points are plotted as dashed vertical lines and the Hill radius as dotted vertical. All three employ different approximations to solar spectra.}
    \label{fig:pwinds}
\end{figure*}

We also benchmark against other relatively fast and simple 1D substellar models such as \citet{ates} (ATES). ATES is a 1D multi-frequency XUV H-He steady-state Godunov-type hydrodynamic code. ATES models the similar ionization and heating/cooling sources to \texttt{Wind-AE}, but ATES also includes free-free and collisional heating/cooling and does not include secondary ionizations (all of which are negligible for HD 209458b with an H-He atmosphere \citep{rmc2009}). ATES includes advection in post-processing, which is limits the accuracy of ionization profiles for low flux planets, but still produces accurate outflow rates. The post processing step also currently requires manual intervention, resulting in a slightly longer runtime than our model.

For the comparison in Fig. \ref{fig:pwinds}, we run ATES for HD 209458b and set the ATES boundary conditions to match the physical BCs computed for the base of the simulation by \texttt{Wind-AE}: $\log_{10}(n_0)=13.04$, $T_{eq}=1535$ K, and H/He number ratio = 0.0629. ATES employs a powerlaw estimate for the stellar SED \citep{ates}, so to match the total flux and the ratio of X-ray to EUV flux, we set the log X-ray luminosity in ATES to 26.89 and log EUV luminosity to 27.8. This difference in SED is the source of the steeper temperature gradient at higher radii and the lower velocity in the ATES profile. The two models are otherwise in good agreement and ATES takes a similar amount of CPU time, but longer wall time, to run.

We also compare to \citet{pwinds}'s \texttt{p-winds}, a 1D multi-frequency steady-state XUV isothermal Parker wind backwards model with multispecies capabilities including line cooling, but not including the X-ray physics detailed in \S\ref{methods:multi}. Because \texttt{p-winds} is isothermal it is incredibly computationally inexpensive, but not a comparable model to the non-isothermal ones listed in this section. We set the input $\dot{M}$ to \texttt{p-winds} equal to the $\dot{M_{/3}}=2.1\times10^{10}$ g s$^{-1}$ computed by \texttt{Wind-AE} and the input isothermal temperature equal to the temperature at the peak of \texttt{Wind-AE}'s temperature profile (T=8616 K). \texttt{p-winds} does not converge for isothermal temperature equal to the average temperature (T=3561 K) of our model. For the \texttt{p-winds} solution, we employ \texttt{p-winds}'s default solar spectrum without any modifications, meaning that integrated flux over 13.6-2000 eV is $F_{\rm{tot}}=1342$ $\ergs$ vs. the $F_{\rm{tot}}=1095$ $\ergs$ in \texttt{Wind-AE}. The results for an H-He atmosphere (0.8:0.2 mass fraction, 0.94:0.6 number fraction) are shown in Figure \ref{fig:pwinds}. Depending on how close in parameter space \texttt{Wind-AE}'s initial guess is and the number of metals present in the wind, \texttt{p-winds} either runs in similar time or is up to 1000 times faster.

\begin{figure*}
    \centering
    \includegraphics[width=0.8\linewidth]{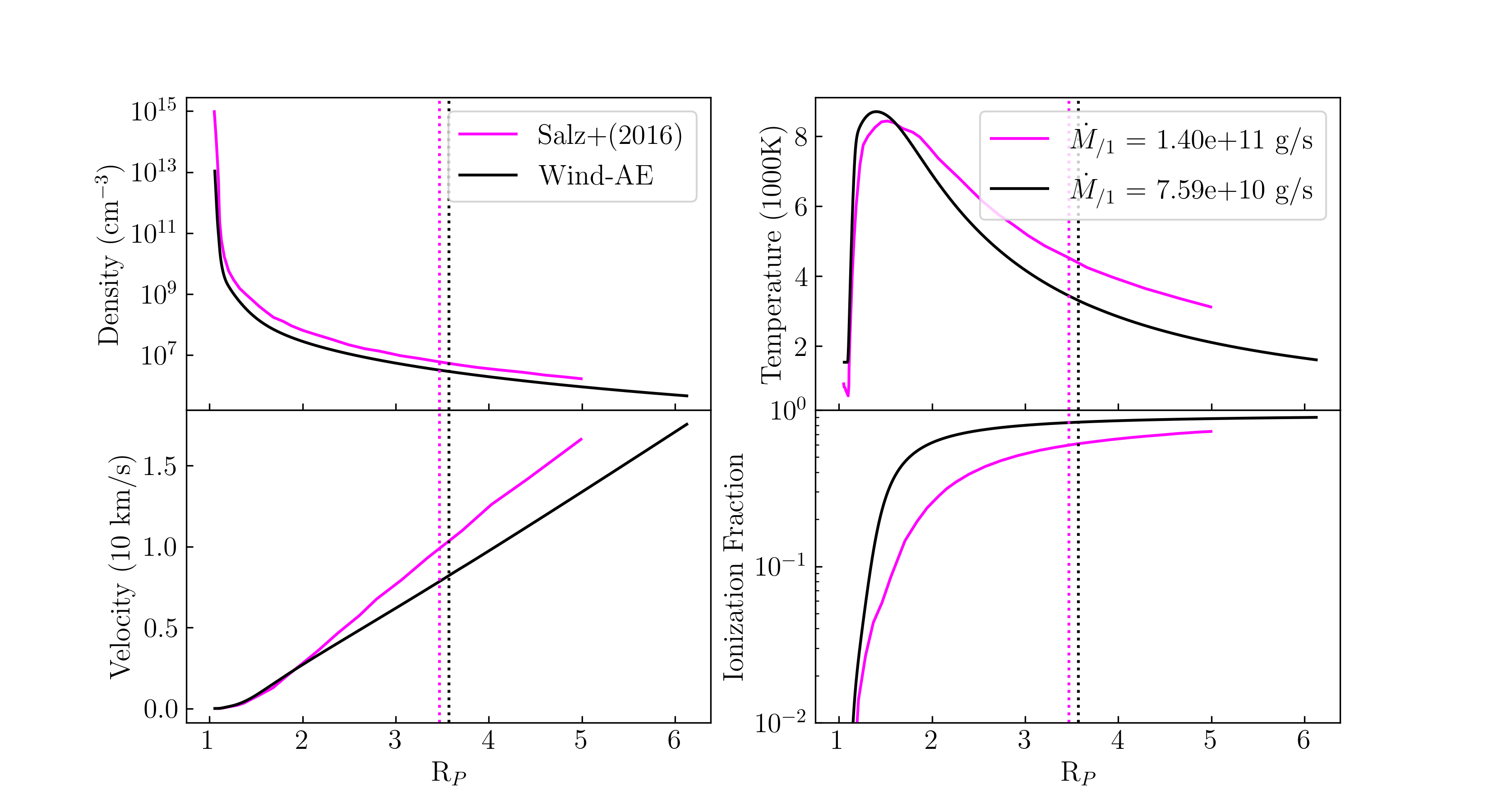}
    \caption{\textbf{Pure-H HD 209458b TPCI Model} - \citet{salz_simulating_2016} (TPCI, magenta) has lower BCs of $T(R_{\rm{min}})=1000$K, $R_{\rm{min}}\sim1.05R_P$, $\rho(R_{\rm{min}})\sim1.6\times10^{-9} g\ cm^{-3}$. We have confirmed that these boundary conditions are not the source of the profile differences.}
    \label{fig:tpci}
\end{figure*}

ATES, \texttt{p-winds}, and \texttt{Wind-AE} all make simplifying assumptions to avoid the expense of non-LTE photochemical computation; however, several 1D models employ the 1D multispecies photoionization solver CLOUDY \citep{CLOUDY} to handle the gas microphysics and photochemistry within the outflow and/or below the wind. \citet{sunbather} (Sunbather) integrates CLOUDY and \texttt{p-winds}, \citet{kubyshkina_precise_2024} (CHAIN) integrates CLOUDY and \citet{kubyshkina_young_2018}, and \citet{salz_simulating_2016} (TPCI) integrates CLOUDY and the 3D MHD model PLUTO \citep{mignone_pluto_2012}. We compare to the GJ 1214b outflow profiles from \citet{kubyshkina_precise_2024} for the latter two here. TPCI is the PLUTO-CLOUDY interface, a steady-state 1D multi-frequency XUV, multispecies solver and contains the same ionization and heating/cooling terms as \texttt{Wind-AE} and also free-free, conductive, and collisional heating/cooling (which are negligible for GJ 1214b). Our outflow profiles are in good agreement for HD 209458b with a pure-H atmosphere (Fig.\ \ref{fig:tpci}) with 1D structure differences likely stemming from differences in the Lisird low activity sun spectrum used \citep{salz_simulating_2016}. 
As in \texttt{Wind-AE}, TPCI is able to self-consistently compute the heating efficiency. 

\begin{figure*}
    \centering
    \includegraphics[width=\textwidth]{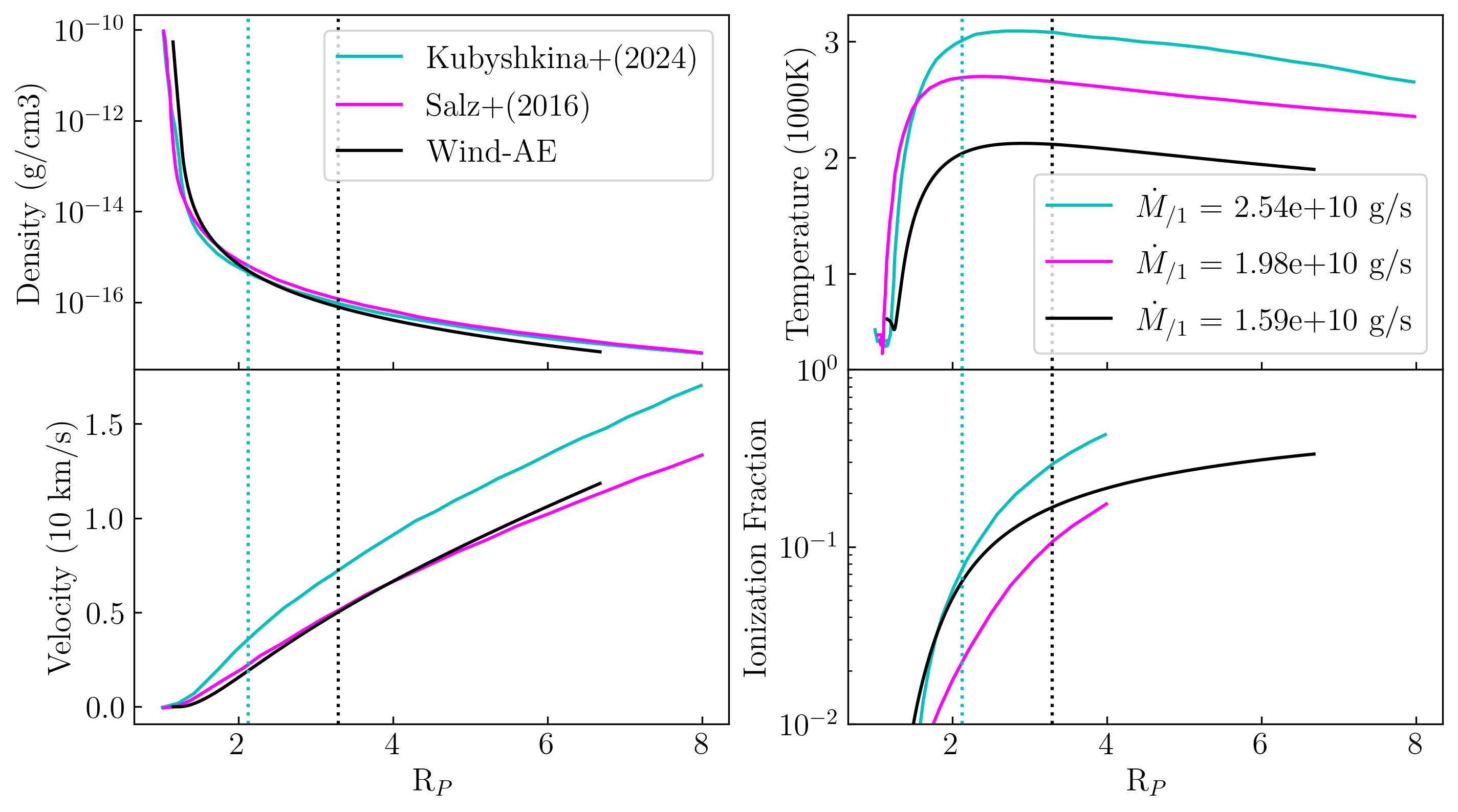}
    \caption{\textbf{H-He GJ 1214b CHAIN and TPCI Models} - GJ 1214b: 6.36 $M_{\oplus}$, 2.69 $R_\oplus$, 0.014 au, 0.18 $M_\odot$, 3.51$\times10^{-3}$ $L_\odot$, 886 $\ergs$. \citet{kubyshkina_2022} CHAIN solar spectrum solution (cyan) and \citet{salz_simulating_2016} TPCI solution (magenta) are reproduced here.}
    \label{fig:TPCI_kub}
\end{figure*}

CHAIN contains the same ionization and heating sources as TPCI, and, additionally, includes H$_3^+$ cooling and secondary ionizations through CLOUDY \citep{kubyshkina_precise_2024}. We model GJ 1214b with the parameters listed in Table 1 of \citet{kubyshkina_precise_2024} and with the scaled solar spectrum with integrated flux $F_{\rm{tot}}=886$ $\ergs$ for the range 13.6-2000eV (Fig.\ \ref{fig:TPCI_kub}). \citet{kubyshkina_precise_2024} attribute the differences from TPCI in the GJ 1214b outflow profiles to differences in the stellar spectra that each model used, resulting in different heating in the lower atmosphere. Despite significant temperature profile differences likely stemming from SED differences, we find good agreement with $\dot{M}_{/1}=1.58\times10^{11}$ g/s. We confirm that lower boundary condition differences are not the source of structure differences for both TPCI and CHAIN. This is consistent with our findings for other low flux hot Jupiters.

\begin{figure*}
    \centering
    \includegraphics[width=\textwidth]{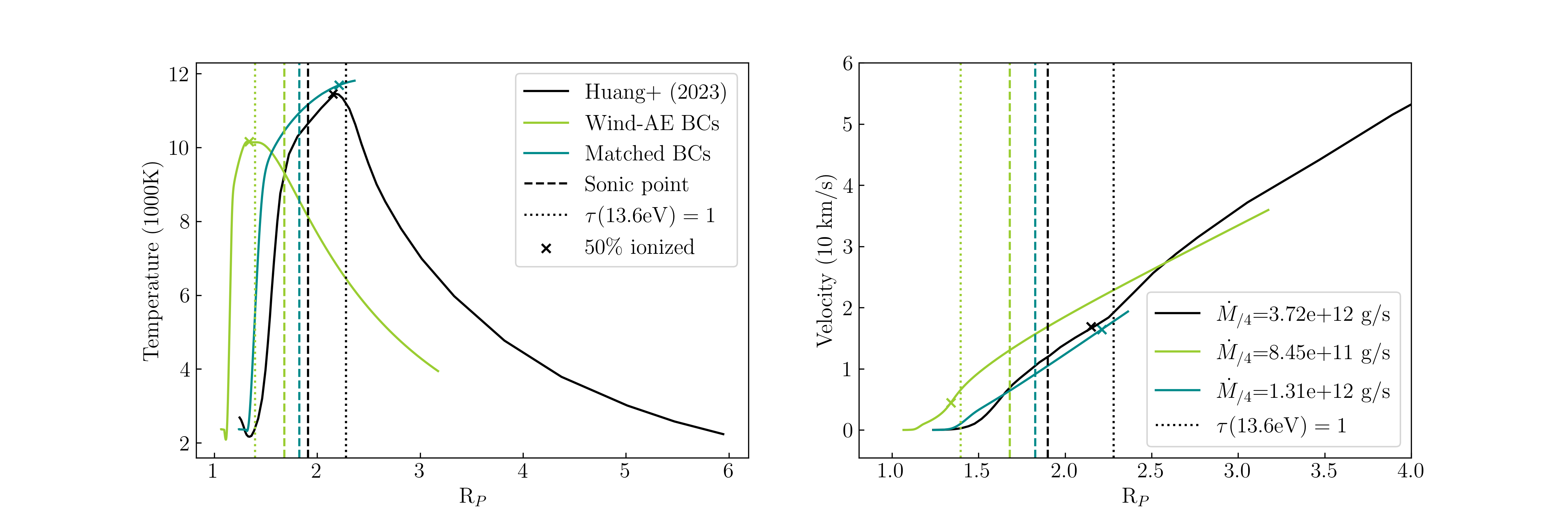}
    \caption{Metals WASP-121 b (\textbf{\citet{huang_hydrodynamic_2023}}) - WASP 121b: 1.18 $M_J$, 1.77 $R_J$, 0.03 au, 1.35 $M_\odot$, 2.6 L$_\odot$, $F_{\rm{tot}}=6.66\times10^{4}$ $\ergs$. Case B (\citet{huang_hydrodynamic_2023}, Fig. 20) with solar abundances of H I, He I, C I, N I, O I, S I, Mg II, Si II, Ca II, Fe II, K I, and Na I is reproduced in black. \texttt{Wind-AE} models for solar abundances of all of the above species except  Fe II, K I, and Na I using our lower BCs (green) and matching BCs (blue) of $R_{\rm{min}}=1.24$ and $\rho(R_{\rm{min}})=1.39\times10^{-11}$ g cm$^{-3}$. Both have the same input stellar spectrum as black.}
    \label{fig:huang}
\end{figure*}

Our lower boundary conditions do turn out to make a significant impact on the profile of WASP-121 b when compared with \citet{huang_hydrodynamic_2023} (Fig. \ref{fig:TPCI_kub}). This model uses the \citet{koskinen_mass_2022}  atmospheric escape model (an update of \citet{koskinen_escape_2013}) with a photochemical hydrostatic model 100 bar and 1 $\mu$bar, where the escape model takes over. This model is extremely comprehensive and includes diffusion, drag, charge exchange, thermal ionizations, and variety of other nuanced physics calculations which our model does not include. When we use the bolometric heating/cooling and molecular layer to compute the lower boundary conditions as in the main body of this text, \texttt{Wind-AE} produces the lime green profile and $\dot{M}_{/4}=8.45\times10^{11}g\ s^{-1}$ in Figure \ref{fig:huang} for an atmosphere that consists of H I, He I, C I, N I, O I, S I, Mg II, Si II, Ca II, and Fe II. To eliminate spectral differences as the source of any discrepancies, we use the same spectrum as \citet{huang_hydrodynamic_2023}.

Notably, when we manually set our lower boundary conditions to match Case B of \citet{huang_hydrodynamic_2023} (H I, He I, C I, N I, O I, S I, Mg II, Si II, Ca II,  Fe II, K I, and Na I), we are able to compute a profile and mass loss rate ($\dot{M}_{/4}=1.3\times10^{12}g\ s^{-1}$ ) similar to that found by \citet{huang_hydrodynamic_2023} ($\dot{M}_{/4}=3.7\times10^{12}g\ s^{-1}$ ) in less than 1/100th of the computational time.

HD 209458b and GJ 1214b are old, low-flux planets whose mass loss is in the energy-limited regime and dominated by EUV flux, with the presence of X-rays having relatively little effect on pure-H and H-He atmospheres. Outflows in the high flux limit, on the other hand are considered recombination limited, young stars exhibit a higher proportion of higher X-ray flux relative to the bolometric and EUV flux, and Lyman-$\alpha$ cooling becomes more significant. To test our model in the high flux limit we compare to two high XUV flux models, the first of which is \citet{owen_planetary_2012}. 

We model a H-He atmosphere for a 1.72 R$_J$, 1 $M_J$ and a 5 R$_J$, 1 $M_J$ planet at 0.1 au around solar-type star with X-ray luminosity of 10$^{30}$ ergs s$^{-1}$ to compare outflow profiles with those in \citet{owen_planetary_2012} Figure 4. We scale our solar spectrum such that the flux in the X-ray between 100 eV and 2000 eV corresponds to $L_X=10^{30}$ ergs s$^{-1}$. The net EUV flux between 13.6-2000 eV is then $F_{\rm{tot}}=3.57\times10^{5}$ $\ergs$. \citet{owen_planetary_2012} make simplifying assumption that the outflow is in ionization equilibrium, which is valid in the high XUV flux limit where the outflow is ``recombination limited". As a result, the \citet{owen_planetary_2012} analytic solution is not frequency dependent, making our choice of a scaled modern-day solar spectrum adequate for the purposes of comparing--though attempts to model a planet around a young XUV active star should take into account the higher ratio of X-ray flux to EUV flux for younger stars.

\begin{figure*}
    \centering
    \includegraphics[width=\textwidth]{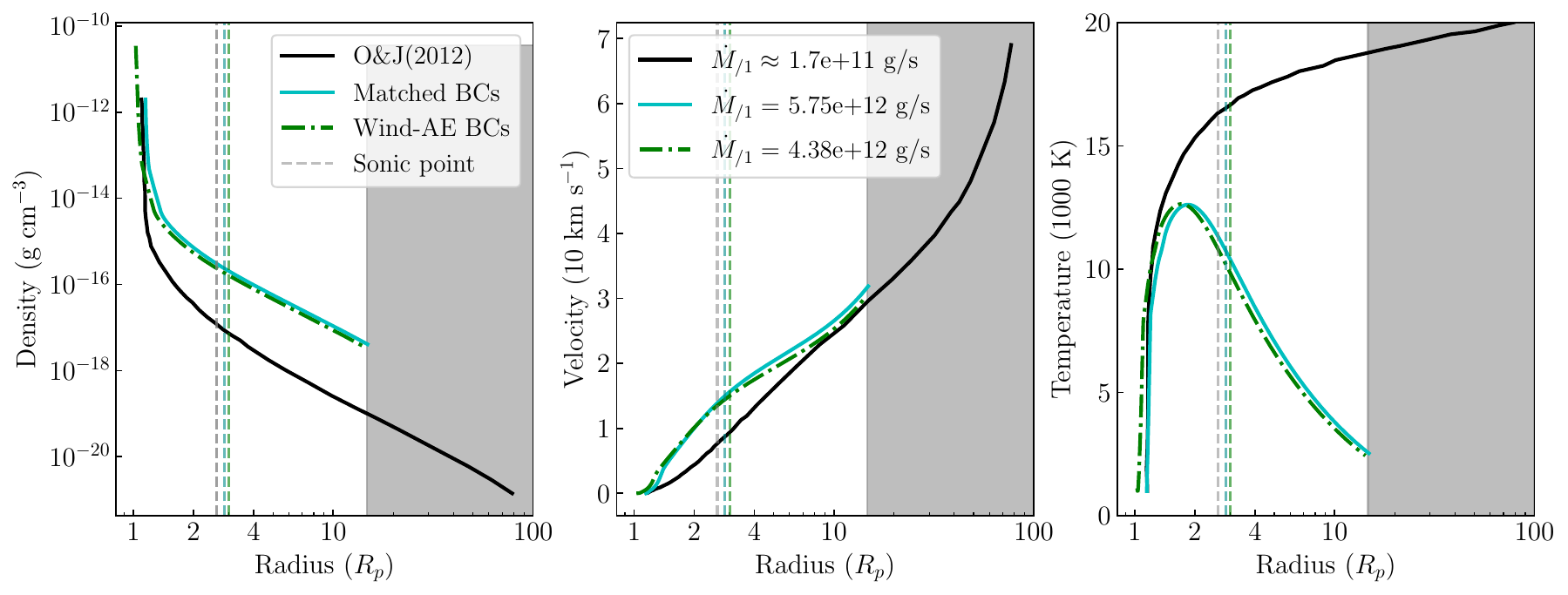}
    \caption{\textbf{Profiles for Planets in the High X-ray Flux Limit} - Reproduction of \citet{owen_planetary_2012} Fig. 4 (black), a model that does not contain PdV or Ly-$\alpha$ cooling. Density, velocity, and temperature results are given for a 1 $M_J$, 1.73 $R_J$ (dark) and 5 $R_J$ (light) planet at 0.1 au irradiated by a 1 $M_{\odot}$ star with X-ray luminosity $L_X=10^{30}$$\ergs$. Wind-AE profiles are plotted for planets with physical BCs (green) and BCs that match those of the black (cyan). To match $L_X=10^{30}$ ergs cm$^{-2}$, our planets are irradiated by a scaled solar spectrum of total flux 13.6-2000 eV of $3.57\times10^{5}$$\ergs$. Sonic points are plotted as dashed vertical lines.}
    \label{fig:james_planet}
\end{figure*}

\begin{figure}
    \centering
    \includegraphics[width=0.5\linewidth]{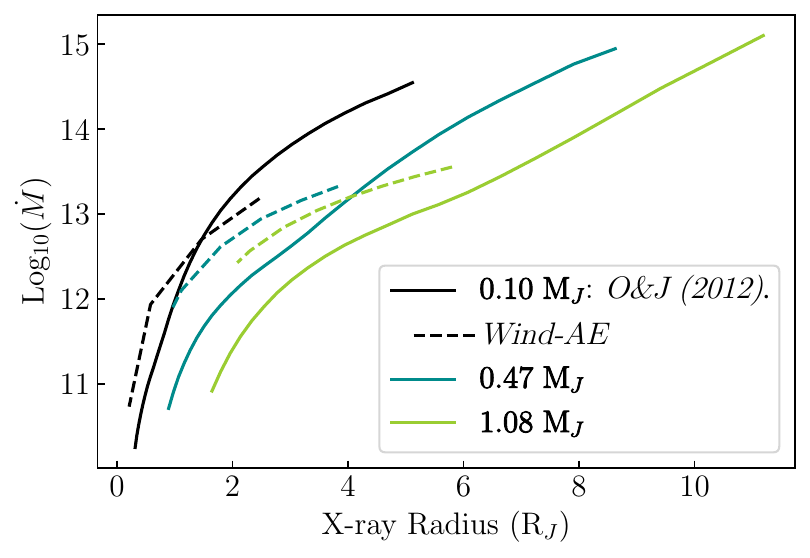}
    \caption{\textbf{\citet{owen_planetary_2012} Mass Loss Grid} - Solid lines are slices at 0.1 $M_J$ (black), 0.47 $M_J$ (dark cyan), and 1.08 $M_J$ (green) through the mass loss grid for planets at 0.1 au irradiated by a 1 $M_{\odot}$ star with X-ray luminosity $L_X=10^{30}$ ergs cm$^{-2}$ from \citet{owen_planetary_2012} Figure 5. Note that the radii in that figure are not optical transit radii, but the approximate $\tau$(1 keV)$=1$ radius. So we plot the log of geometrically-averaged $\dot{M}_{/3}$ in g s$^{-1}$ as a function of $\tau$(1 keV)$=1$ for planets of the same masses (dashed). The dashed solutions do not extend to higher X-ray radii because planets with higher $R_P$'s enter the low escape velocity regime in which the relaxation code is currently not well characterized (see \S\ref{discuss:lowgrav}).}
    \label{fig:james_grid}
\end{figure}

With the inclusion of Lyman-$\alpha$ cooling, PdV cooling, and solving the ionization balance equation, we find that the outflow is much cooler at high altitudes and stays much denser throughout (Fig.\ \ref{fig:james_planet}). Even when we remove the bolometrically-heated/cooled molecular layer and match the lower boundary temperature and density of \citet{owen_planetary_2012} these features remain. As a result, we predict higher geometrically averaged mass loss rates ($\dot{M}_{/3}$) for planets with higher escape velocities and lower mass loss rates for lower escape velocities than predicted in the \citet{owen_planetary_2012} mass loss grid (Fig.\ \ref{fig:james_grid}).

\begin{figure*}
    \centering
    \includegraphics[width=0.7\linewidth]{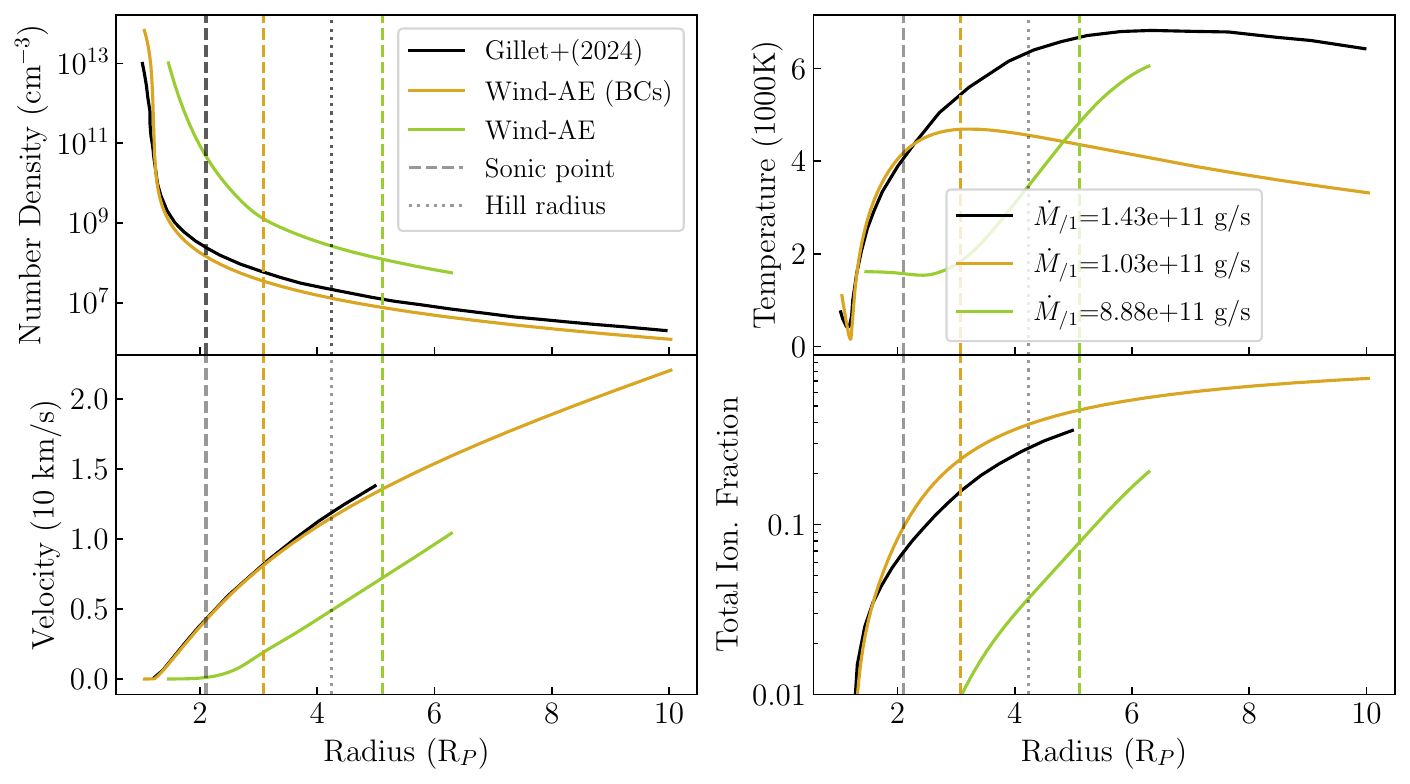}
    \caption{\textbf{Gillet et al. (2024) Secondary Ionization Profiles} - We reproduce profiles from \citet{gillet_self-consistent_2023} Fig. 7 for a 0.05$M_J$, 0.55$R_J$ planet at 0.045 au around a solar-type star for a pure-H outflow modeled using PLUTO with secondary ionizations enabled (blue). \texttt{Wind-AE} profile for the same planet with the same solar spectrum, but with bolometric heating and cooling and $\mu=2.3 m_H$ for $r<\rxuv$ enabled (green) and without (yellow). Gray boundary conditions at $R_{\rm{min}}$=1.1$R_P$ are matched to blue (T=1100K, $\rho$=1.326$\times10^{-10}$g/s, P=12$\mu$bar).}
    \label{ap-fig:gillet}
\end{figure*}

Fig. \ref{ap-fig:gillet} provides a further illustration of the importance of the lower boundary conditions in determining outflow structure.  We match the outflow structure of a low escape velocity planet modeled by \citet{gillet_self-consistent_2023} (black) when we match their boundary conditions (yellow).  However, when we take into account bolometric heating and cooling, the planet is significantly inflated, generating a very different outflow structure. Relative to results that do not employ lower atmosphere models, our boundary condition implementation finds significantly higher $\rxuv$ for low-escape-velocity planets (Fig.\ \ref{ap-fig:gillet}). 

The lower atmospheric model included in \citet{huang_hydrodynamic_2023} is more sophisticated that that used by default in \texttt{Wind-AE}. The fact that \texttt{Wind-AE} provides an excellent match to this work's outflow structure if and only if the lower boundary conditions are matched (Fig. \ref{fig:huang}) indicates that modeling the region below the wind as a simple energetic balance between bolometric heating and cooling with a fixed mean molecular weight has the potential to underestimate the microbar radius and thereby the wind launch radius ($\rxuv$).
Fortunately, \texttt{Wind-AE} can be easily coupled to lower atmosphere models whose outputs of radius, density, temperature, metal abundances, ionization fractions, mean molecular weight, etc., can be easily fed as inputs into \texttt{Wind-AE}. 
The default lower-atmosphere model in \texttt{Wind-AE} is nevertheless an improvement over setting boundary conditions at a fixed multiple of the optical transit radius and is valuable for increasing accuracy of $\rxuv$ when more sophisticated lower-atmospheric modeling is not available.

\section{Boundary Conditions} \label{appendix:bcs}
The upper boundary condition, the sonic point, $R_{\rm{sp}}$, is where $v(R_{\rm{sp}}) = c_s(R_{\rm{sp}})$. This is a natural critical point that emerges from the transonic Parker Wind solution (see \citet{rmc2009} Equations 15-16). A discussion of the lower boundary conditions follows.

\subsection{Bolometric Heating and Cooling}
Though the molecular layer below $\sim10^{-9}$ bar is optically thick to most XUV radiation, it can still be heated by the bolometric flux from the star, which peaks in the optical.
The bolometric heating and cooling naturally enforce an isotherm at the skin temperature below the wind (we do not model molecular line cooling by, e.g., $H_{3+}$). The balance of bolometric heating and cooling allows us to analytically estimate more physical lower boundary condition values at $R_{\rm{min}}$.

We select the lower boundary condition such that the majority of the flux is captured in the wind. For computational efficiency, we choose this to be the 1-microbar radius, but the pressure is customizable. The $\tau(\nu)=1$ surface for the highest energy photons ($\gtrsim$ 1 keV) is below 1 microbar; however, we find that the contributions of these highest energy photons to the mass loss rates in typical systems to be negligible. If concerned with modeling more precise ionization fractions of a species as a function of radius one may set the base of the simulation to a higher pressure. 

The balance between thermal emission and incident stellar bolometric radiation sets the skin temperature, $T_{\rm{skin}}$, in the region below the wind and above the $\tau_{\rm{IR}}=1$ surface.
Above nanobar pressures, the molecules are photodissociated and become atomic in the wind for most planets (though super-Earths may be the exception (see Frelikh et al., submitted). Within the wind, photoionization heating, Lyman-$\alpha$ cooling, and $PdV$ cooling dominate, setting the temperature of the wind throughout the flow.

The addition of bolometric heating and cooling to this investigation allows the relaxation code to physically solve for where the photoionization heating begins to dominate and the wind launches. Since the $\tau(\nu)=1$ surfaces will be different for photons of different frequencies, $\nu$, and will depend on the metallicity and metals present (via the ionization cross section, $\sigma_s(\nu)$), estimating the radius of the wind base is non-trivial. Additionally, the mass loss rate and wind structure are sensitive to the pressure where the wind is launched, because deeper penetration tends to lead to cooler, slower winds with lower mass loss rates (see Fig.\ \ref{fig:multifreq}).

To set the $T_{\rm{skin}}$ isotherm below the wind, the bolometric flux, $F_{*} = \frac{L_*}{4 \pi a^2}$ is computed from the bolometric luminosity, $L_*$, and semi-major axis, $a$. The skin tempearture (\ref{ap-eq:tskin} is obtained by setting the optically thin bolometric heating, $\Gamma_{\rm{bolo}}$ (taken as the sum of direct and indirect bolometric heating), and bolometric cooling, $\Lambda_{\rm{bolo}}$, equal
\begin{align} \label{ap_eq:boloheat}
    \Gamma_{\rm{bolo}}  &= F_{in,\rm{opt}} \kappa_{\rm{opt}} \rho(r)+ F_{in,\rm{IR}} \kappa_{\rm{IR}} \rho(r)\\
    &= F_{*} \rho(r) \left(\kappa_{\rm{opt}} + \frac{1}{4} \kappa_{\rm{IR}}\right) \\
    \Lambda_{\rm{bolo}} &= 2\sigma_{SB} T_{\rm{skin}}^4 \rho(r) \kappa_{\rm{IR}}.
\end{align}
where $F_{in,\rm{opt}}=F_*$ captures the direct bolometric heating from the star and $F_{in,\rm{IR}}=\sigma_{SB}T_{eq}^{1/4} = \frac14 F_*$ is  via the gray slab approximation for re-emission of IR radiation. Then, using the fact that $\Rp$ for most planets given is the \textit{slant path} optical surface, we can solve geometrically using $\tau_{\rm{slant,opt}} = \kappa_{\rm{opt}} l \rho(R_P) = 1$ where $l$ is the slant path length of stellar optical photons through the transiting planet's atmosphere. To first order, $l = \sqrt{8H_{sc}R_P}$, where the scale height $H_{sc}$ is computed using the skin temperature and mean \textit{molecular} weight (as opposed to atomic, as is usually the case in the wind). The default adjustment for the mean molecular weight is a factor of 2.3 to account for molecular hydrogen, but users have the ability to customize this value and are recommended to do so for more metal rich atmospheres. 

\subsection{Computing lower boundary temperature, radius, and density}
The skin temperature, then becomes
\begin{equation} \label{ap-eq:tskin}
    T_{\rm{skin}} = \bigg[ \frac{F_{*} (\kappa_{\rm{opt}} + \kappa_{\rm{IR}} /4)}{2 \sigma_{SB} \kappa_{\rm{IR}}} \bigg]^{1/4}.    
\end{equation}
The exact prefactors in this equation are dependent on geometry and various Eddington coefficients, but the impact of their exact choice is small, thanks to the quartic root. In Equation \ref{ap-eq:tskin}, $\kappa_{\rm{opt}}$=0.004 and $\kappa_{\rm{IR}}$=0.01 are the defaults for optical and infrared opacity, respectively. These values are no longer valid in the atomic, optically-thin wind, so we multiply by an error function to drop $\kappa_{\rm{opt}}$ and $\kappa_{\rm{IR}}$ to zero in the wind. Between the base of the wind and the $\tau_{\rm{IR}}=1$ radius the atmosphere is an isotherm at the skin temperature.

Density at $\Rp$ is then
\begin{equation}
    \rho(R_P) = \sqrt{ \frac{\mu_{mol}G M_P}{8 R_P^3 k_B T_{\rm{skin}}\kappa_{opt}^2}  }.
\end{equation}
For most planets, the pressure at $\Rp$ is of order 10 millibars. However, to ease the computational burden, we place the bottom of the simulation either at 1 microbar or, if the $\tau_{\rm{IR}}=1$ surface to vertically-incident photons, $R_{\rm{IR}}$ is above $\Rp$ we take the base of the simulation to be $R_{\rm{IR}}$. Most XUV rays are absorbed between micro- and nanobar pressures, with winds launching approximately around the 10 nanobar radius. 
At higher pressures than microbar pressures, the balance of bolometric heating and cooling enforces an isotherm at the skin temperature.

Because the $\kappa_{\rm{opt}}$ and $\kappa_{\rm{IR}}$ should drop off as the molecules are thermally- and photodissociated and the mean molecular weight should transition to the atomic, we need a way to enforce this transition occurring before the wind launches at $\rxuv$. To do so, we use a complementary error function, normalized so that its values limit to 1 and 0, because it allows us to modify the rate and location of the drop off as appropriate for a given profile:
\begin{align} 
    x &= \frac{v(r)-v_{\rxuv}}{kH_{sc,0}\frac{\Delta v}{\Delta r}|_{\rxuv}} \nonumber \\
    \mathrm{erfc^\prime}(x) &\equiv \frac{\mathrm{erfc}(x)}{\mathrm{erfc}(x)|_{\mathrm{max}}} \label{ap-eq:erfc}
\end{align}
where $v$ is the local comoving velocity as a function of $r$, $v_{\rxuv}$ is the velocity at $\rxuv$, and $\Delta v/ \Delta r$ is the slope of the velocity in the vicinity of $\rxuv$ so that $x$ is only a function of $r$. Here we elect to use velocity as a function of radius because the steep acceleration from 0 cm/s when the wind launches results in a complementary error function that drops to zero when the wind begins to accelerate at $\rxuv$. Then, $xH_{sc,0}=x\frac{k_BT(\rxuv) \rxuv^2}{\mu_{mol}GM_P}$ is some multiple $k$ of the scale height at $\rxuv$, where $\rxuv$ is computed as the radius at which photoionization heating begins to dominate over $PdV$ cooling. This creates an error function that drops from 1 to 0 over $k$ scale heights at $\rxuv$. The default value of $k$ is 1. The altitude in the atmosphere at which these high energy photons are absorbed (where $\tau(\nu)=1$) depends on the frequency $\nu$ of the incident photon. 
We compute $\rxuv$ and radial extent of the complementary error function postfacto as part of the process of polishing the relaxation solution to self consistency. As such, $\rxuv$ is not affected by our choices for Equation \ref{ap-eq:erfc}.

\begin{figure}
    \centering
    \begin{minipage}{0.45\textwidth}
        \centering
    \includegraphics[width=0.9\textwidth]{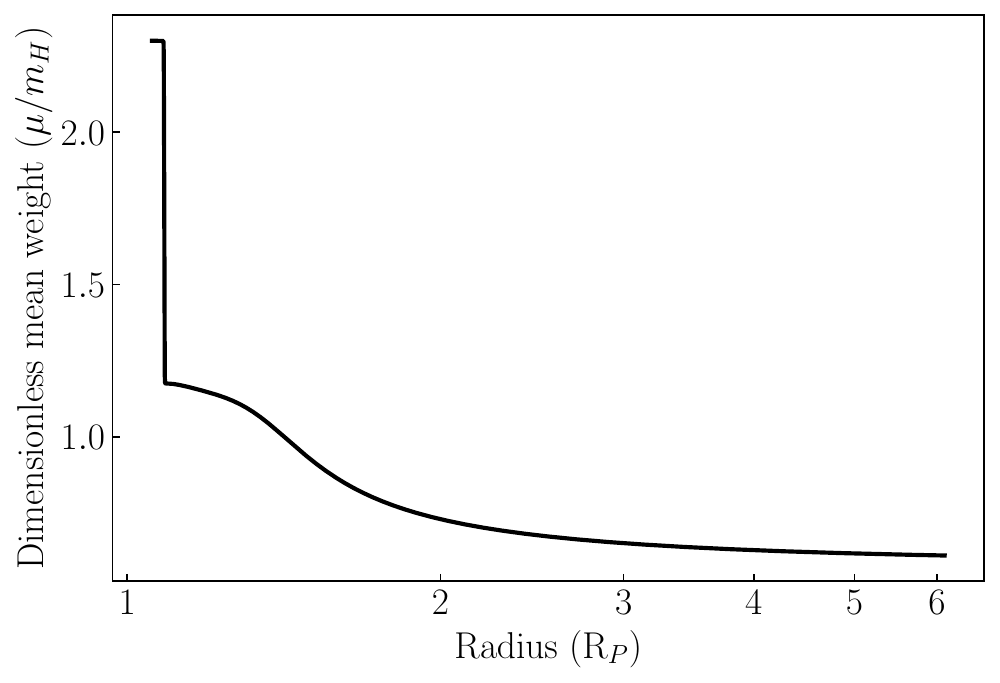}
    \caption{\textbf{Mean Weight Complementary Error Function Transition} - The transition from mean molecular weight $\mu(r<\rxuv)=2.3m_H$ to mean atomic weight as governed by Equation \ref{eq:mutrans} for HD 209458b. Here the erfc decays over 1 scale height at $\rxuv$.}
    \label{ap-fig:mu}
    \end{minipage}\hfill
    \begin{minipage}{0.45\textwidth}
        \centering
        \includegraphics[width=0.9\textwidth]{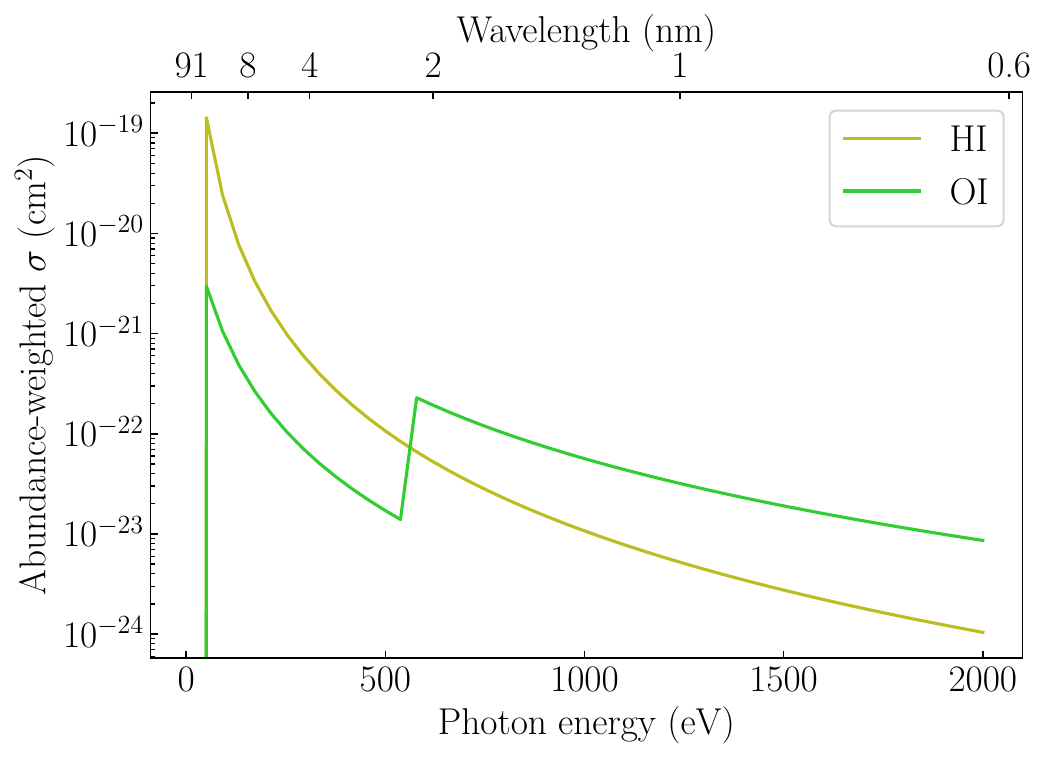} 
        \caption{\textbf{Abundance-weighted Ionization Cross-Sections} - Photoionization cross-sections for HI (olive) and OI (lime green) weighted by their fractional abundance (0.79332 and 0.00669, respectively) at $1\times$ solar metallicity as a function of photon energy (bottom axis) and wavelength (top axis). Frequency-dependent cross sections are derived from coefficients and analytic equations in \citet{verner-recombo,kshell}.}
    \label{ap-fig:sigma}
    \end{minipage}
\end{figure}

\begin{figure}
    \centering

\end{figure}

To compute the simulation lower boundary radius, $R_{\rm{min}}$, and density, $\rho(r=R_{\rm{min}})$, we use the hydrostatic equilibrium equation and the assumption of isothermality of the region between $R_{\rm{IR}}$ and the wind base to derive 
\begin{align} \label{ap-eq:Rmin}
    R_{\rm{min}}(P_{R_{\rm{min}}}) &= \bigg[ \frac{c_s^2}{GM_P} \ln{\bigg(\frac{P_{R_{\rm{min}}}}{P(R_P)}\bigg)} + \frac{1}{R_P}\bigg]^{-1} \\
    \rho(R_{\rm{min}}) &= P_{R_{\rm{min}}}\frac{\mu_{mol}}{k_B T_{\rm{skin}}},
\end{align}
where $P(R_P)$ is the pressure at $\Rp$ and $P_{R_{\rm{min}}}$ is the pressure at the base of the simulation, both in barye ($10^{-6}$ bars). This value should always be between 1 microbar and 1 millibar to capture the contributions from the highest energy photons and to stay in the isothermal portion of the atmosphere. 

For our assumptions of $T(R_{\rm{min}}=T_{\rm{skin}}$ to be reasonable, the atmosphere should be isothermal and hydrostatic between $R_P$ and $R_{min}$. The safest way to guarantee that our $R_{\rm{min}}$ falls within that isothermal region would be to set $R_{min} \geq R_{\rm{IR}}$ to $\tau_{\rm{IR}}=1$ radius to vertically-incident IR photons. Notably, $R_{\rm{IR}}$ is typically below $R_P$, the optical slant path radius of the planet in transit. As such, \texttt{Wind-AE} has a rarely activated condition that, should $R_{\rm{IR}}>R_P$, $R_{\rm{min}}=R_{\rm{IR}}$ and we can similarly solve from hydrostatic equilibrium to obtain
\begin{equation}
    R_{\rm{IR}} = \frac{R_P^2}{H_{sc} \ln\left({\frac{\rho(R_{\rm{IR}})}{\rho(R_P)} e^{R_P/H_{sc}}}\right)},
\end{equation}
where we approximate the scale height at $R_{\rm{IR}}$ to be the scale height at $\Rp$ and $\rho(r=R_{\rm{IR}})$ follows from $\tau_{\rm{IR}} = \rho(R_{\rm{IR}}) \kappa_{\rm{IR}}  H_{sc} = 1$.

\subsection{The high escape velocity limit: $R_{\rm{min}} > \rxuv$}
\begin{figure}
    \centering
    \includegraphics[width=0.5\linewidth]{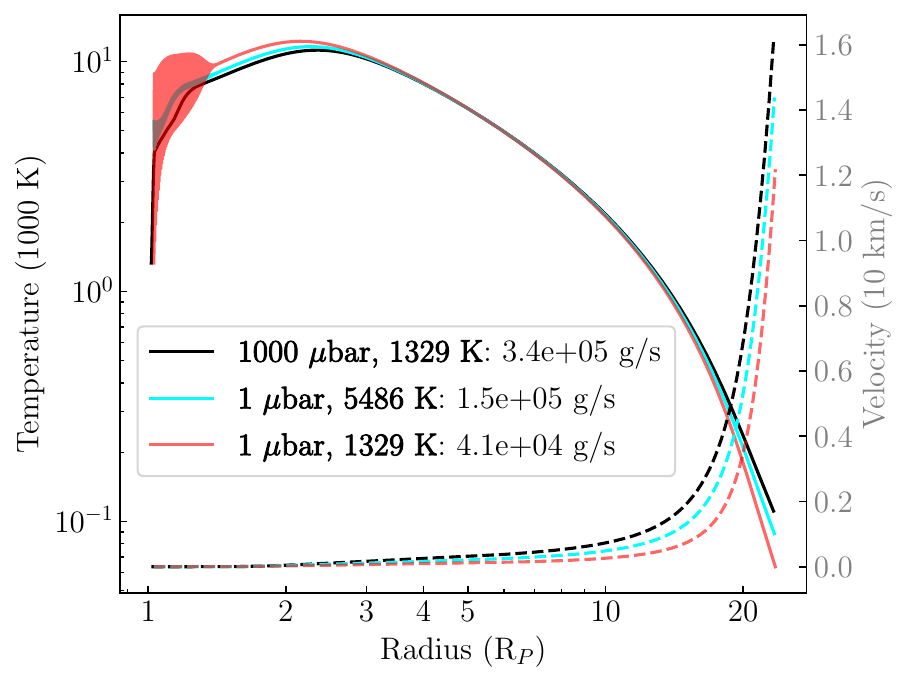}
    \caption{\textbf{Smoothing high-escape-velocity planet lower BC temperature oscillations} - 55 $M_\Earth$, 1.85 $R_\Earth$ at 0.05 au with $F_{\rm{tot}}$=1095 $\ergs$ temperature (right y axis, solid) and velocity profiles (left y axis, dashed). For a lower boundary pressure of 1 microbar, $R_{\rm{min}}=1.028 R_P$ which does not capture the bolometrically heated/cooled region, so $T(R_{\rm{min}}(1\mu\rm{bar}))\neq T_{\rm{skin}}=1324$ K (red) and results in an order of magnitude higher mass loss than the ground truth (black). Averaging the temperature oscillations tells us the $T(R_{\rm{min}}(1\mu\rm{bar}))=5486$ K (cyan) which puts the mass loss rate within a factor of 2. Because neither of these solutions capture the molecular region, there is no higher mean molecular weight or bolometric heating/cooling included. To confirm the validity of this approximation, we lower (black) the base pressure to 1000 microbar ($R_{\rm{min}}=1.016 R_P$) to capture the photoionization base and molecular region, thus making $T(R_{\rm{min}}(1000\mu\rm{bar}))=T_{\rm{skin}}=1324$ K.}
    \label{ap-fig:t_oscill}
\end{figure}

In cases high escape velocity cases and low flux cases such as those discussed in \S\ref{discuss:highgrav}, a larger fraction of XUV photons are absorbed below $R_{\rm{min}}(1\mu bar)$, so setting the base at $R_{\rm{min}}(1\mu bar)$ does not capture $\rxuv$. The correct approach would be to set $R_{\rm{min}}$ to a higher pressure in order to accurately capture the radius at which the wind launches. However, because of the small scaleheight for these planets, this approach is computationally costly and the relaxation method can fail. Therefore, in our model we opt to turn off the bolometric heating/cooling and other molecular layer assumptions, because that layer is not captured at $R_{\rm{min}}(1 \mu bar)$. 

Indeed, the the assumptions we make to derive our lower boundary conditions are no longer valid and holding $T(R_{\rm{min}})=T_{\rm{skin}}$ results in numerical temperature oscillations with radius. Instead, because $R_{\rm{min}}$ does not capture the true base of the wind, $T(R_{\rm{min}})$ should be several thousand degrees kelvin higher than $T_{\rm{skin}}$. Taking the average of the first five radial numerical temperature oscillations near the simulation base gives us a decent estimate of the what the $T(R_{\rm{min}})$ should be for this $R_{\rm{min}}$ located in the middle of the wind.

We confirm that this approximation is adequate by lowering the simulation base to, e.g., $R_{\rm{min}}(100 \mu bar)$, so that we accurately capture the molecular region and $\rxuv$ and compare our oscillation-averaged temperature at 1$\mu bar$ in the shallow solution to the actual temperature at $R(1\mu bar)$ in the deeper solution (Fig.\ \ref{ap-fig:t_oscill}). The density should also slightly change in this case, but we find that the effects of changing $\rho(R_{\rm{min}})$ to a more physically accurate value are secondary. In the highest escape velocity limits, setting $R_{\rm{min}}(1 \mu bar)$ and using the above method to find $T(R_{\rm{min}})$ is no longer sufficient as the planets' scale heights are so small that the region below $R_{\rm{min}}(1 \mu bar)$ is dense enough to absorb higher energy X-rays that contribute to heating and ionizing, thus we may be underestimating temperature and ionization fraction at $R_{\rm{min}}(1 \mu bar)$, as well as the total mass loss rate.


\section{Escape of Lyman-$\alpha$ Photons}\label{appendix:lyaesc}

The planetary winds modeled here are optically thick to Lyman-$\alpha$ radiation since, at line center, the cross-section for Lyman-$\alpha$ absorption by a neutral hydrogen atom is larger than the cross-section for photoionization.  Nevertheless, as illustrated in \citet[][their Appendix C]{rmc2009}, the wind is low enough density that the majority of Lyman-$\alpha$ photons emitted in the wind ultimately scatter into the line wings and escape before their energy can be returned to the thermal bath.  A Lyman-$\alpha$ photon's energy is thermalized if it excites an atom and the atom then experiences a collision, resulting in collisional de-excitation, before the atom has the chance to spontaneously de-excite and re-emit another Lyman-$\alpha$ photon.  

To validate the order-of-magnitude calculation in \citet{rmc2009}, we run a Monte-Carlo calculation of Lyman-$\alpha$ photon escape.  We use the hot Jupiter profile for a H-He HD 209458b provided in Figure \ref{fig:three_planet} and assume spherical symmetry.  We run 1000 photons, each beginning at radius $r = 1.18 R_p$, where the Lyman-$\alpha$ cooling rate peaks for this outflow.  At each step in the calculation, we draw a random direction and a Lyman-$\alpha$ frequency, $\nu$, from a Voigt profile at the local temperature and density, Doppler-shifted into the inertial frame according to the flow velocity, $v_{\parallel}$, along the drawn direction at the point of emission.  We use the integrated optical depth along that direction,
\begin{equation}
\int n_{0,{\rm H}}(r) \sigma_{\rm{abs}}(\nu,r, v_{\parallel}) dl     
\end{equation}
to draw a random distance at which the photon is absorbed.  The number density of neutral hydrogen, $n_{0,{\rm H}}$, and the cross-section for absorption at frequency $\nu$, $\sigma_{\rm{abs}}(\nu)$---which depends through the Voigt profile on temperature and density---are both functions of the radial coordinate $r$. We note that given the chosen direction, the variation of $r$ along the path is determined by geometry. Because our photon frequency is chosen in the inertial frame, the cross-section is also a function of the local bulk velocity along the chosen direction, $v_{\parallel}$, evaluated at each $r$.  

Once the photon is absorbed we calculate, using gas conditions at its new radial distance, the timescale, $t_{\rm col}$ on which it is expected to be de-excited by collision with an electron \citep[cross-section $2\times 10^{-15}$ cm$^2$;][]{Brackmann1958}, a proton \citep[cross-section $2\times 10^{-14}$ cm$^{2}$ including charge-exchange;][]{Hunter1977} or other species. For other species, we use the electron cross-section which is a reasonable approximation for typical collisions with a neutral hydrogen atom.  More detailed treatment of individual species is not merited because electron collisions dominate.  Given a spontaneous decay rate of $A_{21} = 6.265\times 10^8$ s$^{-1}$ for hydrogen Lyman-$\alpha$, the probability that the photon is thermalized before re-emission is $1-e^{-1/(A_{21}t_{\rm col})} \approx (A_{21}t_{\rm col})^{-1}$.  We draw a random number to determine whether the photon is thermalized at this step.  Because our example wind profile transitions to the bolometrically-heating regime below $1.1R_{p}$ and we do not trust detailed model conditions below this radius, we treat any photon that reaches a radius smaller than $1.1R_{p}$ as having escaped from our modeled flow.  The energy from these photons may be radiated away by bolometric or molecular radiation at depth, or they may be thermalized at depth, heating the gas below our simulation region and contributing to the physical processes that determine appropriate lower boundary conditions for our simulations. We consider a photon to have escaped the wind outward into space if it reaches a radius of $10R_p$.

We find that 72.4\% of our modeled photons escaped the wind outward to space and 27.5\% were lost through the lower boundary, with only 0.1\% thermalized in our simulation domain, demonstrating that a majority of the photons indeed escape.  We re-ran the calculation starting photons at $1.11R_p$, near the base of the wind simulation where Lyman-$\alpha$ excitation via secondary electrons is most important.  Though this starting point is very near to our auto-thermalization distance, 52.2\% of the 1000 modeled photons escaped outward to space, 47.5\% escaped through the lower boundary, and 0.3\% were thermalized in the simulation domain. 

In examples with higher incident flux, Lyman-$\alpha$ cooling is typically more important.  We therefore repeated this calculation for a planet with the same mass and radius but an incident flux 100 times larger.  We found that photons emitted from the peak of the Lyman-alpha cooling region escaped outward 65.6\% of the time, escaped through the lower boundary 33.4\% of the time, and were thermalized in the domain 1\% of the time.  For this example, no bolometric heating region is modeled, and photons started just above the lower boundary of the simulation escaped outward 40.1\% and inward 59.3\% of the time, with 0.6\% of photons thermalized in the domain.  

In both cases, a majority of the Lyman-$\alpha$ cooling radiation indeed escapes from the outflow.  For secondary electron energy that goes into Lyman-$\alpha$ excitation, the ultimate escape fraction depends on the true fate of photon energy that diffuses below the base of our simulation.  As discussed in Section \ref{methods:multi:equations}, our choice to treat this energy as escaping does not make a substantial difference to our results. 

\section{Multispecies \& Multi-frequency Versions of Finite Difference Equations}\label{appendix:finite}
To solve for Equations \ref{eq:mass_continuity} - \ref{eq:ion_eq_generic}, we update equations (9)-(13) of \citet{rmc2009} to include our multispecies and multi-frequency assumptions.

\begin{align}
    E_{1,j}&\equiv\Delta_j\rho-\frac{d\rho}{dr}\Delta_j r \nonumber\\
    &=\Delta_j\rho+\rho\left(\frac2r+\frac1v\frac{d v}{dr}\right)\Delta_j r =0 \label{ap-eq:rho}\\
    E_{2,j}&\equiv\Delta_jv-\frac{dv}{dr}\Delta_j r\nonumber\\
    &=\Delta_jv-\frac{v}{v^2-\gamma \lambda k T/\mu} \left[\frac{2\gamma\lambda kT}{\mu r}-\frac{(\gamma-1)Q}{\rho v} - \frac{GM_P}{r^2}+  \frac{GM_*}{(a-r)^2} + \frac12 \frac{GM_*(a-r(1+M_P/M_*))}{a^3}   \right]\Delta_jr=0 \label{ap-eq:v}\\
    E_{3,j}&\equiv\Delta_jT-\frac{dT}{dr}\Delta_j r\nonumber\\
    &=\Delta_jT-\left[(\gamma-1)\left(\frac{Q}{\rho v}\frac{\mu}{k} + \frac{T}{\rho} \frac{d\rho}{dr}\right)+\frac{T}{\mu} \frac{d\mu}{dr} \right]= 0
   \label{ap-eq:T} \\
   E_{4+s,j}&\equiv\Delta_j\Psi_s+\frac{d\Psi_s}{dr}\Delta_j r\nonumber\\
    &=\Delta_j \Psi_s -\frac{m_s}{Z_s \rho v}\left(\mathcal{R}_s 
 -\mathcal{I}_s \right)\Delta_j r = 0 \label{ap-eq:Ys}\\
    E_{(4+N_{sp})+s,j}&\equiv\Delta_jN_{\mathrm{col}}-\frac{dN_{\mathrm{col}}}{dr}\Delta_j r\nonumber\\
    &=\Delta_j N_{\mathrm{col}} + \Psi_s\frac{ Z_s\rho}{m_s} \Delta_j r =0 \label{ap-eq:Ncol}
\end{align}
where $j$ is the point along the radial grid, $Q=\Gamma+\Lambda$, and $Z_s$ is the mass fraction of species $s$. We write $N_{sp}$ as short-hand for $N_{\mathrm{species}}$ the number of species, $s$, in the wind, where $s$ ranges from $[0,N_{\mathrm{species}}-1]$. Therefore, the number of equations in the system of equations becomes 3+2$N_{\mathrm{species}}$, hence the computational expense of modeling more metals.

The final equation is given in terms of $N_{\mathrm{col},s}$, rather than $\tau$ as in \citet{rmc2009}, because $\tau_\nu=\sum_s \sigma_{\nu,s} N_{\mathrm{col},s}$ and individual species may ionize at different rates as a function of radius, so we must track the column density of each species individually in order to calculate $\tau$. 

The most important changes are summarized here. For a more detailed discussion of numerical and analytic changes, see \citet{McCann_2021}.

\section{Spectrum}\label{appendix:spectrum}
We avoid the cost of running a high resolution spectrum by fitting a polynomial to the input stellar spectrum. Any observed spectrum---such as the \texttt{FISM2} solar spectrum from the LISIRD database we use for this paper---or realistic simulated XUV spectrum will vary widely in flux and shape across the spectral range, as well as be high resolution, making fitting polynomials difficult. If the the spectral qualities can be well approximated by low degree polynomial(s), though, it is inexpensive to accurate perform numerical integrations using Gauss-Legendre quadrature. Thus, we use the smoothing and binning algorithm \citep[discussed in more detail in][\S2.3.5]{McCann_2021}.

Logarithmic fits and/or the least squares method would not locally (or, potentially, even globally) conserve energy along the spectrum, thus we employ a Savitzky-Golay filter, which smooths evenly-spaced noisy data with a rolling polynomial.
First we smooth the peaks the troughs of the spectrum by multi-passing the spectrum through the Savitzky-Golay filter.  The effect of running a Savitzky-Golay filter on small segment is similar to running a single pass filter on a larger wavelength range, but it distorts the data less than a standard, larger single pass filter and better preserves the area under the smoothed spectrum. The filtered spectrum is then renormalized to conserve total energy in each bin. Next a 5th degree polynomial is fit to the filtered spectrum, again rescaling to preserve energy in each bin. The polynomial is calculated by used a spline with with an infinite smoothing factor, which relaxes the spline to a single bin interval. 

Binning the spectrum allows us to run the multipass filtering in fewer smoothing passes and allows us to more accurately preserve the spectrum shape, especially at the ionization wavelengths of species present in the model. Subbinning, in particular, allows us to fit the spectrum with more low order polynomials, as opposed to fewer polynomials that would have to be higher order and would be more difficulty to accurately and cheaply integrate using Gauss-Legendre quadrature. For our bins, we choose bin width $2r$ centered at wavelength $\lambda_0$ such that the error over $\lambda\in[\lambda_0-r,\lambda_0+r]$  is less than $\epsilon$. Using the analytic \citet{verner-recombo} cross section relations, we can take the logarithmic derivatives of the Verner cross sections, $\sigma_{\lambda,s}$ for species $s$ and use the chain rule to derive the bin halfwidth, 
\begin{equation}
    r\leq  \left( \frac{6\epsilon}{||\sigma_{\lambda,s}^{(3)}||_{\infty}} \right)^{1/3}.
\end{equation}
Bin edges are also placed at the ionization energies and K-shell ionization energies of the species present, unless one of the ionization energies of an existing species is within 2 nm of an existing species' ionization energy.

The physical effects of smoothing a spectrum are also mitigated by using the above method. Ionizing energy is conserved since the peaks of the spectrum are smoothed and distributed locally---meaning there will be an equal amount of higher and lower than the peak ionization energy photons in the wind. That being said, at the edges of the spectrum, where there are not necessarily symmetric peaks over which to smooth, this method may over or underestimate the number of higher or lower energy photons. However, we take this smoothed approximation for a high resolution spectrum to be sufficient for our work.

\begin{figure}
    \centering
    \includegraphics[width=0.5\linewidth]{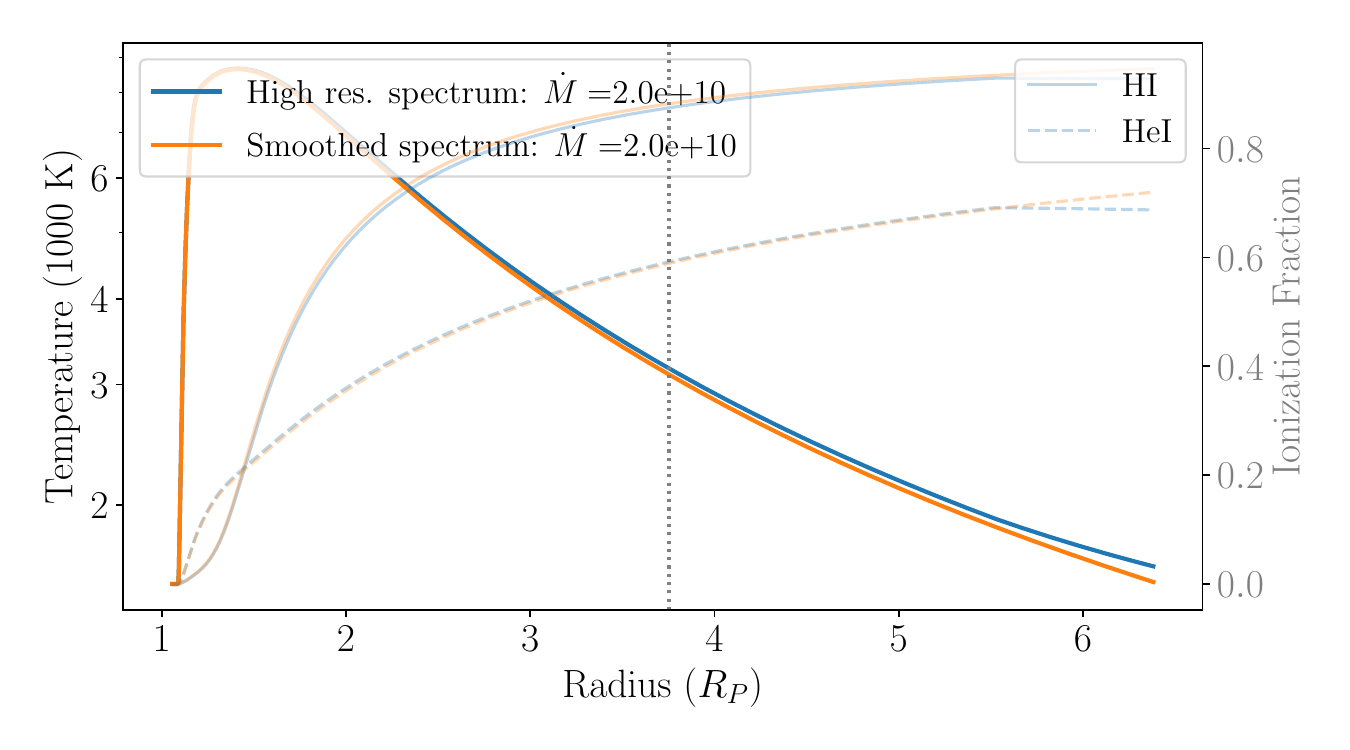}
    \caption{\textbf{High resolution scaled solar spectrum vs. smoothed - In blue, solutions using the FISM2 LISIRD solar spectrum scaled to an integrated flux of 1095 $\ergs$ and, in orange, solutions using the smoothed spectrum normalized to the same integrated flux (as in Figure \ref{fig:spectrum}). Heavier, darker lines are the temperature profile and lighter, thinner lines are the ionization fraction.}}
    \label{ap-fig:hires}
\end{figure}

\section{Metal Line-Cooling}\label{appendix:metal_line}
The metal line cooling rates for CII, CIII, OII, and OIII are computed via the emmisivity equation (Eq.\ \ref{eq:metal_line_cool}). Where $A'$ in that equation is the correctly scaled version of $A$ such that $\Lambda$ is the cooling rate per unit volume. \texttt{CLOUDY} \citep{CLOUDY} is then used to compute the coefficients $A$, $T_{line}$ (line transition temperature), and $n_c$ (critical number density).

\begin{figure}
    \centering
    \includegraphics[width=0.5\linewidth]{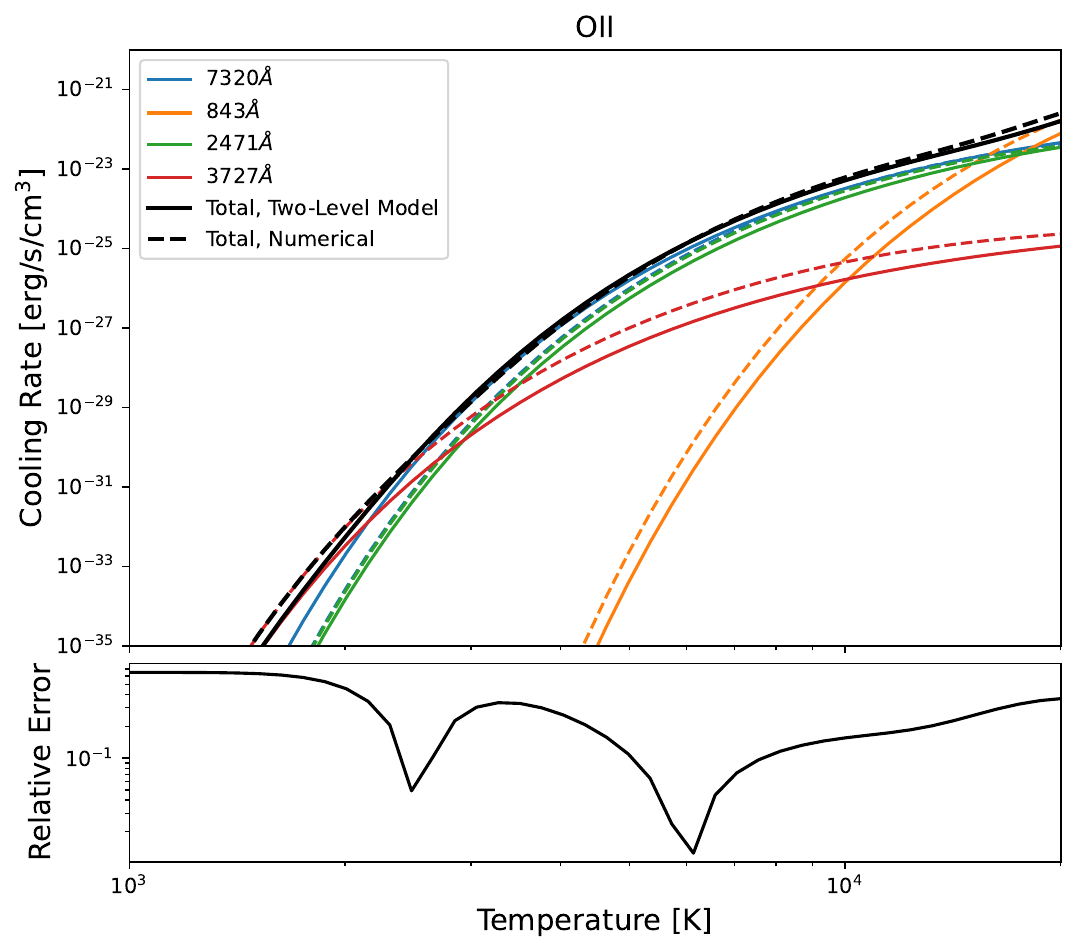}
    \caption{\textbf{Top:} the colored lines show a comparison of our two-level atom model fits to the cooling rates (solid) against the numerically computed cooling rates (dashed) for the individual transitions of OII. The black lines shows a comparison of the two level model for the 4 most relevant lines compared to the full numerical cooling rate including all the energy levels and transitions (still under the assumption of a statistically collisional equilibrium). The lines are shown for an electron density of 5.4556$\times10^8$~cm$^{-3}$. \textbf{Bottom:} the relative error between the total cooling rate computed from our two-level atom approach and the total numerically computed cooling rate, demonstrating our accuracy is at the 10-30\% level in the temperature range of interest.}
    \label{fig:two_level_compare}
\end{figure}

 We use {\sc chianti}\citep{chianti1,chianti8} to identify all the relevant lines and extract the values which are listed below. These fits provide total cooling rates that are typically good to 10-30\% and at worst a factor of two in extreme regions of the parameter space. We show a comparison of the two-level model fits compared to the numerical solution including all the energy levels and transitions (still under the assumption of a statistical collisional equilibrium) for OII in Figure~\ref{fig:two_level_compare}, demonstrating the fit's accuracy in the temperature range of interest.

\begin{table}[ht!]
\parbox{.45\linewidth}{
\centering
    \begin{tabular}{c|c|c|c}
         Line & $A'$ & $T_{\rm line}$ & $n_c$   \\
         \hline
         157$\mu$m & 1.783E-20  &  91.2 & 1.388E+01 \\
         2326\AA &  1.215E-10  & 61853.9& 1.210e+09 \\
         1334\AA & 2.413E-03 & 107718.1 & 3.740E+15 
    \end{tabular}
    \caption{Properties for the cooling functions of CII - all values in cgs units.}
    \label{tab:CII_table}
}
\hfill
\parbox{.45\linewidth}{
\centering
    \begin{tabular}{c|c|c|c}
         Line & $A'$ & $T_{\rm line}$ & $n_c$   \\
         \hline
         1910\AA &  3.84E-10& 75460.8 & 1.314E+9\\
         977\AA & 1.791E-03& 147263.9 & 7.172E+14\end{tabular}
    \caption{Properties for the cooling functions of CIII - all values in cgs units.}
    \label{tab:CIII_table}
}
\end{table}

\begin{table}[ht!]
\parbox{.45\linewidth}{
\centering
    \begin{tabular}{c|c|c|c}
         Line & $A'$ & $T_{\rm line}$ & $n_c$   \\
         \hline
         843\AA &  5.786E-4& 172421.6 & 1.322E15\\
         2471\AA & 3.812E-13& 58225.3 & 4.488E7\\ 
         3727\AA & 4.299E-16 & 38575.0 & 5.365E3 \\
         7320\AA & 3.769E-13 & 53063.6 & 3.110E7 
    \end{tabular}
    \caption{Properties for the cooling functions of OII - all values in cgs units.}
    \label{tab:OII_table}
    }
\hfill
\parbox{.45\linewidth}{
\centering
   \begin{tabular}{c|c|c|c}
         Line & $A'$ & $T_{\rm line}$ & $n_c$   \\
         \hline
         52$\mu$m &  3.138E-18  & 277.682 & 2.549E3 \\
         5000\AA & 3.387E-14 & 28728.6 & 9.667E5 \\ 
         166\AA & 6.560E-10 & 86632.4 & 1.476E10 \\
         83.5\AA & 1.752E3 & 172569.7 & 5.406E21 
    \end{tabular}
    \caption{Properties for the cooling functions of OIII - all values in cgs units.}
    \label{tab:OIII_table}
}

\end{table}

\end{document}